\documentclass[numberedappendix, fleqn, twocolumn, twocolappendix]{openjournal}

\usepackage{amsmath}
\usepackage{hyperref}
\usepackage{graphicx}
\usepackage{grffile}
\hypersetup{
  colorlinks=true, 
  linkcolor=blue, 
  citecolor=blue,  
  filecolor=blue, 
  urlcolor=blue}

\newcommand{\vecParam}[1]{\boldsymbol{#1}}
\newcommand{\cf}{cf}
\newcommand{\prob}{P}  
\newcommand{\params}{\vecParam{\theta}} 
 
\newcommand{\data}{\mathbf{d}}
\newcommand{\dataElem}[1]{d_{#1}}
\newcommand{\D}{\textrm{d}}
\newcommand{\eg}{e.g.}
\newcommand{\ie}{i.e.}
\newcommand{\psf}{S}
\newcommand{\Pq}{P_{\mathrm{q}}}
\newcommand{\chiSq}{\tilde{\chi}^2}
\newcommand{\chiSqMax}{{\tilde{\chi}^2_{\max}}}

\newcommand{\chiSqPos}{\tilde{\chi}^2_{\mathrm{pos}}} 
\newcommand{\chiSqMaxPos}{\tilde{\chi}^2_{\mathrm{pos, max}}}
\newcommand{\cutPq}{P^{\mathrm{th}}_{\mathrm{q, min}}}
\newcommand{\cutChiSq}{\tilde{\chi}^{2, \mathrm{th}}}
\newcommand{\cutChiSqMax}{\tilde{\chi}^{2, \mathrm{th}}_{\max}}
\newcommand{\mjy}{\mu {\rm{Jy}}}
\newcommand{\normal}{{\textrm{N}}}
\newcommand{\figcite}{(shown as in Fig.~\ref{fig:fiducial_quasar}).}

\begin{document}

\title{An automated method for finding the most distant quasars
  \vspace*{-15mm}}
\shorttitle{Finding the most distant quasars}

\author{
$\!$Lena Lenz$^{1}$,
Daniel J.\ Mortlock$^{1,2}$\thanks{E-mail: {\tt mortlock@ic.ac.uk}},
Boris Leistedt$^{1}$,
Rhys Barnett$^{1}$
and
Paul C.\ Hewett$^{3}$
}
\affiliation{$^1$Department of Physics, Imperial College London, London SW7 2AZ, United Kingdom \\
$^{2}$Department of Mathematics, Imperial College London, London SW7 2AZ, United Kingdom\\
$^{3}$Institute of Astronomy, University of Cambridge, Madingley Road, Cambridge CB3 0HA, United Kingdom}
\email{mortlock@ic.ac.uk}
\shortauthors{L.~Lenz et al.}

\begin{abstract}
Upcoming surveys such as {\em Euclid}, the Vera~C.~Rubin Observatory's Legacy Survey of Space and Time (LSST) and the {\em Nancy Grace Roman Telescope} ({\em Roman}) will detect hundreds of high-redshift ($z \gtrsim 7$) quasars, but distinguishing them from the billions of other sources in these catalogues represents a significant data analysis challenge. We address this problem by extending existing selection methods by using both i) Bayesian model comparison on measured fluxes and ii) a likelihood-based goodness-of-fit test on images, which are then combined using the $F_\beta$ statistic (where $\beta$ is a parameter which can be tuned to prioritise completeness). The result is an automated, reproduceable and objective high-redshift quasar selection pipeline. We test this on both simulations and real data from the cross-matched Sloan Digital Sky Survey (SDSS) and UKIRT Infrared Deep Sky Survey (UKIDSS) catalogues. On this cross-matched dataset we achieve an area under the curve (AUC) score of up to $0.81$ and an $F_3$ score of up to $0.79$; or, if the completeness is fixed to be 0.9, then we can obtain an efficiency of 0.15. This is sufficient to be applied to the {\em Euclid}, LSST and {\em Roman} data when available.
\end{abstract}


\maketitle


\section{Introduction}
\label{sec:intro}

A quasar is the most luminous form of an active galactic nucleus (AGN), in which gas accretes onto a super-massive black hole (SMBH) of up to $M \simeq 10^{10} M_{\odot}$ \citep{Lynden_Bell_1969, Onken_etal:2020}.  With luminosities of up to $\sim \! 10^{15} L_{\sun}$ \citep{Wolf_etal:2024}, the most powerful quasars can be detected at cosmological distances: the most distant known at present \citep{Banados_etal:2018, Yang_etal:2020, Wang_etal:2021} have unambiguous spectroscopic redshifts of $z \simeq 7.5$, and are seen as they were just $\sim 700$~Myr after the Big Bang\footnote{We adopt the \cite{Planck:2020} best-fit cosmological model throughout.}. There have also been some more indirect identifications of lower luminosity AGNs at higher redshifts of up to $z \simeq 11$ \citep{Bogdan_etal:2024, Maiolino_etal:2024}, but such objects are extremely difficult to study at the level of individual objects, even with the James Webb Space Telescope (JWST).  The most luminous high-redshift quasars constrain the evolution of the first SMBHs \citep{Weedman:1986, Mclure_Dunlop:2004} and act as probes of cosmological reionization (reviewed by, \eg, \citealt{Mortlock:2016}), so are among the most scientifically valuable individual astronomical sources (reviewed by, \eg, \citealt{Fan_etal:2023}).

Unfortunately, luminous high-redshift quasars are extremely rare: there are just $\sim \! 2.5 \times 10^{-2} \, $deg$^{-2}$ quasars with redshifts of $z > 7$ that are brighter than\footnote{We use AB magnitudes \citep{Oke_Gunn:1983} throughout.} $J = 23$, and there may be be only a single quasar at $z \gtrsim 9$ with absolute magnitude brighter than\footnote{We characterise the intrinsic brightness of a quasar in terms of its emission at a rest-frame wavelength of $1450$~\AA.} $M_{1450} < -26$ in the observable Universe \citep{Fan_etal:2019b}.  Moreover, at redshifts of $z \gtrsim 6$ there is sufficient neutral hydrogen (H{\sc{i}}) to completely absorb photons blueward of Ly$\alpha$, rendering all astronomical sources at these distances effectively invisible at optical wavelengths \citep{Gunn_Peterson:1965}.  Even to just detect luminous high-redshift quasars in appreciable numbers hence requires wide-field surveys covering many thousands of square degrees at near-infrared (NIR) wavelengths. By meeting these desiderata the Sloan Digital Sky Survey (SDSS; \citealt{York_etal:2000}) yielded the first $z > 6$ quasar \citep{Fan_etal:2001} and then the UKIRT Infrared Deep Sky Survey (UKIDSS; \citealt{Lawrence_etal:2007}) yielded the first $z > 7$ quasar \citep{Mortlock_etal:2011}.  The next major steps in this field will come from the {\em Euclid} \citep{Laureijs_etal:2011, Mellier_etal:2024} Wide Survey \citep{Scaramella_etal:2022}, which should reach $z \gtrsim 8$ \citep{Barnett_etal:2019}, the {\em Nancy Grace Roman Space Telescope} ({\em Roman}, \citealt{Akeson_etal:2019}), and the Vera~C.~Rubin Observatory's Legacy Survey of Space and Time (LSST, \citealt{Ivezic_etal:2008}), which should complete the census at $z \simeq 7$ \citep{Tee_etal:2023}.

While just detecting high-redshift quasars in significant numbers requires considerable observational resources, this is not useful unless it is also possible to reliably distinguish them from much higher numbers of other astronomical sources in these catalogues\footnote{There is a subtle but important distinction between i) the `detection' of a rare source, which just means that is included somewhere in a (potentially) large catalogue, and ii) its `identification', which implies that its unusual nature has been established with some reasonable confidence.}.  Even though the vast majority of sources detected by surveys like {\em Euclid} and LSST can be trivially rejected from any such search on the basis of their colours, high-redshift quasars are still hugely outnumbered by other astronomical populations with similarly red colours at optical-NIR wavelengths. Most relevant are M, L, T and Y dwarfs in our Galaxy, which have a number density at $J = 23$ of $\sim \! 3.6 $ deg$^{-2}$ \citep{Stern_etal:2007, Wang_etal:2016}, and early-type galaxies (ETGs) at $1 \lesssim z \lesssim 2$, which have a number density at $J = 23$ of $\sim \! 1.4 \times 10^3 $ deg$^{-2}$ \citep{Barnett_etal:2019}. A further difficulty is that non-astronomical artefacts in the images can masquerade as sources with unusual measured properties and so be selected in any search for rare objects. While it is possible to obtain a follow-up spectroscopic observation to definitively confirm or reject any individual candidate, finite observational resources place strong limits on the size of useful candidate lists, motivating the considerable effort that has been put into developing efficient selection techniques\footnote{The focus here is the identification of quasars whose spectral energy distributions (SEDs) are not significantly affected by the presence of dust. The SEDs of the target population. therefore, are understood and possess distinctive characteristics that can be used in selection-schemes.} (reviewed in detail in Section~\ref{sec:classification}).  With {\em Euclid} and LSST expected to catalogue billions of astronomical objects, the selection pipelines will have to be more fully automated than has previously been the case.  Particularly in the case of {\em Euclid}, which will go beyond $z \gtrsim 8$, the presumed decrease in quasar numbers means that traditional astronomical methods such as heuristic colour cuts and visual inspection of images will only be able to play a small part in effective candidate selection procedures.

In this paper we address the problem of high-redshift quasar selection in upcoming wide-field surveys using a combination of traditional (Bayesian) statistical techniques and machine learning (ML) methods.  We start by casting the various existing and potential selection methods in a systematic classification framework (Section~\ref{sec:classification}).  We then describe our selection method, which combines Bayesian model comparison on photometric data and goodness-of-fit criteria applied to images in an automated pipeline (Section~\ref{sec:method}).  To assess the likely utility of this method on future {\em Euclid} and LSST data we test its performance on simulated data (Section~\ref{sec:simulations}) and on real data from SDSS and UKIDSS (Section~\ref{sec:dataset}). We then conclude by discussing the prospects for applying our approach to {\em Euclid}, {\em Roman} and LSST data (Section~\ref{sec:conclusion}).


\section{High-redshift quasar selection as a classification problem}
\label{sec:classification}

The selection of high-redshift quasar candidates from wide-field survey data can usefully be seen as a specific instance of a general classification problem.  The question of how to approach this task can then be divided into separate components: feature selection (Section~\ref{sec:features}); choice of summary quantities/statistics (Section~\ref{sec:statistics}); and optimisation of selection boundaries (Section~\ref{sec:boundaries}).


\subsection{Features}
\label{sec:features}

High-redshift quasars can be distinguished from the more numerous contaminants by exploiting any of several distinct observable signatures (\ie, features):

\begin{itemize}

\item
\textbf{Photometry/colours:}
Quasars have self-similar optical-NIR spectral energy distributions (SEDs) that distinguish them from most other astronomical sources (\eg, \citealt{Temple_etal:2021}).  This is in part due to their intrinsic blue continua and broad emission lines, although at $z \gtrsim 6$ the most distinctive spectral feature is the redshifted Ly~$\alpha$ break \citep{Gunn_Peterson:1965} which makes all such distant sources optical drop-outs, visible only in the NIR.  Optical-NIR spectra can hence provide classification with near certainty, although such observations can only be obtained for hundreds or thousands of targets.  It is hence necessary to use photometric data, at least initially, selecting sources with extremely red optical-NIR colours in filters spanning the (redshifted) Ly~$\alpha$ break.

\item
\textbf{Radio and X-ray emission:}
Some quasars emit significantly at radio wavelengths \citep{Matthews_etal:1963, Sandage_Wyndham:1965, Schmidt:1969} and most emit X-rays \citep{Hopkins_etal:2006}, neither of which is the case for MLTY dwarfs \citep{Williams:2018} and ETGs \citep{Tantalo_etal:2010}.  The fraction of quasars which are radio-loud might be as high as $10$ per cent but decreases with redshift \citep{Wilson_Colbert:1995, Kellerman_etal:1989, Krolik:1999, Padovani:2011, Padovani:2017, Rankine_etal:2021}. Using radio data can significantly increase the efficiency of a search for this subset of the high redshift quasar population \citep{Gloudemans_etal:2022} and the detection of X-rays from optical drop-outs has similarly been used to argue that these sources are high-redshift quasars (\eg, \citealt{Bogdan_etal:2024}).

\item
\textbf{Proper motion:}
High-redshift quasars are at cosmological distances (\ie, many Gpc) and so are effectively stationary on the sky.  The same is true for ETGs, but MLTY dwarfs of comparable magnitudes can have proper motions of up to an arcsec per year, which is detectable if the data (possibly from different surveys) have been taken months or years apart \citep{Lang_etal:2009}.

\item
\textbf{Morphology:}
Quasars outshine their host galaxies by several magnitudes and are sufficiently small intrinsically that, given their distances, they appear as point-sources \citep{Peterson:1997}. This is also the case for Galactic dwarfs but not for ETGs, which have extended profiles and elliptical morphologies and can hence be distinguished on this basis \citep{Lang_etal:2009, Almeida_etal:2007, Barnett_etal:2019}. Also relevant here is that the possibility that high-redshift quasars are multiply-imaged by gravitational lensing \citep{Wyithe_Loeb:2002}: colour-based selection of point-like sources can be biased against the identification of such systems \citep{Fan_etal:2019a}, but their distinctive morphological appearance can be exploited \citep{Byrne_etal:2024}.

\item
\textbf{Sky position:}
While quasars and galaxies are distributed isotropically across the sky, MLTY dwarfs are more common closer to the Galactic plane, where source crowding can also be a problem. Given that the rarity of high-redshift quasars necessitates the use of wide-field surveys, it is impossible to restrict searches to the Galactic poles, so some exploration of more contaminated regions of the sky is inevitable.  While the Galactic latitude (and longitude) cannot be used as a definitive classifier, the probability a candidate is a quasar does depend on position on the sky, particularly for all-sky surveys.

\item
\textbf{Variability:}
Many quasars are variable sources across their entire emission spectrum, fluctuating by up to a magnitude on time-scales ranging from days to years \citep{Smith_Hoffleit:1963, Vanden_Berk_etal:2004, Meusinger_etal:2011, MacLeod_etal:2012}. Changing-look quasars can even change between type I and type II \citep{Ruan_etal:2016, Potts_etal:2021}. The variability in the emission from Galactic dwarfs from star spots has smaller amplitude \citep{Artigau_etal:2009, Goulding_etal:2012} and ETGs are (effectively) constant sources.

\end{itemize}

\noindent
In the method described in Section~\ref{sec:method} below we mainly utilised photometry/colours, proper motion and morphology as surveys like {\em Euclid} and LSST will be most able to exploit these features.


\subsection{Classification methods}
\label{sec:statistics}

Any combination of the above features could be applied in a search for high-redshift quasars, although even for a given feature set there are many different ways to define a classifier.

The most common approach in high-redshift quasar searches is to start with multi-band photometric data from wide-field optical and NIR surveys.  The simplest selection method is to apply heuristic photometric selection based on, \eg, linear cuts in magnitude/colour space \citep{Koo_Kron:1982, Warren_etal:1994, Fan_etal:2001, Richards_etal:2002, Willott_etal:2007, Venemans_etal:2007, Venemans_etal:2013, Banados_etal:2016} and the approach could be considered the traditional default.  Magnitude and colour cuts are easy to reproduce and interpret due to their simplicity.  The specific cuts will depend on the target redshift range, primarily based on the observed wavelength of the redshifted Ly$\alpha$ break.  However, colour cuts can introduce biases in which objects are selected \citep{Wagenveld_etal:2022} and the number/rate of false positives can be unacceptably high \citep{Glikman_etal:2008}.  Further, colour cuts cannot properly differentiate between sources with different photometric uncertainties -- while the imposition of a magnitude limit can play some role here, sources with very different signal-to-noise-ratios, $S/N$, could end up being classified in the same way.  In addition, there is no differentiation between marginal candidates close to the cuts and the more promising objects that would pass any sensible selection criteria \citep{Warren_etal:1987, Mortlock_etal:2012}.

If the different population models and measurement processes were known well then the optimal procedure would be to use Bayesian model comparison to obtain posterior probabilities for each source and then select/rank on this quantity alone.  In practice the population and noise models are known only partially, motivating a more pragmatic approach in which measure of goodness-of-fit (with the quasar predictions) or outlier identification (relative to the dominant contaminating population) is used to approximate the full answer.  \citet{Richards_etal:2004}, \cite{Richards_etal:2009} and \cite{Richards_etal:2015} used kernel density estimation (KDE) to define population models from which they calculated heuristic class probabilities for each source, although measurement uncertainties were included only in an approximate manner.  \citet{Mortlock_etal:2012} adopted parametrised models of the quasar and contaminant populations and included the measurement uncertainties by calculating marginal likelihoods and hence the (posterior) probability that each source is a quasar, an approach adopted and adapted by \cite{Matsuoka_etal:2016}, \cite{Pipien_etal:2018}, \cite{Barnett_etal:2019} and \cite{Barnett_etal:2021}.  Importantly, both these approaches incorporate the fact that non-quasars vastly outnumber the target population to ensure that a source only has a high quasar probability if the data represent sufficient evidence to override the prior default.  \citet{Bovy_etal:2011b} and \citet{Bovy_etal:2012} combined aspects of both these approaches by using extreme deconvolution \citep{Bovy_etal:2011} to specify a data-driven model of each underlying population as a Gaussian mixture model (GMM), a formalism \citet{Nanni_etal:2022} then extended into the high-redshift regime.  Another approach, conceptually intermediate between traditional colour-cuts and probabilistic selection is to use either the likelihood or a goodness-of-fit statistic (\eg, $\chi^2$) with the imbalance between the target and contaminant populations incorporated into the threshold value adopted \citep{Kirkpatrick_etal:2011, Reed_etal:2017, Reed_etal:2019}.  

ML techniques are increasingly and successfully being used as general classifiers in astronomy \citep{Carballo_etal:2008, Yeche_etal:2010, Doert_Errando:2014, Carrasco_etal:2015, Hernitschek_etal:2016, Schindler_etal:2019, Xin_etal:2019, Kang_etal:2019, Bailer-Jones_etal:2019, Ma_etal:2019, Clarke_etal:2020, Wenzl_etal:2021}. However, they come at the cost of acting as `black boxes' whose results can be difficult to interpret; and they can depend very sensitively on arbitrary inputs and imbalances in training sets.  Examples of work-arounds that are potentially applicable to high-redshift quasar searches include: \citet{Bailer-Jones_etal:2008}, who controlled the prior of quasars and their contaminant populations such that they do not have to adapt their training set; and \citet{Wagenveld_etal:2022}, who combined aspects of the methods presented by \citet{Mortlock_etal:2012} and \citet{Bovy_etal:2011b} to obtain a GMM-likelihood using ML methods, while also including informative priors.

The above approaches all focus primarily on making the most effective use of the photometric data, but the other features listed in Section~\ref{sec:features} have also been utilised, albeit mostly just via simple heuristic cuts such as selecting only non-extended sources or radio-loud sources (\eg, \citealt{Gloudemans_etal:2022}). More sophisticated approaches have necessarily been adopted when using multi-epoch data to select according to variability and/or motion \citep{Lang_etal:2009, Richards_etal:2009, Graham_etal:2014, Peters_Richards:2015,  Heintz_etal:2018, Shu_etal:2019}.


\subsection{Selection boundaries}
\label{sec:boundaries}

For fixed values of any tuneable classification parameters, the performance of any binary classification method can be fully characterised in terms of: the number of quasars correctly selected, \ie, true positives, TP; the number of contaminants incorrectly selected, \ie, false positives, FP; the number of contaminants correctly rejected, \ie, true negatives, TN; and the number of quasars incorrectly rejected, \ie, false negatives, FN.  The irrelevant absolute numbers of inputs from the two classes can be factored out by working with fractional summaries, for which standard choices are:

\begin{itemize}

\item
The fraction of selected sources which are true positives, 
\begin{equation}
\,\,\,\,\,\,\,\,\,\,\,\,\,\,\,
  \mathrm{precision} = \mathrm{\frac{TP}{TP + FP}},
\end{equation}
also known as the true positive rate (TPR) and, in astronomy, as efficiency.

\item
The fraction of target sources which are selected, 
\begin{equation}
\,\,\,\,\,\,\,\,\,\,\,\,\,\,\,
  \mathrm{recall} = \mathrm{\frac{TP}{TP + FN}},
\end{equation}
also known as sensitivity and, in astronomy, as completeness.

\item
The fraction of contaminants which are selected, 
the false positive rate,
\begin{equation}
\,\,\,\,\,\,\,\,\,\,\,\,\,\,\,
  \mathrm{FPR} = \mathrm{\frac{FP}{FP + TN}}.
\end{equation}

\end{itemize}

A good classifier would have both precision/TPR and recall close to 1, and FPR close to 0, although in all but the most trivial cases there is some trade-off between these desiderata.  For a classifier with one or more free parameters there is freedom to tune the performance; but, with no generic way to balance these considerations, this can only be done by accounting for the specifics of the problem.  In high-redshift quasar searches this comes down to an inevitably subjective assessment of the scientific value of a new (\eg, record-breaking) discovery and the available observational resources for follow-up observations.  In more traditional searches this has been set by somewhat arbitrary choices of magnitude and colour cuts; and even when taking a more principled statistical approach there is still a choice required to set the minimum quasar probability or maximum chi-squared value.  One result of this ambiguity is a lack of systematic comparisons between the (very) different selection methods, and even studies which focussed on the selection methods (\eg, \citealt{Barnett_etal:2019, Barnett_etal:2021, Nanni_etal:2022}) have typically used fixed threshold choices.

Here we aim to take a more systematic approach by adopting two techniques from the statistics/ML literature: receiver operating characteristic (ROC) curves (Section~\ref{sec:roc}, and already common in astronomy); and the $F_\beta$ performance measure (Section~\ref{sec:fbeta}).


\subsubsection{ROC curves}
\label{sec:roc}

The standard graphical method of illustrating the performance of a classifier with a tuneable parameter (here a magnitude limit, $S/N$ cut, or probability threshold) is a ROC curve, the locus of (FPR, TPR) values for different values of the parameter.  (Examples of these appear below in Section~\ref{sec:dataset}.) The most common single numerical summary is the area under the curve (AUC), with a perfect classifier having an AUC of 1 and a poor classifier an AUC of 1/2, corresponding to a random or arbitrary classification.  ROC curves and AUC summaries have been used across astronomy, and represent a better way to characterise high-redshift quasar selection methods.

However, in any situation in which the aim of the classification procedure is to separate a small number of rare/valuable objects from more numerous contaminants the AUC is not necessarily useful: as the number of positive samples in such a data-set is very small, classifying nearly every candidate as a contaminant (negative label) would give a high AUC, essentially independent of the actual performance of the classifier.  A more subtle metric than the ROC curve and the AUC is hence needed in this situation.


\subsubsection{F$_\beta$ performance measure}
\label{sec:fbeta}

One way of ensuring that a classifier assessment prioritises recall is by using the $F_\beta$ performance measure \citep{Baeza-Yates_Ribeiro-Neto:2011}, defined as 
\begin{equation}
  F_{\beta} = (1 + \beta^2 ) \, 
  \frac
  {\mathrm{precision} \times \mathrm{recall}}
  {\beta^2 \times \mathrm{precision} + \mathrm{recall}},
\end{equation}
where recall is prioritised more strongly for higher values of $\beta$.  A perfect classifier has an $F_{\beta} = 1$, like the AUC; but if the performance is poor then $F_{\beta} \rightarrow 0$, not 1/2, as for the AUC.  For $\beta=1$, the equation above becomes the harmonic mean between precision and recall, 
\begin{equation}
    F_1 = \frac{2}{1/\mathrm{precision} + 1/\mathrm{recall}}. 
\end{equation}
We will focus on $\beta > 1$, but will still include $\beta=1$ for reference, so typically show results for $\beta = 1$, $\beta = 2$ and $\beta = 3$.


\section{High-redshift quasar selection method}
\label{sec:method}

\begin{figure*} 
\centering
\includegraphics[trim={2.75cm 1.1cm 2.2cm 1.5cm}, clip=True]{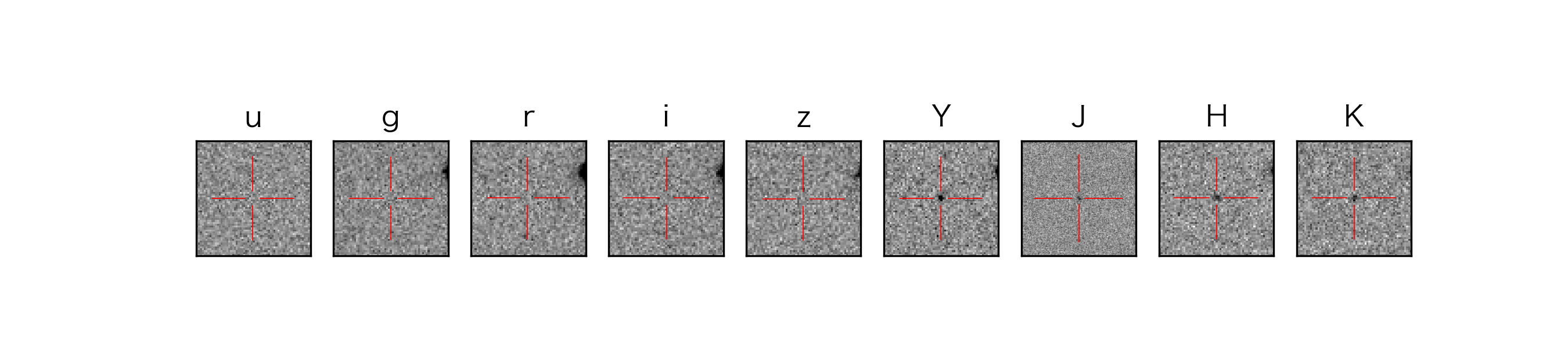}
\includegraphics[trim={2.75cm 1.3cm 2.2cm 2.0cm}, clip=True]{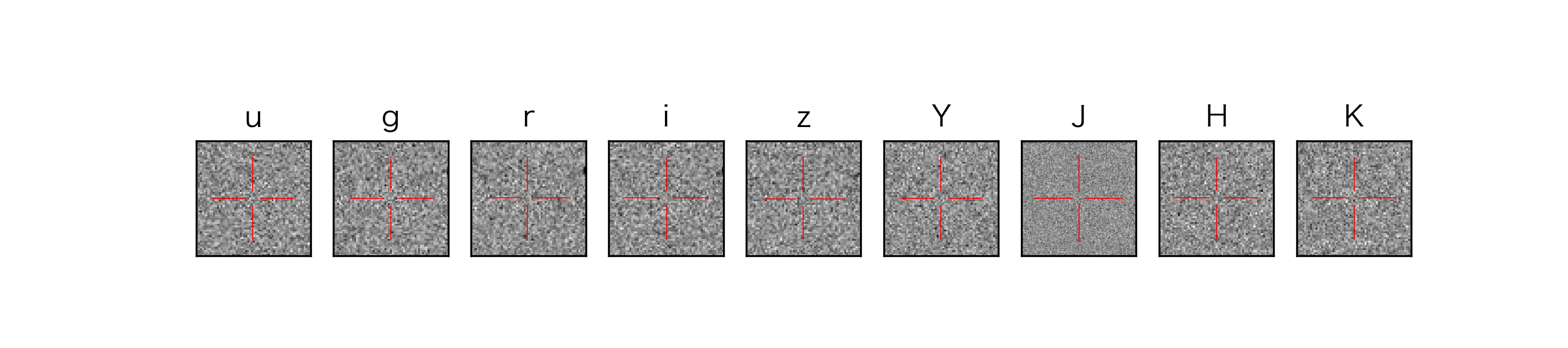}
\caption{The redshift $z = 7.09$ quasar J112001.48+064124.3 as measured by SDSS ($u$, $g$, $r$, $i$ and $z$) and UKIDSS ($Y$, $J$, $H$, $K$) in the top row and with the best-fit quasar model subtracted in the bottom row.  North is up and east is to the left; the images are $l_{\rm plot} = 24\farcs4$ along a side.}
\label{fig:fiducial_quasar}
\end{figure*}

\begin{table}
\renewcommand{\arraystretch}{1.4}
\centering
\caption{Approaches used to select high-redshift quasars.}
\label{tab:statistical_methods_summary}
\begin{tabular}{|c|c|c|c|}
    \hline
    data & features exploited & output & Eq. \\
    \hline
    \hline
    catalogues & photometry; sky position & $\Pq$ & \ref{eq:p_model_i} \\
    \hline
    images & morphology; motion; variability & $\chiSq$ & \ref{eq:chi_squared} \\
    \hline
\end{tabular}
\end{table}

\begin{figure}
    \centering
\includegraphics[width=7cm]{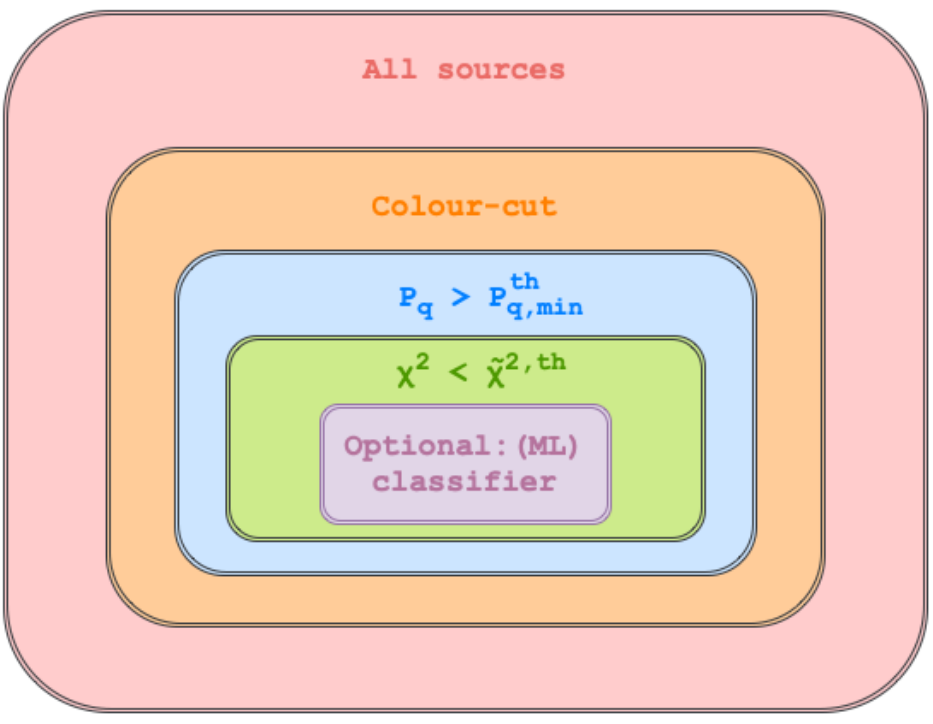}
    \caption{The sequential candidate selection process.}
    \label{fig:sequential_select}
\end{figure}

Our aim here is to develop a high-redshift quasar selection method that automatically exploits as many as possible of the data features described above in Section~\ref{sec:classification}, as well as to formalise this more strongly as a classification problem. As the particular focus is on {\em Euclid}, {\em Roman} and LSST, we consider the data that such surveys will provide for all sources: pixelized images obtained using broad-band optical and NIR filters.  It is implicit that the images will, between them, have sufficient optical and NIR coverage to bracket the Ly$\alpha$ break over the redshift range of interest (but if this condition is not satisfied then the result will simply -- and correctly -- be that no sources are selected as candidate high-redshift quasars).  The upper row of Fig.~\ref{fig:fiducial_quasar} shows a fiducial example of such data:  $u$, $g$, $r$, $i$ and $z$ images from SDSS and $Y$, $J$\footnote{Due to microstepping \citep{Dye_etal:2006} the pixel scale of the UKIDSS $J$-band images is half that of the UKIDSS $Y$, $H$ and $K$ images.}, $H$ and $K$ images from UKIDSS of the redshift $z = 7.09$ quasar J112001.48+064124.3.  Of the features listed in Section~\ref{sec:features} such images are sufficient to exploit photometry/colours, morphology and sky position; and, if there are observations at different epochs (not in general the case for the {\em Euclid} Wide Survey, but relevant the {\em Euclid} Deep Fields and for LSST), also proper motion and variability.  

It is only the radio and/or X-ray emission -- or the lack thereof -- which is completely inaccessible without cross-matching to surveys at these wavelengths/energies.  Such external surveys have been used in many high-redshift quasar searches (\eg, \citealt{McGreer_etal:2006, Banados_etal:2021, Gloudemans_etal:2022, Bogdan_etal:2024}), but this inevitably means some reduction in the effective search area.  As the focus here is a generic selection procedure that can be uniformly applied to an entire sources catalogue from, \eg, {\em Euclid} or LSST, we only consider data from the primary survey here.

Ideally, we would use a Bayesian approach in which the target quasars and the contaminant populations were compared at the pixel-level across all available observations \citep{Lang_etal:2009}, but this would require detailed models for the colours, morphology, motion and variability of all populations (including non-astrophysical artefacts), which is extremely difficult to define fully (\eg, \citealt{Anders_etal:2006}).  Hence, we (inevitably) also use catalogue-level data, \ie, photometry and astrometry of detected sources.  This latter qualification is important as it presumes that i) detection is done in the NIR and ii) for optical non-detections forced photometry is obtained (not just upper limits, \cf\ \citealt{Mortlock_etal:2012,Nanni_etal:2022}).  This is particularly relevant for {\em Euclid} given that its optical $I_E$ images will be deeper than its NIR images, implying optical detection as the natural default; details of {\em Euclid}'s NIR-driven catalogue generation methodology is given Section~7.4 of \cite{Mellier_etal:2024}.

While working with a single data model for the images (Section~\ref{sec:data_model}), we adopt a compromise by calculating the two quantities outlined in in Table~\ref{tab:statistical_methods_summary}: the probability that a source is a quasar evaluated on photometric catalogue data (Section~\ref{sec:bayesian_model_comparison}); and a goodness-of-fit statistic from pixel-level image data (Section~\ref{sec:pixel}). These summaries are then combined by defining a selection threshold in this space (Section~\ref{sec:threshold}). These statistical algorithms are then embedded in a survey-specific end-to-end high-redshift quasar pipeline that goes from input coordinates to these summary outputs (Section~\ref{sec:pipeline} and Fig.~\ref{fig:sequential_select}).


\subsection{Data model}
\label{sec:data_model}

For any given candidate the data we consider is a set of $B > 1$ small, square `postage stamp' images that are (approximately) centred on the source of interest.  These are constructed to have the same angular side-length, $l$, although the exact geometry is not particularly important, provided the images are large enough to allow a reliable background measurement.  The images are indexed by $b \in \{1, 2, \ldots, B\}$, which implicitly indicates the passband of the image.  Several images could have been obtained with the same filter, either due to overlaps (relevant for all of SDSS, UKIDSS, the {\em Euclid} Wide Survey and {\em Roman}) or planned repeat observations (SDSS Stripe 82, the {\em Euclid} Deep Fields and LSST).  We assume the noise is independent between different observations, so our data model can be specified in terms of individual images (Section~\ref{sec:data_model_images}) and single-image flux measurements (Section~\ref{sec:data_model_fluxes}).


\subsubsection{Images}
\label{sec:data_model_images}

The pixels in image $b$ have an angular side-length of $\theta_{\textrm{px},b}$ and the number of pixels along each side of the postage stamp, $N_b = l / \theta_{\textrm{px},b}$, is hence potentially band-dependent as well. 

The point-spread function (PSF) is assumed to either be known from external meta-data or fit as part of the pipeline (Section~\ref{sec:pipeline}), in which case we adopt a \cite{Moffat:1969} profile unless a survey explicitly provides a different model. The radial PSF, $\psf_b(r)$, is defined such that the the contribution to the flux in pixel $(i, j)$ from a point-source of flux $F_{s,b}$ at angular position $(x_s, y_s)$ is given by integrating over the pixel area to obtain
\[
f_{s,b,i,j} 
\]
\[
=
F_{s,b}
  \int_{x_i - \theta_{\textrm{px},b}/2}^{x_i + \theta_{\textrm{px},b}/2} \!\!\!\!\!\!\!\!\!\!\! \D x 
  \int_{y_j - \theta_{\textrm{px},b}/2}^{y_j + \theta_{\textrm{px},b}/2} \!\!\!\!\!\!\!\!\!\!\! \D y 
  \,\,\, \psf_b\!\!\left[\sqrt{(x-x_s)^2 + (y-y_s)^2} \, \right]  
\]
\begin{equation}
  \simeq F_{s,b} \, 
  \psf_b\!\!\left[\sqrt{(x_i-x_s)^2 + (y_j-y_s)^2} \, \right] 
  \, \theta_{\textrm{px},b}^2 , 
\label{eq:pixel_flux_approx}
\end{equation}
where $(x_i, y_j)$ is the angular position of the pixel centre.  The total signal in a pixel is the sum of the contributions from all $S$ (point-)sources in/near the field, along with a constant band-specific background $\mu_b$ to give
\begin{equation}
\label{eq:summed_sources}
f_{b,i,j} = \mu_b \, \theta_{\textrm{px},b}^2 + \sum_{s = 1}^S f_{s,b,i,j}
\end{equation}
\[
\simeq 
  \left\{
  \mu_b
  + 
  \sum_{s = 1}^S
  F_{s,b} \,
  \psf_b\!\left[\sqrt{(x_i-x_s)^2 + (y_j-y_s)^2} \, \right] 
  \right\} \theta_{\textrm{px},b}^2.
\]
While the focus is primarily on point-sources, we explore the possibility of ETGs with detectable extension \citep{Barnett_etal:2019} using a modified effective radial profile in Section~\ref{sec:simulations}.

Any actual measured image is, of course, affected by a range of different noise-like processes, including Poisson photon-counting noise, thermal fluctuations, detector read-out noise, telluric emission (\eg, \citealt{Bernstein:2002,Le_etal:2015}).  Any of these effects could in principle be modeled and included in the formalism, but the sources under consideration here are typically detected at only moderate $S/N$ in the NIR (and not at all in the optical images), meaning the dominant contribution is fluctuations in the observed background.  The data model for the measured flux in pixel $(i,j)$ is hence taken to be
\begin{equation}
\label{eq:noise_model}
    d_{b,i,j} = f_{b,i,j} + n_{b,i,j},
\end{equation}
where $f_{b,i,j}$ is given in Eq.~\ref{eq:summed_sources} and $n_{b,i,j}$ is the noise in the pixel. Assuming the noise follows a normal distribution\footnote{Here $\normal(x; \mu, \sigma^2) = \exp[-1/2 \, (x - \mu)^2 / \sigma^2] / (2 \pi \, \sigma)$ is a normal density in $x$ with mean $\mu$ and variance $\sigma^2$.} with variance $\sigma_{\textrm{px},b}^2$ and is independently and identically distributed (i.i.d.)\ for all pixels in a given image, the likelihood of the full image data $\data_b$ is 
\begin{equation}
\prob(\data_b | F_{1:S}, x_{1:S}, y_{1:S}, \mu_b)
= 
\prod_{i = 1}^{N_b}
\prod_{j = 1}^{N_b}
\normal\!\left(d_{b,i,j}; f_{b,i,j}, \sigma_{\textrm{px},b}^2\right).
\label{eq:image_likelihood}
\end{equation}
This likelihood forms the basis for the goodness-of-fit test described in Section~\ref{sec:pixel} as well as the image simulations used in Section~\ref{sec:simulations}, although with estimated/reported values $\hat{\sigma}_{\textrm{px},b}$ and $\hat{\mu}_b$ used in place of the (unknowable) true values.

None of the algorithms described below rely on specific functional forms (\eg, for fast numerical evaluations) so, if justified by the properties of a particular data-set, it would be possible to extend the above data model by, \eg, including photon noise from bright sources or a heavy-tailed distribution for the background pixel noise to be more robust to outliers.  This would require a semantic change, as the intuitive $\chi^2$ statistic used in Section~\ref{sec:pixel} would have to be replaced with a scaled log-likelihood.


\subsubsection{Fluxes}
\label{sec:data_model_fluxes}
 
The first stages of the selection sequence illustrated in Fig.~\ref{fig:sequential_select} use measured fluxes (or, equivalently, magnitudes) rather than images.  In most cases these will have been provided in the catalogues produced by the survey pipelines, but where necessary we perform forced photometry (Section~\ref{sec:pipeline}).  Either way, the measured $b$-band flux, $\hat{F}_b$, and its associated uncertainty, $\sigma_b$, can be seen as summary statistics of the full image-level data\footnote{We use reported fluxes as a data summary, rather than as an estimator, so the bias identified by \cite{Portillo_etal:2020} is not relevant here (although the magnitude of this effect would anyway be too small to have a significant impact here).}.  Assuming these values come from PSF-weighted sums of the pixel values, the assumption of Gaussian noise propagates through to the fluxes, yielding a normal sampling distribution of the form
\begin{equation}
\label{eq:normal_flux}
\prob(\hat{F}_b | F_{\textrm{s}}, \sigma_b)
  = \normal(\hat{F}_b; F_{\textrm{s}}, \sigma_b^2).
\end{equation}
This is used explicitly as the likelihood in Eq.~\ref{eq:likelihood_fluxes}, although (again) using estimated/reported values $\hat{\sigma}_b$ for the uncertainty. This distribution could (also again) be replaced with a more robust heavy-tailed distribution if it were required.


\subsection{Catalogue-level Bayesian model comparison}
\label{sec:bayesian_model_comparison}

In an astronomical context, Bayesian model comparison \citep{Burnham:2002, Sivia_Skilling:2006, Neath_etal:2017} is a way of probabilistically classifying a source as being a member one of several different astronomical populations, properly taking into account absolute numbers and demographics of the populations, along with the measurement uncertainties.  This methodology was initially applied to high-redshift quasar searches using catalogue-level flux data by \cite{Mortlock_etal:2012} and then refined by \cite{Barnett_etal:2019} and \cite{Barnett_etal:2021} and made publicly available as the Python package {\tt Pq\_server}\footnote{\url{https://github.com/rhysrb/Pq_server}}. We adopt this method here, adding more filters and removing the requirement for a $J$-band detection with a view to fully exploring the upcoming {\em Euclid} and LSST data.

Our starting point is that, as described in Section~\ref{sec:data_model}, for each candidate we have measured fluxes, $\hat{F}_{1:B} = (\hat{F}_1, \hat{F}_2, \ldots, \hat{F}_B)$, and estimated/reported uncertainties, $\hat{\sigma}_{1:B} = (\hat{\sigma}_1, \hat{\sigma}_2, \ldots, \hat{\sigma}_B)$, in each of $B$ images.  We consider three populations: high-redshift quasars, q; MLTY dwarfs, s; and ETGs, g, each of which is defined by a surface density $\Sigma_t(\params_t)$ as a function of its internal parameters, $\params_t$ (where $t \in \{\textrm{q}, \textrm{s}, \textrm{g}\}$).  We base our population models on those described in \cite{Mortlock_etal:2012}, \cite{Barnett_etal:2019} and \cite{Barnett_etal:2021}, but with several refinements as detailed in Appendix~\ref{sec:models}.

The probability that a source is a quasar can then be written as \citep{Mortlock_etal:2012, Barnett_etal:2019}
\begin{align}
\label{eq:p_model_i}
\Pq 
  & = 
  \prob(t = \textrm{q} | \hat{F}_{1:B}, \params_\textrm{q}, \params_\textrm{s}, \params_\textrm{g})
  \nonumber \\
  & = \frac{W_{\textrm{q}}(\hat{F}_{1:B}, \params_\textrm{q})}
  {W_{\textrm{q}}(\hat{F}_{1:B}, \params_\textrm{q}) + W_{\textrm{s}}(\hat{F}_{1:B}, \params_\textrm{s}) + W_{\textrm{g}}(\hat{F}_{1:B}, \params_\textrm{g})},
\end{align}
where the weight associated with population $t$ is given by integrating over all the parameter values as
\begin{equation}
\label{eq:likelihood_bayes}
  W_t(\hat{F}_{1:B}, \params_t) = 
  \int \D \params_t \, 
  \Sigma_t(\params_t) \, 
  \prob (\hat{F}_{1:B} | \params_t, t)
  ,
\end{equation}
with $\prob (\hat{F}_{1:B} | \params_t, t)$ the likelihood.
Given the data model described in Section~\ref{sec:data_model}, and in particular the assumption of i.i.d.\ normally distributed noise (Eq.~\ref{eq:normal_flux}), the likelihood is 
\begin{equation}
\label{eq:likelihood_fluxes}
\prob (\hat{F}_{1:B} | \params_t, t)
  = \prod_{b = 1}^B
  \normal\left[
  \hat{F}_b; 
  F_b(\params_t, t), 
  \hat{\sigma}_b^2
  \right],
\end{equation}
where $F_b(\params_t, t)$ is the (true) flux in band $b$ of a source of type $t$ with parameters $\params_t$.  We numerically evaluate the integrals in Eq.~\ref{eq:likelihood_bayes} by evaluating the integrand on a grid of parameter values and then combining these using Simpson's rule.  In the process of numerically evaluating the integral in Eq.~\ref{eq:likelihood_bayes} we also get best-fit values of parameters under each population model, $\hat{\params}_t$, which we then use in the image-level goodness-of-fit test described in Section~\ref{sec:pixel}.


\subsection{Image-level goodness-of-fit}
\label{sec:pixel}

A high value of $\Pq$ calculated as above means only that the quasar model explains the photometric data better than the explicitly considered alternatives (\ie, MLTY dwarfs and ETGs). It does not, however, imply that the quasar hypothesis is a good fit to the data in absolute terms, particularly at the image level.  While it is implausible to specify a full set of image-level models for the various contaminant populations -- which would have to also include non-astrophysical artefacts -- the fact that high-redshift quasars are stationary point-sources makes it possible to use this as an effective `null hypothesis' for a goodness-of-fit test on the images themselves.  

In terms of the data model described in Section~\ref{sec:data_model} this null hypothesis is that there is a high-redshift quasar centred in each of the $B$ available images.  For the $b$'th image this is a point-source at position $(x, y) = (0, 0)$ with flux $F_b = F_b(\hat{\params}_{\textrm{q}}, \textrm{q})$ as implied by the best-fit model parameters $\hat{\params}_{\textrm{q}}$ (Section~\ref{sec:bayesian_model_comparison}) and with a profile given by the PSF, $S_b(r)$.  Given an estimated background $\hat{\mu}_b$, Eq.~\ref{eq:image_likelihood} yields the likelihood as 
\begin{equation}
\label{eq:likelihood_model}
\prob(\data_b | \hat{\params}_{\textrm{q}}, \textrm{q}, \hat{\mu}_b)
\end{equation}
\[
= \! \prod_{i,j}
\normal\!\left\{d_{b,i,j}; \! 
  \left[ F_b(\hat{\params}_{\textrm{q}}, \textrm{q}) \, S_b\!\!\left(\sqrt{x_i^2 + y_j^2}\!\right) \!
  + \hat{\mu}_b\right] \! \theta_{\textrm{px},b}^2, 
  \hat{\sigma}_{\textrm{px},b}^2\right\}\!,
\]
where $\theta_{\textrm{px},b}^2$ is the angular area of each pixel and $\hat{\sigma}_{\textrm{px},b}$ is the reported/estimated pixel noise level.  We include only pixels in a small annulus around the source (see Section~\ref{sec:pipeline}), so the product is, in general, not over the entire $N_b \times N_b$ postage stamp image but over a smaller region with $N$ pixels.  Exploiting the Gaussian form of the likelihood, we then simplify this to a $\chi^2$ statistic by taking the (natural) logarithm of Eq.~\ref{eq:likelihood_model} and omitting constant terms to obtain
\begin{align}
\chi^2_b 
  & = 
  - 2 \, \log \left[\prob(\data_b | \hat{\params}_{\textrm{q}}, \textrm{q}, \hat{\mu}_b) \right] - \textrm{constant}
  \\
  & = \! \sum_{i, j} \! \left\{ \!
    \frac{d_{b,i,j} 
    -
    \left[ F_b(\hat{\params}_{\textrm{q}}, \textrm{q}) \, S_b\!\left(\sqrt{x_i^2 + y_j^2}\right)
  + \hat{\mu}_b\right] \theta_{\textrm{px},b}^2}
  {\hat{\sigma}_{\textrm{px},b}} 
  \! \right\}^{\!\!2} \!\!\!
  \nonumber
  .
\end{align}
(If a different noise model were adopted then this would have to be retained as a more general log-likelihood; while the mathematical form of the equivalent expression might be more complicated, this would not have any algorithmic or computational implications.)  Then dividing through by the number of pixels in the annulus gives our first test statistics, the band-specific reduced $\chi^2$, given by
\begin{equation}
\label{eq:chi_passband}
\chiSq_b = \frac{1}{N} \, \chi^2_b.
\end{equation}

Rather than working with the full set of $B$ statistics, we extract several key quantities which we have found to be most useful.  First, we average across all images/bands to produce our second, highest level goodness-of-fit statistic,
\begin{equation}
\label{eq:chi_squared}
    \chiSq = \frac{1}{B} \, \sum_{b=1}^B \chiSq_b,
\end{equation}
which is the single best summary for many astrophysical contaminants which diverge from the quasar prediction in all bands (\eg, due to being slightly resolved).  We also calculate the maximum $\chi^2_b$ value, $\chiSqMax = \max (\chi^2_{1:B})$ in order to retain sensitivity to cases where most images are consistent with the null model but there is one which is a poor fit (as is more commonly the case with non-astrophysical artefacts).

We also repeat the above analysis for a quasar position that is fit to the data in the images rather than using catalogued positions.  In cases with near neighbours or some evidence of motion the resultant statistics, $\chiSqPos$ and $\chiSqMaxPos$, can differ significantly.


\subsection{Selection threshold}
\label{sec:threshold}

We can apply thresholds to both the flux-level probabilities and the goodness-of-fit statistics to define our final selection.  In Section~\ref{sec:sim_results}, we test which combination of thresholds gives the best results on our cross-matched dataset.


\subsection{Data processing pipeline}
\label{sec:pipeline}

\begin{figure}
    \centering
\includegraphics[scale=0.27]{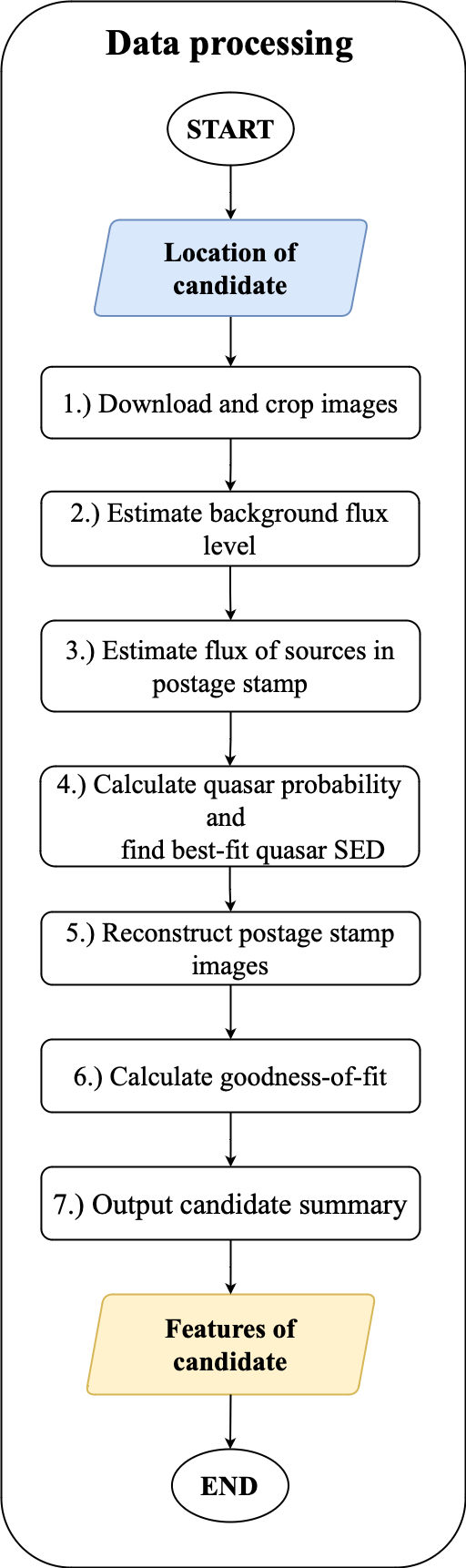}
    \caption{Flowchart of the data processing pipeline; the steps are described in more detail in Section~3.5.}
    \label{fig:flowchart}
\end{figure}

\begin{table*}
    \centering
    \renewcommand{\arraystretch}{1.4}
    \caption{Tuneable parameters for the data processing pipeline described in Section~\ref{sec:pipeline}.}
    \label{tab:code_options}
    \begin{tabular}{|c|c|c|}
    \hline
    parameter & description & SDSS-UKIDSS default value \\
    \hline
    \hline
    $l$ & angular size of each postage stamp & $34\farcs8$ \\
    \hline
    $l_{\text{plot}}$ & angular size of plotted postage stamp image & $24\farcs4$ \\
    \hline
    $R_{\mathrm{clip}}$ & angular radius for sigma-clipping & $\,\,8\farcs4$ \\
    \hline
    $\alpha$ & sigma-clipping level & $\,\,\,\,3.0$   \\
    \hline
    $R_{\mathrm{keep}}$ & angular radius within which no other source is reconstructed & $\,\,1\farcs2$ \\
    \hline
    $R_{\mathrm{flux}}$ & angular radius for fitting the flux & $\,\,2\farcs6$ \\
    \hline
    $R_{\chiSq}$ & angular radius for calculating $\chiSq_b$ & $\,\,1\farcs2$ \\
    \hline
    \end{tabular}
\end{table*}

The high-redshift quasar selection method described above is implemented in a multi-stage pipeline which, for a single candidate, goes from the sky coordinates through to summary quantities.  This pipeline could in principle be run on every catalogued source, but in practice we first make a pre-selection based on conservative magnitude and colour cuts which reject the large fraction of sources with measured photometry completely unlike the target quasars. This pipeline broadly follows the steps taken by standard astronomical image analysis software such as {\tt SourceExtractor} \citep{Bertin_Arnouts:1996} and {\tt PhotUtils} \citep{Bradley_etal:2020} but allows the low-level control required for this particular analysis (which is quite different from making large-scale reliable catalogues). While the implementation is inevitably survey-specific, with several tuneable parameters (Table~\ref{tab:code_options}), the data-processing sequence is generic.  The main steps are illustrated in Fig.~\ref{fig:flowchart} and outlined below, with the default values of the various tuneable quantities as appropriate for the SDSS-UKIDSS data analysed in Section~\ref{sec:dataset}.


\subsubsection{Download and crop images}
\label{sec:download}

Given the basic input of a position on the celestial sphere (\ie, right ascension and declination), we query the relevant survey databases to obtain all available images that include this location.  Images are discarded if specific error flags are set that indicate any data processing issues. This process is specific to each survey/database, but is more a technical than scientific task.

We then cut out the images to postage stamps of size $l \times l$ pixels, typically with a size of $\sim \! 30$\,arcsec (Table~\ref{tab:code_options}). The reason we use such a relatively large area is so we can reconstruct sources near the target in case they are contributing additional flux. (The postage stamp images used in the plots are typically smaller than this so that it is possible to visually resolve individual pixels when undertaking visual inspection.) 

While it is expected that artefacts such as satellite trails and diffraction spikes will have been flagged or removed by the low-level survey data analysis pipelines, rare object searches inevitably include the small fraction of these which are missed.  So as a precaution we identify -- and hence reject -- images which contain such artefacts by using the \cite{Canny:1968} edge detection filter and then applying a \cite{Hough:1959} line transform, implemented as {\tt probabilistic\_hough\_line} from the Python {\tt sklearn} library.  This is to detect and reject images that likely contain artefacts.


\subsubsection{Estimate background flux level}
\label{sec:background}

To estimate of the local background flux level, $\hat{\mu}_b$, we mask known sources and then apply sigma-clipping to remove outliers.  We adopt an iterative algorithm in which we calculate the implied background from the $N$ retained pixels as $\hat{\mu}_b = \sum_{i,j} d_{b,i,j} / (N \, \theta_{\textrm{px},b}^2)$ and then remove all pixels for which $|f_{b,i,j} - \hat{\mu}_b \, \theta_{\textrm{px},b}| > \alpha \, \sigma_{\textrm{px},b}$, with $\alpha = 3$ the default clipping level.  (Technically, $\sigma_{\textrm{px},b}$ must also be estimated from the data but it is assumed that any uncertainty is negligible, so it is treated as a known quantity.)  The algorithm is applied on a circular region, centred on the target, of angular radius $R_{\mathrm{clip}}$, for which the default value is $8\farcs4$.  We use this radius as it gave the best classification results: large enough to not be dominated by the central source but small enough that we retrieve the local background level.  This works well except if there is a bright source that takes up a large part of the postage stamp. In this case we increase the area for the background calculation until $80$~per~cent of the available area is identified as part of the background by the sigma-clipping algorithm, \ie, until the majority of the contained pixels represents the background.


\subsubsection{Estimate flux of sources in postage stamp}
\label{sec:fluxest}

We require an estimate of the fluxes of all sources within the postage stamp in order to subtract off their contributions in the next step.  In most cases these are provided by the input surveys, but the initial selection of sources with unusual colours means that there is an over-representation of cases with undetected faint sources or unresolved companions, so we fit these directly (\ie, effectively performing forced photometry).

We fit for the flux and/or position of such sources by considering a circular area of angular radius $R_{\mathrm{flux}}=2\farcs4$ and then adopt a least-squares estimate of the flux and/or the source position using the PSF from above and the pixel likelihood given in Eq.~\ref{eq:image_likelihood}.  For sources in the postage stamp at radius $r > R_{\text{keep}}$, we only fit the flux, whereas for the central source of interest we separately fit the flux and the position to account for potential proper motion between different observations.

Following on from the data model, the maximum likelihood estimate for the flux for a point-source at known position $(x_s, y_s)$ is given by
\begin{equation}
\hat{F}_b 
  = 
  \frac{ 
  \sum_{i, j} \psf_{b,i,j} \, \theta_{\text{px},b}^2 \, ( \dataElem{b,i,j} - \hat{\mu}_b \,\theta_{\text{px},b}^2 ) 
  }
  {\sum_{i, j}  \psf_{b,i,j}^2 \, \theta_{\text{px},b}^4 },
\end{equation}
where $S_{b,i,j} = \psf_b[ \sqrt{(x_i-x_s)^2 + (y_j-y_s)^2}]$ is the value of the (assumed known) band $b$ PSF at the centre of the pixel $(i, j)$.  For the position and the combination of flux and position, we use a non-linear least-squares fit by comparing the observed pixel data to the generated pixel data as a function of flux and/or position.


\subsubsection{Calculate quasar probability and find best-fit quasar SED}
\label{sec:pqcalc}

Using the measured fluxes from the input catalogue(s), if available, or from our own fits we calculate the probability that the source is a high-redshift quasar, $\Pq$, along with the best-fit quasar parameters, $\hat{\params}_{\textrm{q}}$, as described in Section~\ref{sec:bayesian_model_comparison}.


\subsubsection{Reconstruct postage stamp images}
\label{sec:reconstruct}

Using the best-fit quasar model we now reconstruct each postage stamp image as in Eq.~\ref{eq:pixel_flux_approx} with $S$ sources in the postage stamp, along with photometry and positions for any other sources in the image.  We do not reconstruct sources that are within a radius of $R_{\mathrm{keep}}$ from the image centre to avoid obtaining falsely good fits to the quasar model.

An illustration of this procedure working in an almost ideal case is shown in Fig.~\ref{fig:fiducial_quasar}, which shows that the residual after subtracting the best-fit model of J1120$+$0641 is consistent with uncorrelated Gaussian noise across the pixels.


\subsubsection{Calculate goodness-of-fit}
\label{sec:gof}

The individual band-by-band reduced $\chiSq_b$ (Eq.~\ref{eq:chi_passband}) and the overall reduced average $\chiSq$ (Eq.~\ref{eq:chi_squared}) values are calculated as described in Section~\ref{sec:pixel}. We do this both with the quasar assumed to be at the reference location on the image and at the best-fit position (\ie, mean image centroid).  We calculate the goodness-of-fit starting from the centre of the postage stamp over a circular region of radius $r= 1\farcs2$. If the radius is too large, the resulting $\chi^2$ value is overly influenced by the background pixels and hence is less informative; if the radius is too small, it does not encompass the central source and, in particular, becomes insensitive to any change in position.

By not reconstructing sources at $r \leq R_{\text{keep}}$, we obtain large reduced chi-squared values for moving sources and for sources that are in the immediate vicinity of the target.
Blended or merged sources will also result in large reduced chi-squared values since the point-source model will not be a good fit.


\subsubsection{Output candidate summary}
\label{sec:features_summary}

\begin{table}
    \centering
    \renewcommand{\arraystretch}{1.4}
    \caption{Summary outputs from the data processing pipeline.}
    \label{tab:stored_info}
    \begin{tabular}{|c|c|}
    \hline
    quantity & definition \\ 
    \hline
    \hline
    RA & right ascension (J2000) \\
    \hline
    dec & declination (J2000) \\
    \hline
    $\chiSq$ & residual chi-squared $\eqref{eq:chi_squared}$ \\
     \hline
    $\chiSq_{1:B}$ & chi-squared in all bands $\eqref{eq:chi_passband}$\\
    \hline
    $\chiSqMax$ & maximum chi-squared\\
    \hline
    $\chiSqPos$ & $ \chiSq $ with fitted source centroid \\
    \hline
    $\chiSqMaxPos$ & $\chiSqMax$ with fitted source centroid \\
    \hline
    $\hat{z}$ & best-fit quasar redshift estimate \\
    \hline
    $\Pq$ & quasar probability \eqref{eq:p_model_i}\\
    \hline
    $F_{1:B}$ & PSF flux in all bands (Jy) \\
    \hline
    $\sigma_{F,1:B}$ & PSF flux uncertainty in all bands (Jy) \\
    \hline
    $m_{1:B}$ & magnitude in all bands \\
    \hline
    $\sigma_{m,1:B}$ & magnitude uncertainty in all bands \\
    \hline
    $F_{1:B}^{\mathrm{db}}$ & database flux in all bands (Jy) \\
    \hline
    $\sigma_{F,1:B}^{\mathrm{db}}$ & database flux uncertainty in all bands (Jy) \\
    \hline
    $m_{1:B}^{\mathrm{db}}$ & database magnitude in all bands \\
    \hline
    $\sigma_{m,1:B}^{\mathrm{db}}$ & database magnitude uncertainty in all bands \\
    \hline
    $\mathrm{S/N}_{1:B}$ & signal-to-noise ratio in all bands \\
    \hline
    $\mu_{1:B}$ & background pixel flux in all bands (Jy) \\
    \hline
    $\sigma_{\textrm{px},1:B}$ & background pixel flux uncertainty in all bands (Jy) \\
    \hline
    \end{tabular}
\end{table}

Finally, we produce summary information for each candidate, reporting the quantities listed in Table~\ref{tab:stored_info}. These features encompass information on the photometry (fluxes and magnitude), background level, fitting residuals and flux-based quasar probability.  It is the last two of these we use to select or reject candidates as described in Section~\ref{sec:threshold}.


\section{Tests on simulated data}
\label{sec:simulations}

We first assessed the performance of the high-redshift quasar selection pipeline described in Section~\ref{sec:method} by applying it to simulated images of high-redshift quasars (Section~\ref{sec:simulations_quasars}), MLTY dwarfs (Section~\ref{sec:simulations_dwarfs}) and ETGs (Section~\ref{sec:simulations_etgs}).  The results are summarised in Table~\ref{tab:simulations}.

The first set of simulations tests if our method correctly gives high probabilities and low $\chiSq$ values to simulated quasars. By varying the flux levels we can detect at what point our method breaks down.  The second set of simulations tests if our method can detect a moving source, \eg, a local star, by obtaining large values in $\chiSq_b$ in at least one passband $b$.
Lastly, simulating ETGs as extended sources should again give low quasar probabilities and large $\chiSq$ values.

\begin{figure}
    \centering
    \includegraphics[scale=0.27]{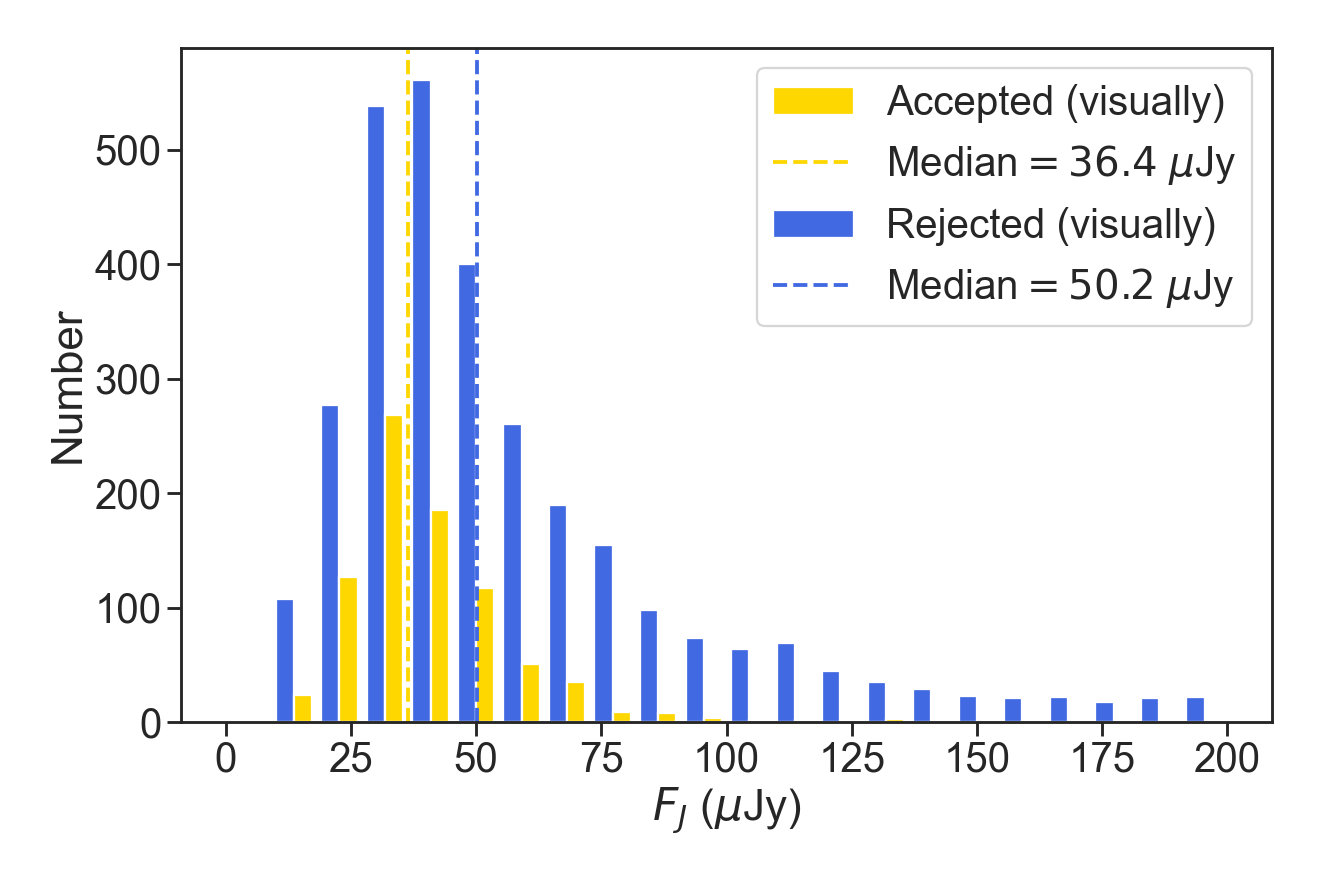}
    \caption{Distribution of flux in the $J$-band taken from our validation set (see Section~\ref{sec:validation_set}). The orange histogram corresponds to the visually accepted candidates, and the blue histogram to the visually rejected ones.}
    \label{fig:histo_Fj}
\end{figure}

We use the $J$ band as a reference, simulating sources with $J$-band fluxes between $10~\mjy$ and $100~\mjy$. The flux in the other bands is then calculated from the SED templates with $J$ as the reference band.  Before attempting any automated classification we undertake visual inspection of these simulated sources, examining the images for motion, non-point-like morphology and artefacts (although there are none in these simulations), thus mimicking the visual inspection of the real candidates described in Section~\ref{sec:dataset}.  These results are summarised in Fig.~\ref{fig:histo_Fj}, showing most of the viable candidates are, as expected, unambiguously detected but only with moderate $S/N$.

The minimum flux in $J$ of the visually accepted candidates is $F_{J, \text{min}}^{\text{accepted}} =12~\mjy$, motivating a minimum of $10~\mjy$ in our simulations.  Having a set of simulated faint sources helps us to analyse how the performance decreases when going to lower $S/N$.  We expect the difficulty in distinguishing quasars from contaminants to to increase at low $S/N$s for the residual chi-squared calculation, whereas the quasar probability will depend more subtly on the relative numbers of the contaminant populations.

The results are summarised in Table~\ref{tab:simulations} and in Figs~\ref{fig:simulated_quasars}-\ref{fig:simulated_ETG}.  For visual clarity and to aid comparison in these plots we i) take a uniform distribution in $\Pq$ between $0.5$ and $1.0$ for sources with $\Pq \geq 0.5$ and ii) spread out the low $\Pq$ values down to the (obviously) meaninglessly low value of $10^{-100}$.  Whenever $\chiSq$ is small for the simulated contaminants, the quasar probability $\Pq$ is close to zero; and, conversely, when $\Pq$ is large then $\chiSq$ is large too. Thus, even the most quasar-like of our simulated contaminant sources are a very bad fit to our quasar model.  The exception are simulated non-stationary MLTY dwarfs, where both $\Pq$ and $\chiSq$ can have perfectly acceptable quasar values when they have moved too far away from the centre. In that case, it is as if there was zero flux in the SDSS bands, which then looks similar to a Gunn-Peterson trough and hence gives a high quasar probability. Realistically, we expect surveys to be close enough in time that the source would not move outside $R_{\chiSq}$ and hence would still be detected using our procedure.

\begin{table}
    \centering
    \renewcommand{\arraystretch}{1.4}
    \caption{Classification results for simulated sources.}
    \label{tab:simulations}
    \begin{tabular}{|c|c|c|c|c|c|}
    \hline
    & & \multicolumn{2}{c|}{$\min \{\chiSq\}$ pair} & \multicolumn{2}{c|}{$\max \{\Pq\}$ pair} \\ 
    \hline
    source & environment & $\chiSq$ & $\Pq$ & $\chiSq$ & $\Pq$\\
    \hline
    \hline
    $\!\!$quasars$\!\!$ & isolated                       & $0.7$ & $1.0$ & $0.9$ & $1.0$  \\
    $\!\!$quasars$\!\!$ & nearby source                  & $0.8$ & $1.5\!\times\!10^{-3}$ & $1.0$ & $1.0$ \\
    \hline 
    dwarfs              & stationary                     & $0.8$ & $1.3\!\times\!10^{-14}$  & $1.1$ & $2.9\!\times\!10^{-3}$ \\
    dwarfs              & $\!\!\!$non-stationary$\!\!\!$ & $0.8$ & $8.8\!\times\!10^{-4}$ & $1.0$ & $1.0$ \\
    \hline 
    ETGs                & isolated                       & $0.8$ & $4.3\!\times\!10^{-5}$ & $1.0$ & $6.2\!\times\!10^{-5}$ \\
    \hline
    \end{tabular}
\end{table}

\subsection{Quasars}
\label{sec:simulations_quasars}

\begin{figure*} 
    \centering
    \includegraphics[trim={2.75cm 1.1cm 2.2cm 1.5cm}, clip=True]{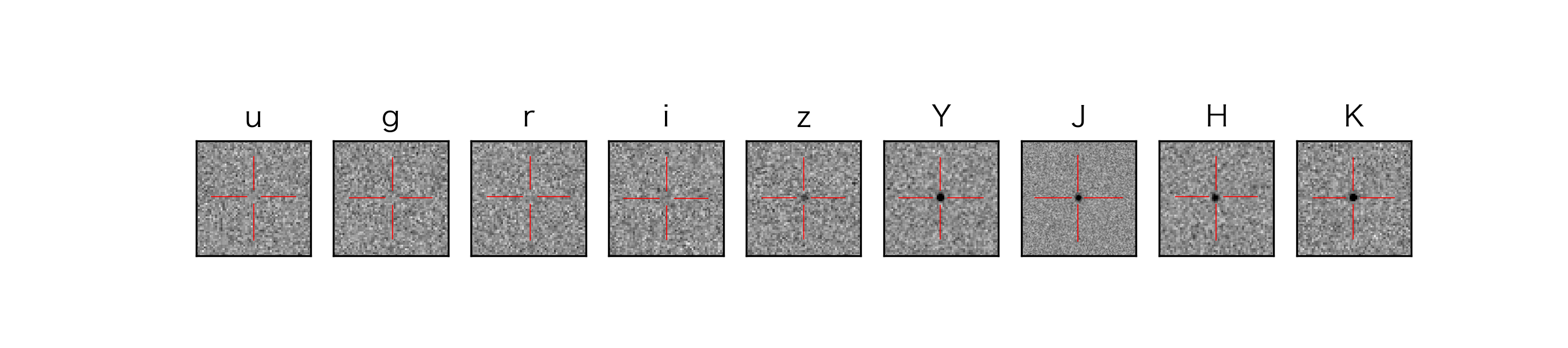}
    \includegraphics[trim={2.75cm 1.1cm 2.2cm 2.0cm}, clip=True]{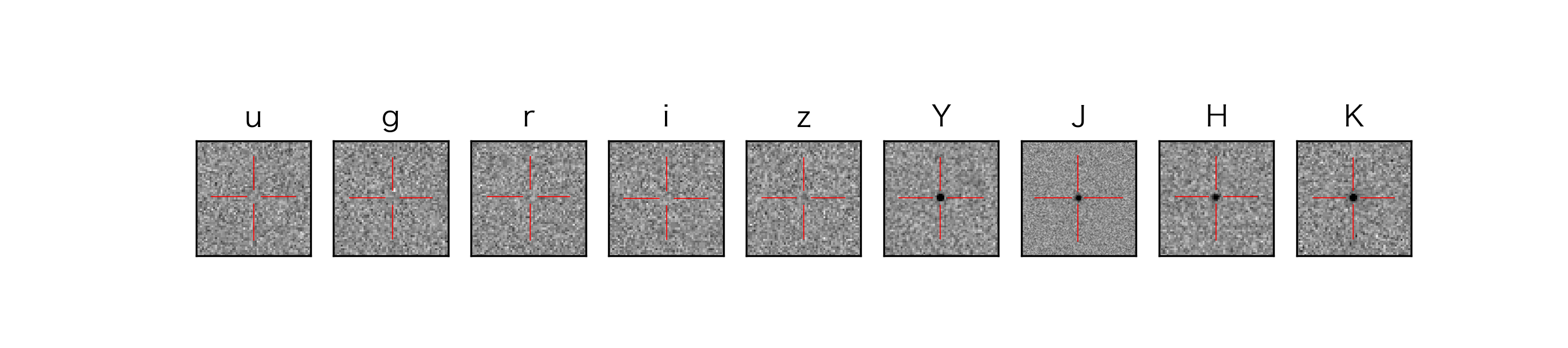}
    \includegraphics[trim={2.75cm 1.1cm 2.2cm 2.0cm}, clip=True]{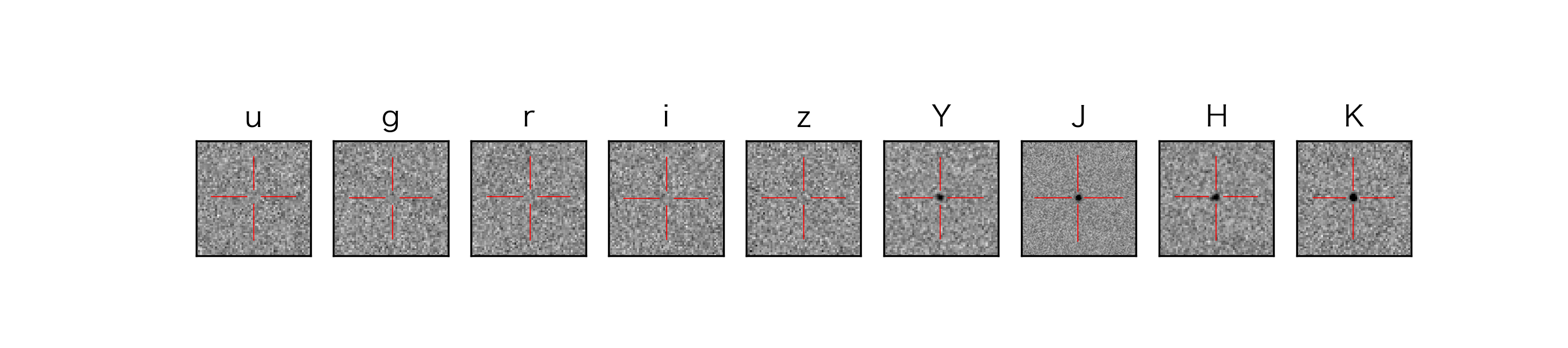}
    \includegraphics[trim={2.75cm 1.1cm 2.2cm 2.0cm}, clip=True]{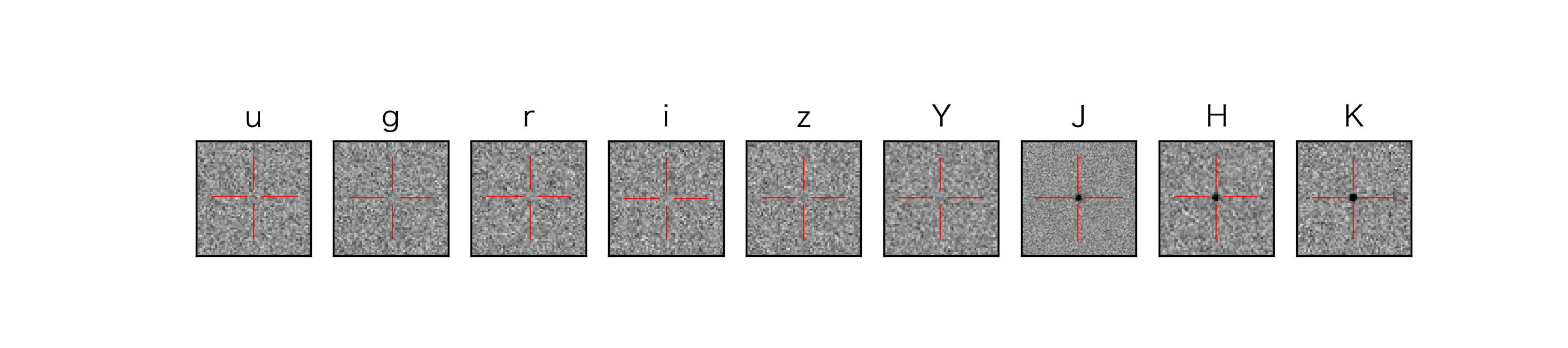}
    \caption{Simulated  quasars of magnitude $J$-band flux $F_J = 0.1~\mjy$, standard line equivalent width and a red continuum at redshifts of, from top to bottom, $z = 6.5$, $z = 7.0$, $z = 7.5$ and $z = 8.0$ \figcite}
\label{fig:simulation_jflux_faint}
\end{figure*}

\begin{figure*} 
    \centering
    \includegraphics[trim={2.75cm 1.1cm 2.2cm 1.5cm}, clip=True]{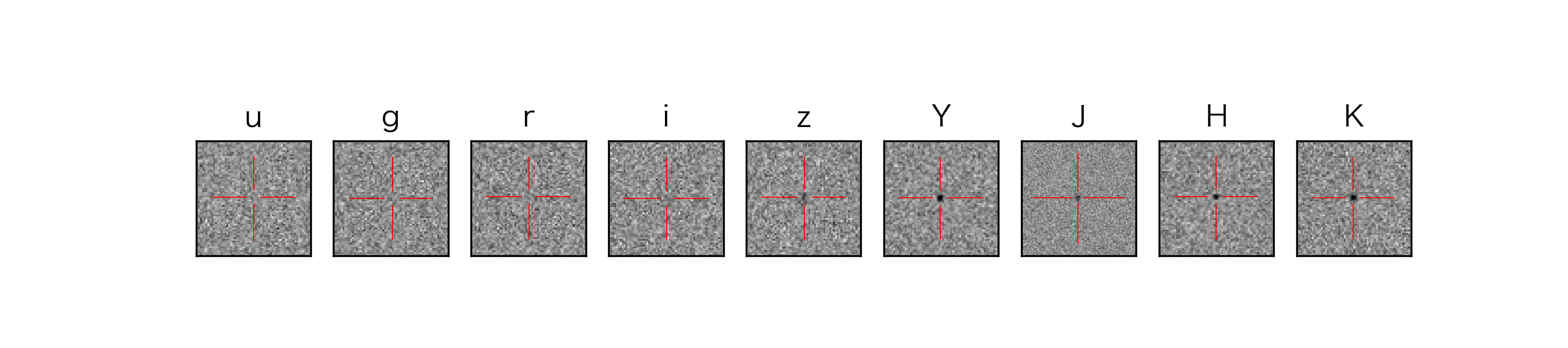}
    \includegraphics[trim={2.75cm 1.1cm 2.2cm 2.0cm}, clip=True]{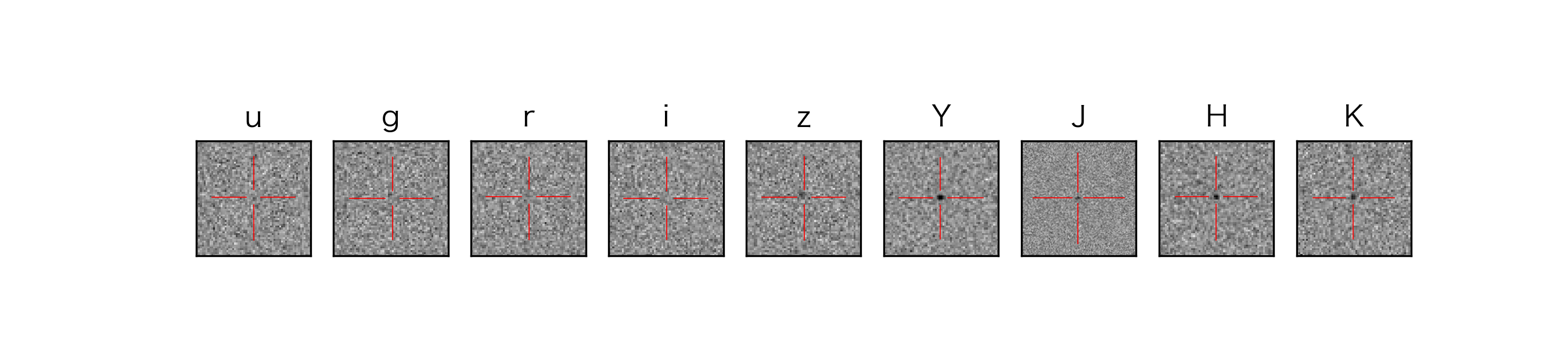}
    \includegraphics[trim={2.75cm 1.1cm 2.2cm 2.0cm}, clip=True]{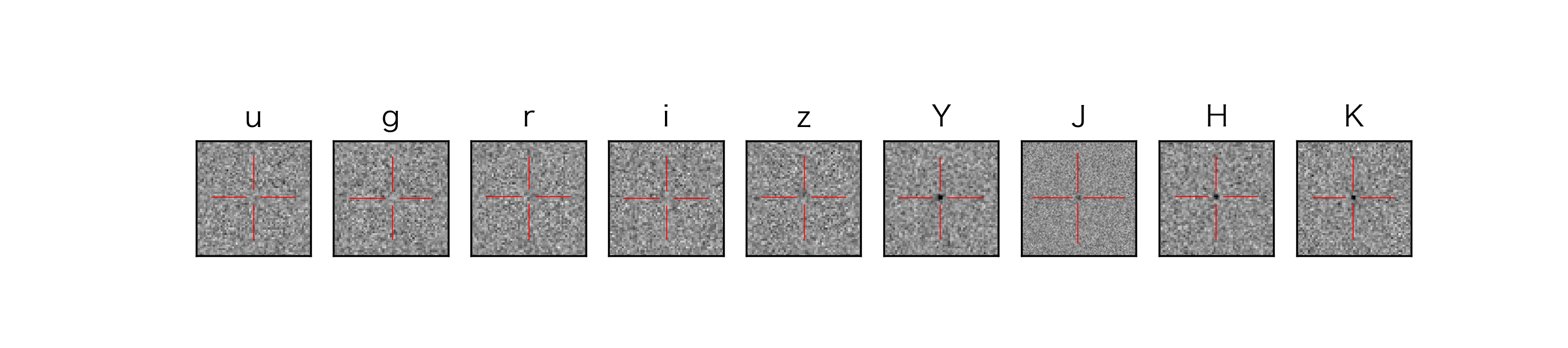}
    \includegraphics[trim={2.75cm 1.1cm 2.2cm 2.0cm}, clip=True]{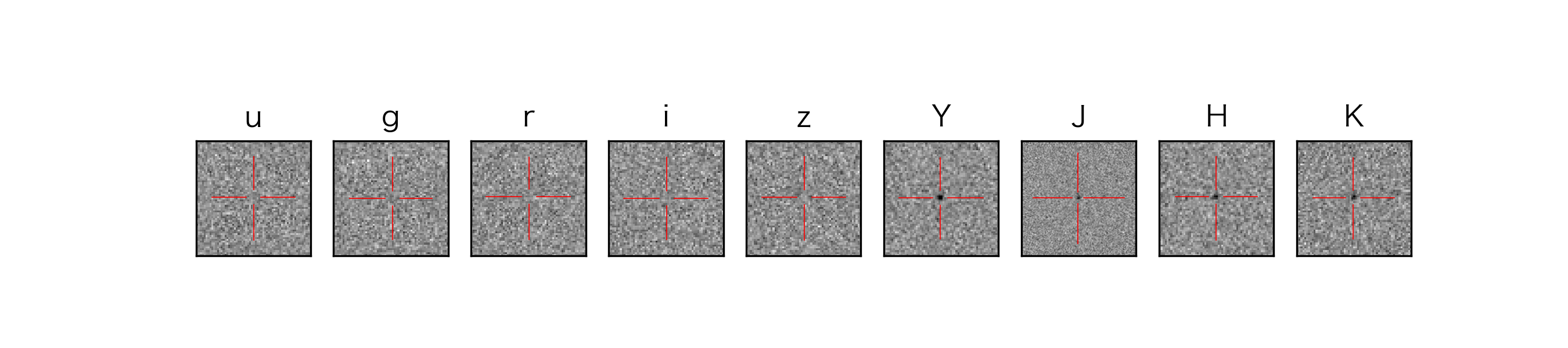}
    \caption{A simulated source at redshift $z=6.0$ with correct quasar template colours but with positional offsets between the SDSS and UKIDSS bands of, from top to bottom, 0.0~arcsec, 0.6~arcsec, 1.1~arcsec and 1.7~arcsec \figcite}
\label{fig:simulation_moving_sdss}
\end{figure*}

The most important input into our simulations are the objects we are searching for: high-redshift quasars.  As described in more detail in Appendix~\ref{section:model_q}, these are taken to be stationary point-sources with SEDs taken from \cite{Temple_etal:2021}.  Examples of the resultant simulated images are shown in Figs~\ref{fig:simulation_jflux_faint}~and~\ref{fig:simulation_moving_sdss}. Including bright nearby sources will make this more difficult, and we expect some sources with low probabilities and high chi-squared values in that case.  Simulating bright nearby sources also allows us to check how it affects the background pixel flux estimate and the overall source flux estimate.

\begin{figure}
    \centering
    \includegraphics[scale=0.39]{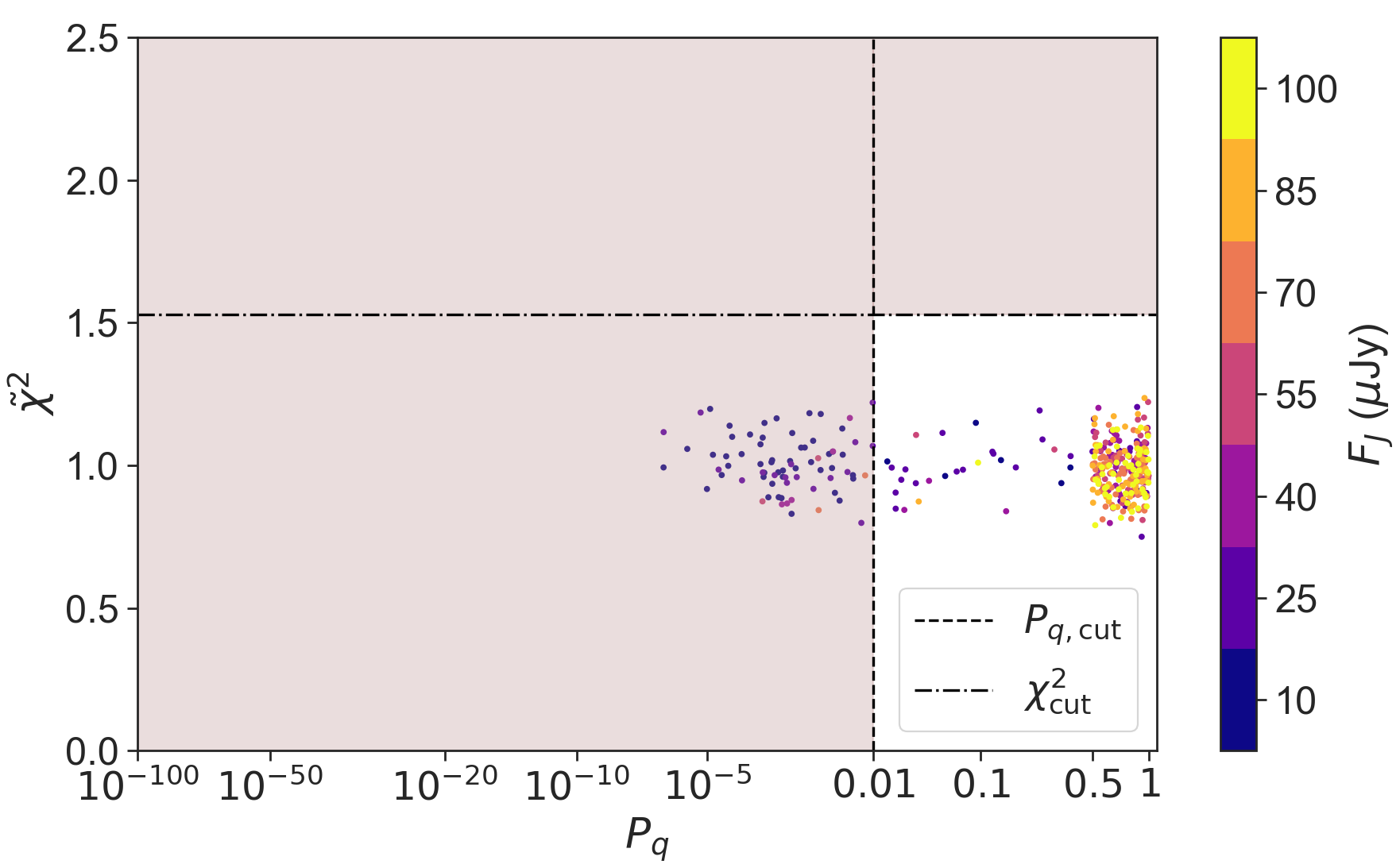}
    \caption{Classification results for $378$ simulated quasars with $J$-band fluxes between $10~\mjy$ (indicated by the colour bar) and $100~\mjy$ and redshifts between $z = 5.2$ and $z = 8.5$. For visual clarity we i) take a uniform distribution in $\Pq$ between $0.5$ and $1.0$ for sources with $\Pq \geq 0.5$ and ii) spread out the low $\Pq$ values down to the (obviously) meaninglessly low value of $10^{-100}$; the same axis range is used across this series of plots to aid comparison. Sources in the shaded regions are rejected.}
    \label{fig:simulated_quasars}
\end{figure}

\begin{figure}
    \centering
    \includegraphics[scale=0.39]{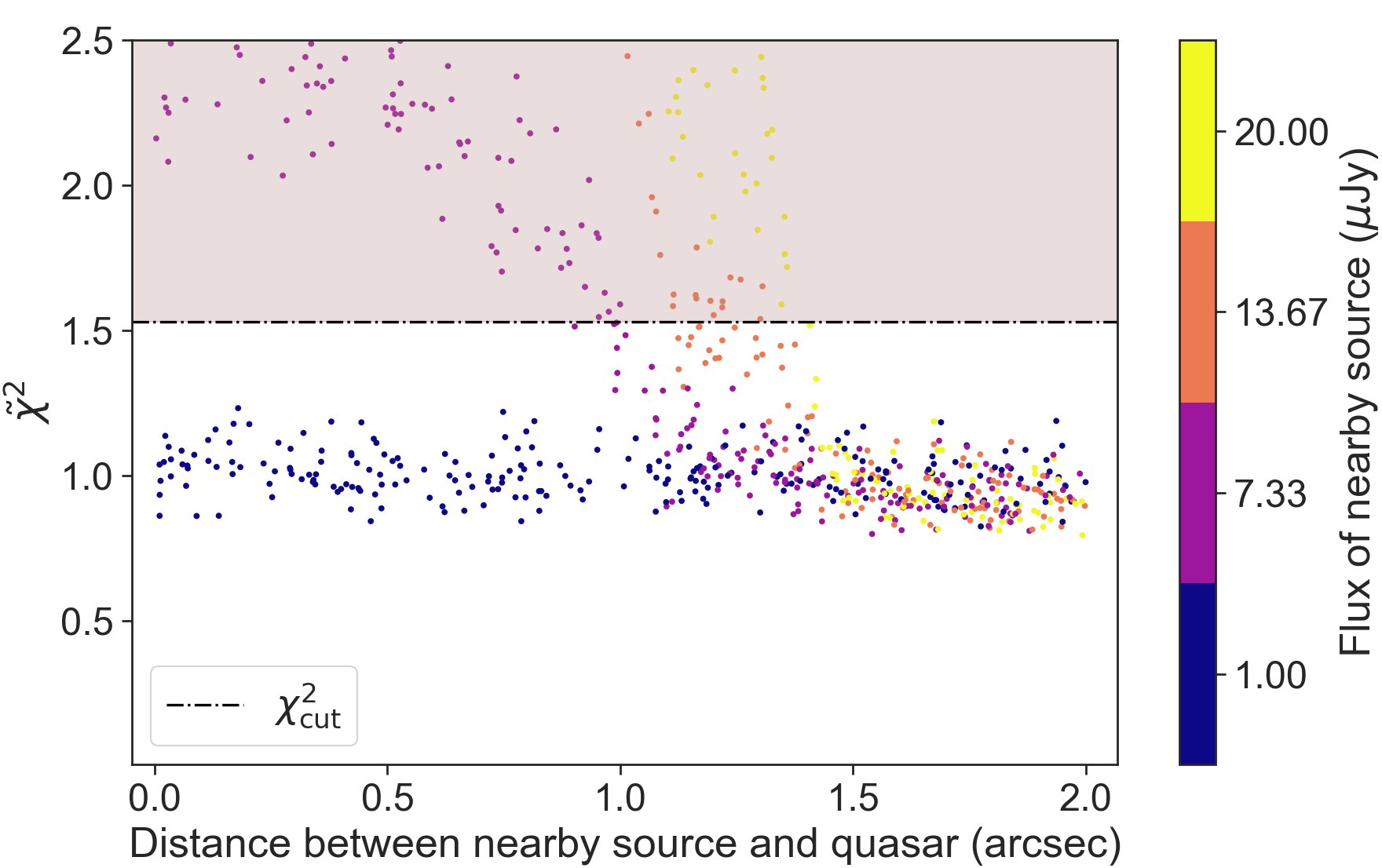}
    \caption{Classification results for $950$ simulated quasars with a nearby source of different fluxes between $1~\mjy$ and $20~\mjy$ (indicated by the colour bar) at different distances from the central source.  Sources in the shaded regions are rejected.}
    \label{fig:simulated_quasars_nearby_source}
\end{figure}

The results for simulated quasars are shown in Fig.~\ref{fig:simulated_quasars}.  As expected, the residual chi-squared values are close to $1$ or below. If $\chiSq < 1$, then the uncertainties are smaller than expected by our model.  While $\chiSq$ stays around $1$ for most $J$-band flux levels, the probability starts to decrease for faint sources with $F_J \lesssim 25$  $\mu$Jy. This makes sense, as the smaller the $S/N$, the more difficult it gets to extract the source signal. Therefore, we might miss a small portion of faint quasars.  However, in some cases a low probability is expected. This is because in the redshift range we use for our simulations, a low flux in $J$ can result in simulated negative fluxes in the bands bluewards of $J$. This is especially the case for the simulated quasars at the highest redshift end. 

The results for simulated quasars with a nearby source can be seen in Fig.~\ref{fig:simulated_quasars_nearby_source}. The $J$-band flux is varied and the fluxes in the other bands are calculated using the template SEDs. The nearby source is of constant flux over all passbands, in the range $1 ~\mjy < F_J < 20~\mjy$. We do not include sources brighter than $20~\mjy$ as these have even larger reduced chi-squared values and therefore are a clear rejection. The distance between the source and the central quasar is up to $2$ arcsec.  As expected, a constant flux of only $1~\mjy$ gives a low $\chiSq$ independent of where the source is located. This is because the flux is within the uncertainty of the background pixel flux.  At $F = 7.3~\mjy$, we already see $\chiSq > 2$, which starts to go down at around a distance of $0.75$~arcsec.  Generally, the reduced chi-squared values decrease with increasing distance of the source from the central quasar.  This makes sense, as the further the nearby source moves away from the central source, the less impact it has on the central flux and chi-squared calculation.  As the sources moves outside $R_{\chiSq}=1\farcs2$, the radius for the $\chiSq$ and flux calculation, it has less of an impact on the central source which means that the central flux can be recovered.  The $\chiSq$ value reaches $\sim\! 1$ at a distance of  $\sim\! 1.5$~arcsec for the source fluxes we considered.

In our pipeline we only reconstruct sources that are at least $R_{\text{keep}}$ away from the central source. The limit allows us to include moving sources while not letting actual nearby objects impact the central source calculation.


\subsection{MLTY dwarfs}
\label{sec:simulations_dwarfs}

We simulate MLTY dwarfs by generating potentially moving point-sources with colours of MLTY dwarfs.  Since the images of SDSS and UKIDSS were acquired several years apart, a local star appears at a different location in each survey.  We therefore need to simulate motion of MLTY dwarfs.  In our case, we use the overall source locations as reported in the database of UKIDSS LAS DR10PLUS.  For this reason we only simulate offset locations in the SDSS bands. This offset is uniformly distributed between $0$ arcsec and $2$ arcsec. Had we taken the source locations from SDSS, we would have simulated the offsets in UKIDSS instead.

We test if our $\chiSq$ method can reveal sources with apparent motion and flag up such candidates.  We use all available MLTY dwarf types from \cite{Barnett_etal:2019}, from M0 up to T8. 

\begin{figure}
    \centering
    \includegraphics[scale=0.385]{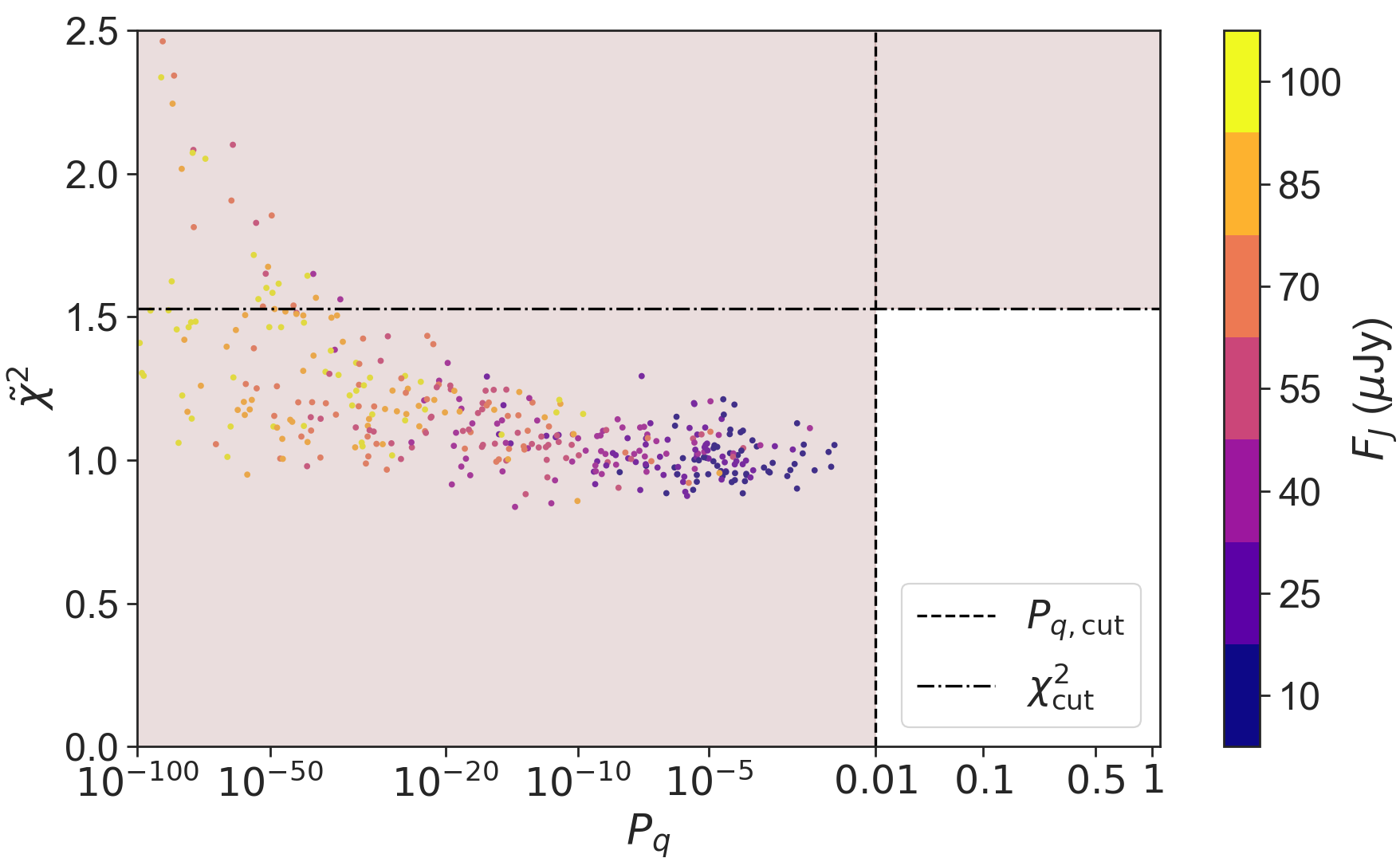}
    \includegraphics[scale=0.39]{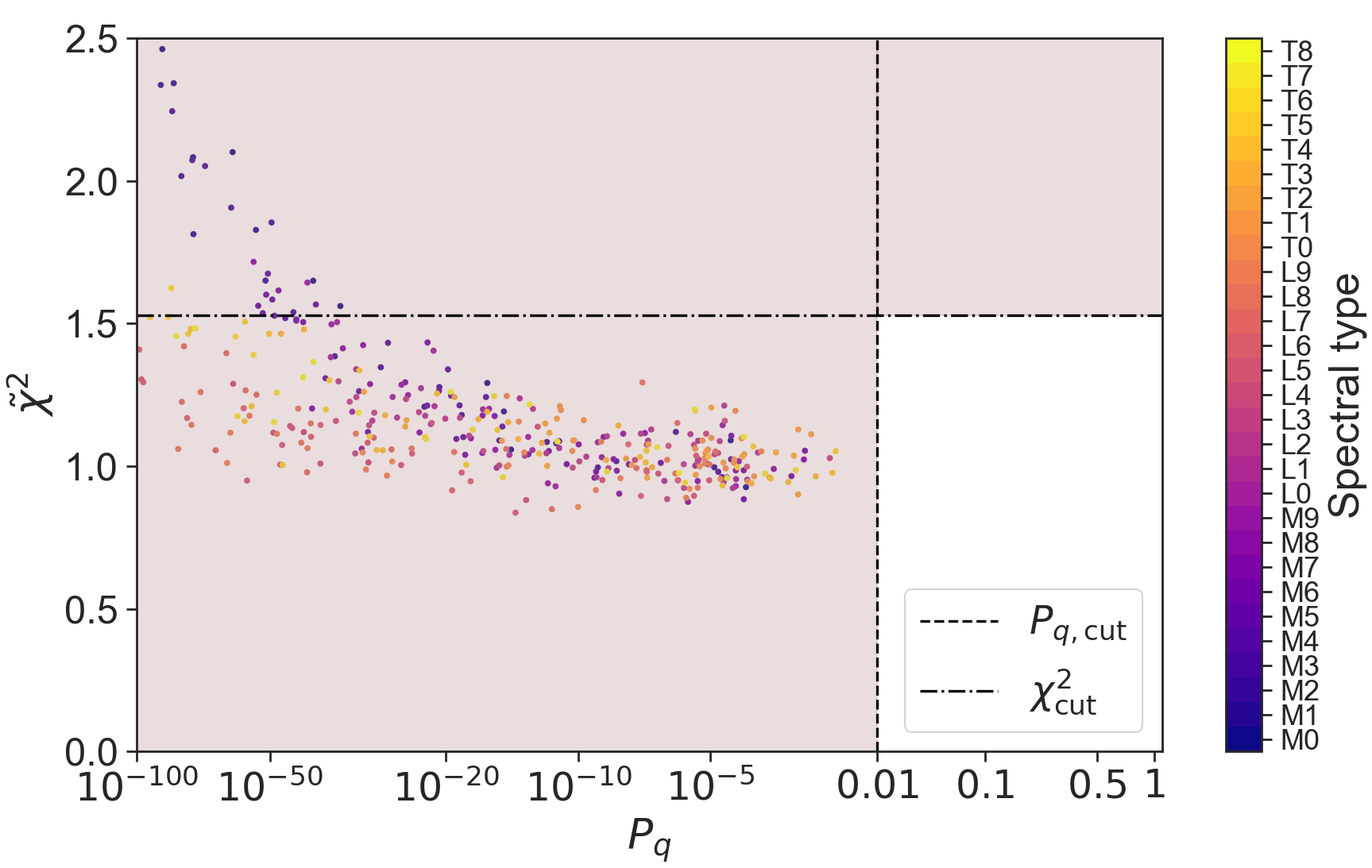}
    \caption{Classification outputs for $406$ simulated stationary MLTY dwarfs with the source location is fixed in the centre across all bands. The MLTY SED models have $J$-band fluxes between $10~\mjy$ and $100~\mjy$ (top panel) and types $M0$ to $T8$ (bottom panel).  Sources in the coloured regions are rejected.}
    \label{fig:simulated_MLTY_stationary}
\end{figure}

Figure~\ref{fig:simulated_MLTY_stationary} shows the simulated stationary MLTY dwarfs as a function of $J$ band flux and spectral type respectively.  All simulated sources have a quasar probability below $10^{-3}$.  Some sources have low $\chiSq$ values, particularly the ones with faint $J$-band photometry of $F_J = 10~\mjy$.  Again, this is expected as at such low flux levels a tiny difference in the predicted quasar SED will not account for a large difference in $\chiSq$, especially if the source flux is within the uncertainty of the background pixel flux.  In terms of spectral types, T and L type dwarfs have lower residual chi-squared values, while M dwarfs give the worst fit. The plots have been cropped to be standardised and show the more interesting areas, but the residual chi-squared values for MLTY dwarfs extend upto $\chiSq \simeq 20$.  However, no sources simultaneously have $\chiSq \sim 1$ and $\Pq > 10^{-3}$.  Hence, the combination of $\Pq$ and $\chiSq$ can easily distinguish between quasars and MLTY dwarfs when the sources follow our MLTY SED templates.

\begin{figure}
\centering
    \includegraphics[scale=0.39]{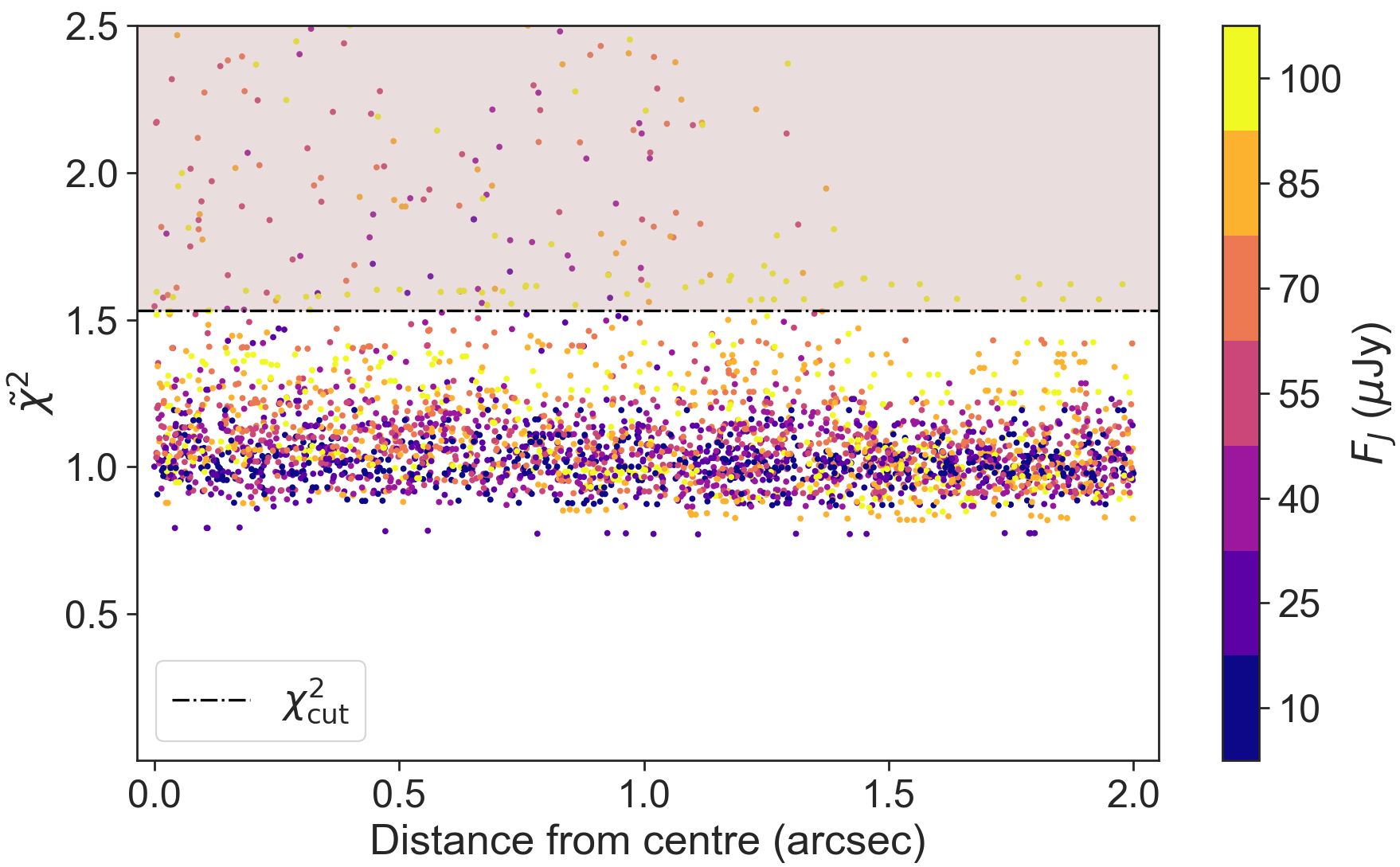}
    \includegraphics[scale=0.39]{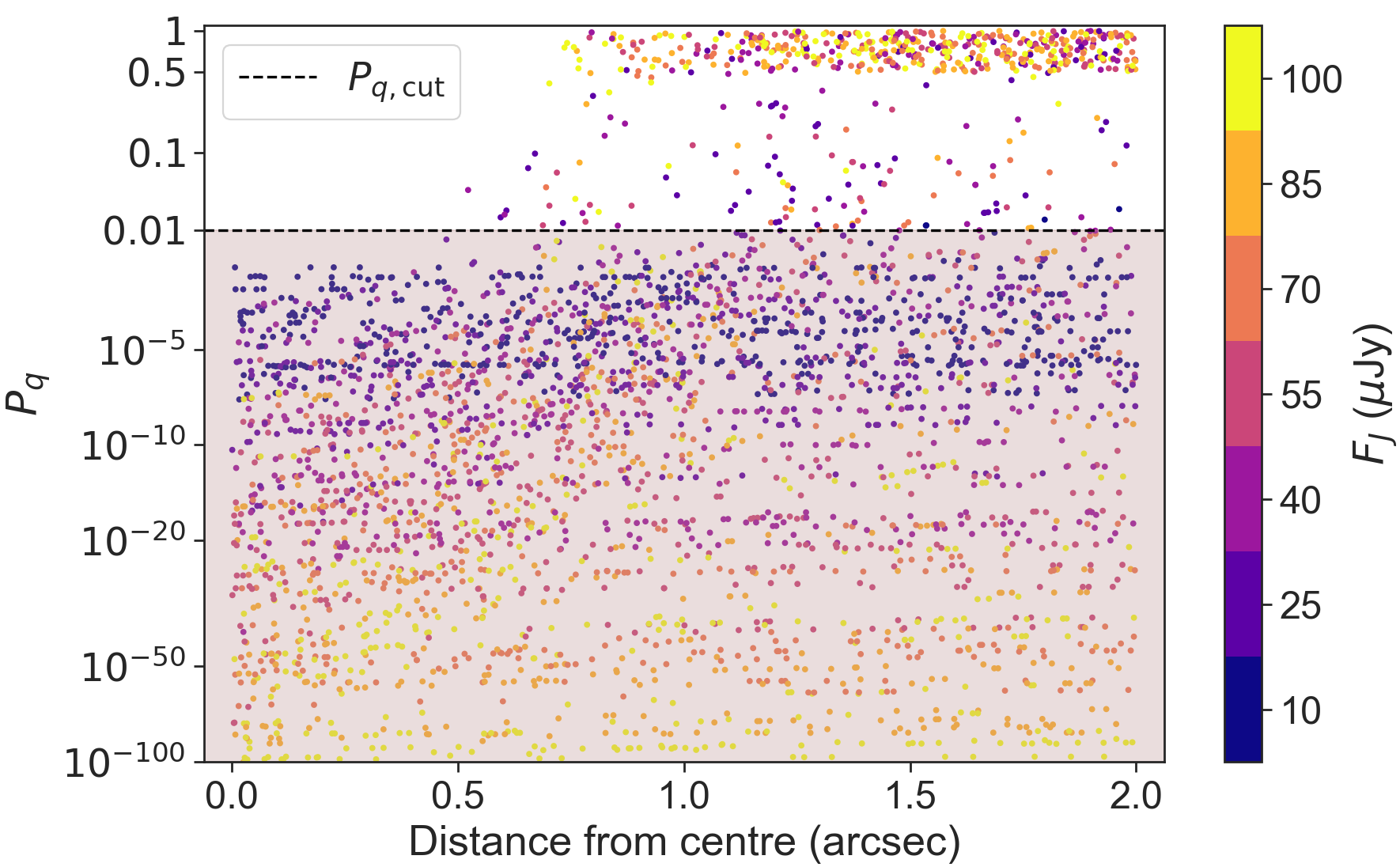}
    \caption{Classification outputs for $3587$ simulated non-stationary MLTY dwarfs, where the source location is offset in the optical SDSS bands from the NIR UKIDSS position, shown in terms of both $\chi^2$ (top panel) and evolving $\Pq$ (bottom panel).  Sources in the coloured regions are rejected.}
    \label{fig:simulations_MLTY_sdss_offset}
\end{figure}

The results for the non-stationary MLTY dwarfs can be seen in Fig.~\ref{fig:simulations_MLTY_sdss_offset}.  As expected, the brighter the source, the higher the chi-squared value.  For dwarfs located in the centre, the maximum quasar probability is around $10^{-3}$. This holds for the faintest dwarfs of $F=10$ $\mu$Jy only. For brighter MLTY dwarfs, the probability tends to zero.

For a subset of dwarfs, the quasar probability starts to increase once the MLTY dwarf moves away from the centre. It gets close to a probability of $1$ at around a distance of $0\farcs8$ from the central quasar, which corresponds to $2$ pixels on the postage stamp.

Once the source has moved outside the radius $R_{\chiSq}=1\farcs2$ within which the chi-squared calculation takes place, $\chiSq$ returns to normal values.  This is because our pipeline would essentially calculate zero flux for such sources, as the central source has moved too far outwards. In this low flux regime it is difficult to make accurate predictions about what source population the object belongs to.

\subsection{Early-type galaxies}
\label{sec:simulations_etgs}

ETGs are simulated as extended sources as described in Appendix~\ref{section:model_g}.  As can be seen in Fig.~\ref{fig:simulated_ETG}, the quasar probabilities are close to zero as expected, with a maximum probability of $\Pq = 6.2 \times 10^{-5}$.  The residual chi-squared $\chiSq$ increases with increasing flux levels as expected.  The $\chiSq$ are large except in the case for  simulations of faint ETGs.  This occurs below a threshold of $\sim\! 25$ $\mu$Jy, where it becomes more difficult to distinguish between quasars and contaminants. 

\begin{figure}
    \centering
    \includegraphics[scale=0.39]{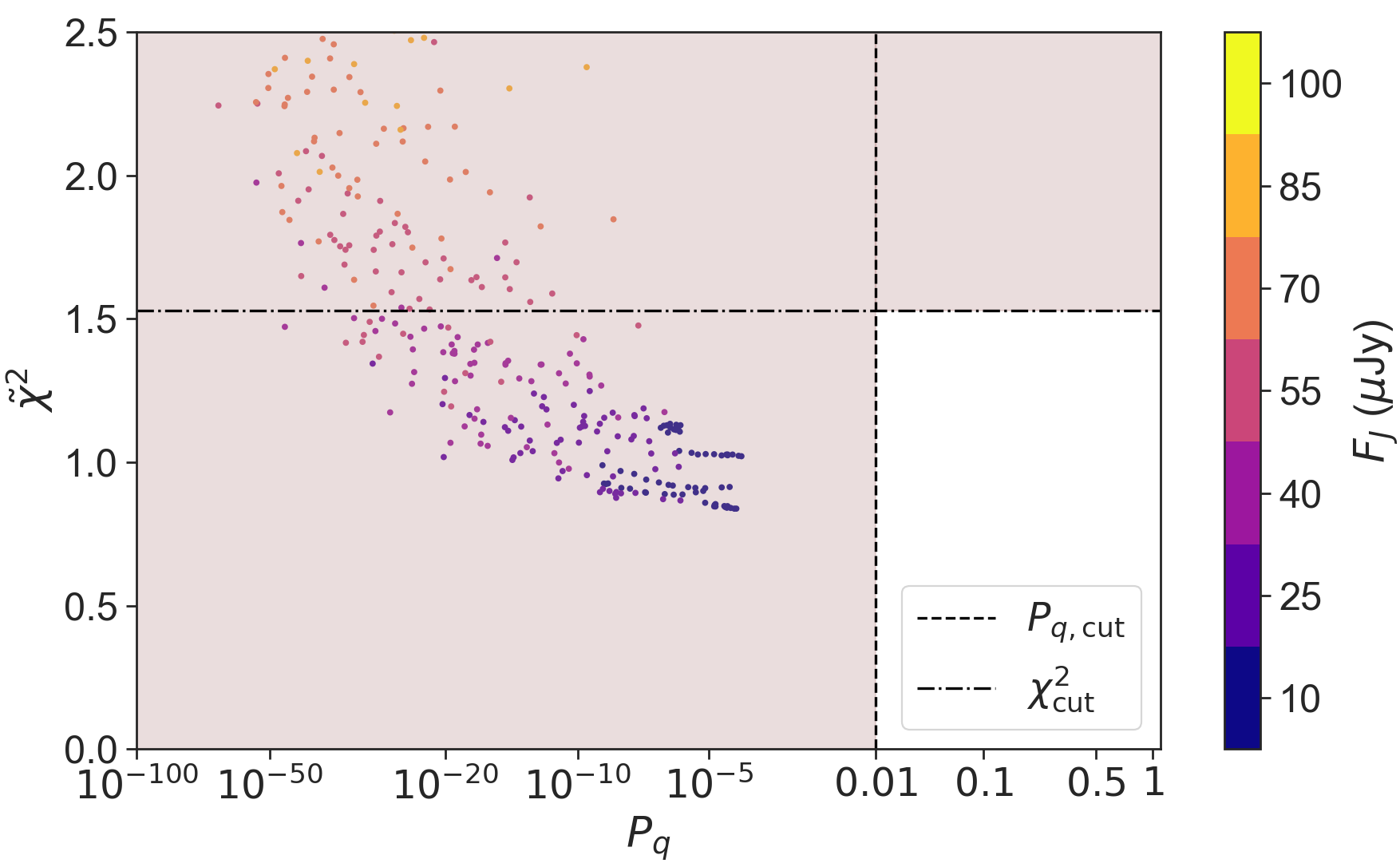}
    \caption{Classification outputs for a set of $420$ simulated ETGs as a function of quasar probability, residual chi-squared and flux in $J$. These lie in the redshift range between $1 \leq z \leq 2$ with formation redshifts of $z=3$ and $z=10$.  Sources in the coloured regions are rejected.} 
    \label{fig:simulated_ETG}
\end{figure}

Overall, this shows that our method can distinguish between ETGs and quasars whenever the source follows the colours of either of our templates and is sufficiently bright.


\section{Tests on real data}
\label{sec:dataset}

The tests of the high-redshift quasar selection method on simulated data presented in Section~\ref{sec:simulations} are unavoidably optimistic, in particular regarding the image-level goodness of-fit tests: real astronomical images diverge from the ideal in numerous ways, with a range of artefacts (\eg, figure~3 of \citealt{Dye_etal:2006}) that are inevitably selected in rare object searches (\eg, Appendix~A of \citealt{Reed_etal:2015}).  Hence, it is important to test on existing real data, even if different from the data-sets for which our method is intended (\ie, {\em Euclid} and LSST).  We use the cross-matched catalogues from SDSS and UKIDSS (Section~\ref{sec:manual_survey_data} which were used as the basis of the high-redshift quasar searches described in \cite{Venemans_etal:2007}, \cite{Mortlock_etal:2009}, \cite{Mortlock_etal:2011} and \cite{ Mortlock_etal:2012} and so provides a ready-made validation set of 4,300 visually inspected candidates (Section~\ref{sec:validation_set}).

\begin{figure}
    \centering
    \includegraphics[scale=0.3, trim={1.01cm 0.5cm 0.5cm 0.5cm}, clip=True]{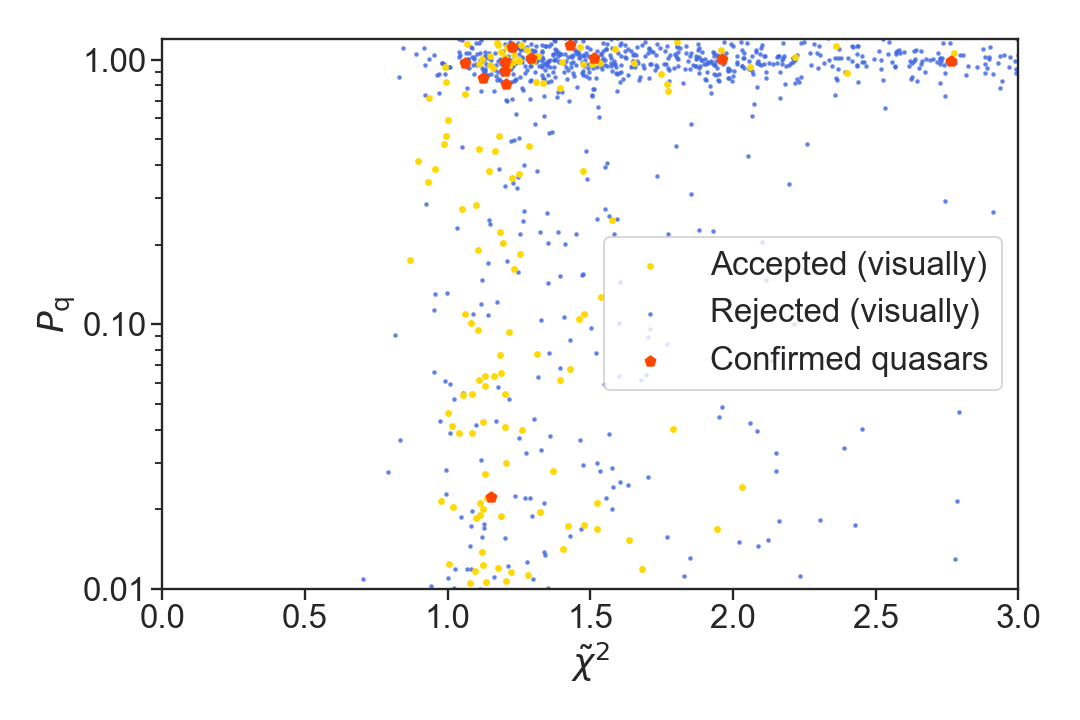}
    \caption{Scatter plot of $\chiSq$ vs $\Pq$ for visually inspected candidates from the SDSS-UKIDSS validation data-set.  The orange crosses correspond to visually accepted candidates; the blue dots correspond to visually rejected candidates; and the red dots indicate confirmed $6 \lesssim z \lesssim 7$ quasars.  A small scatter is added for candidates with $\Pq>0.9$ to increase visibility, hence some appear with $\Pq > 1$.}
    \label{fig:scatter_visual}
\end{figure}


\subsection{Survey data}
\label{sec:manual_survey_data}

We largely follow standard procedures to download SDSS and UKIDSS images which, between them provide photometry in the $u$, $g$, $r$, $i$, $z$, $Y$, $J$, $H$ and $K$ bands. A few specific points are relevant for each survey:

{\bf SDSS:} We obtain optical data from SDSS DR12.  PSF information is provided.  We do not include SDSS images which are flagged as any of: {\tt PSF\_FLUX\_INTERP}; {\tt BAD\_COUNTS\_ERROR}; {\tt COSMIC\_RAY}; {\tt BADSKY}; {\tt MAYBE\_EGHOST}; {\tt MAYBE\_CR}; {\tt EDGE}; {\tt INTERP\_CENTER} and {\tt DEBLEND\_NOPEAK}.

{\bf UKIDSS:} We obtain NIR data from the UKIDSS LAS Data Release 10 PLUS (DR10PLUS).  UKIDSS provides the FWHM of the PSF, which we use to construct a \cite{Moffat:1969} profile.  For UKIDSS, we require that {\tt ppErrBits < 256} and {\tt errBits < 8}.


\subsection{Validation data-set}
\label{sec:validation_set}

We use a labelled set of $\sim\! 4,300$ sources visually-inspected as part of the high-redshift quasar search described by \citet{Mortlock_etal:2012}. These sources have been selected via magnitude and colour cuts, in many cases early in the search process, so they are not a uniform sample.  They were examined for flags that would indicate that the source is extended, has  proper motion or unreliable photometry.  The labels are binary, where a positive (accepted) label corresponds to the source resembling a quasar in all of the bands, and a negative (rejected) label corresponds to a source that either looks like a contaminant or to an image with data processing issues around the nominal source position.

From the pool of candidates, only $\sim\! 800$ have a positive label, out of which $14$ are spectroscopically confirmed\footnote{There are also a few candidates for which a spectrum confirmed the sources as a Galactic star; these are not treated separately.} quasars at $z > 5.7$ \citep{Fan_etal:2000, Fan_etal:2001, Fan_etal:2004, Goto:2006, Venemans_etal:2007, Mortlock_etal:2009, Mortlock_etal:2011, Jiang_etal:2016}.  This gives a ratio of approximately $1:6$ between probable quasars and contaminants.

Further, only $\sim\! 450$ sources have $\Pq > 0.01$, and $\sim\!330$ have $\Pq > 0.1$. Thus, requiring $\cutPq = 0.01$ already reduces the pre-selected candidate list by a factor of $\sim 10$.  Requiring $\cutChiSq = 1.5$ reduces the candidate list to $\sim 2300$. Requiring $\cutChiSq = 1.5$ and $\cutPq = 0.1$ gives $\sim\! 160$ sources, less than half the number from just requiring $\cutPq = 0.1$.


\subsection{Selection criteria}
\label{sec:thresholds}

The main utility of the various summary quantities described above is, for each candidate, to make a selection decision as to undertake visual inspection and/or follow-up observations.  We accept a candidate if it satisfies these three criteria:
\begin{itemize}
    \item $\Pq > \cutPq$, a cut in quasar probability to remove sources for which the photometry is better explained by a non-quasar SED.
    \item $\chiSq < \cutChiSq$, a cut in the residual average chi-squared value across all bands to remove moving and extended sources.
    \item $\chiSqMax < \cutChiSqMax$, a cut in the residual maximum single-band chi-squared value to remove sources for which the photometry has been affected by image artefacts.
\end{itemize}
The values for the three thresholds  (\ie, $\cutPq$, $\cutChiSq$ and $\chiSqMax$) must be chosen in such a way that they effective balance selection efficiency and sample completeness (Section~\ref{sec:classification}).  For the validation data described above we use the $F_\beta$ statistic (Section~\ref{sec:fbeta}) as detailed futher in Section~\ref{sec:sim_results}.

One could also select the thresholds manually based on the distribution of summary statistics, as illustrated in Fig.~\ref{fig:scatter_visual}, which shows a scatter plot of $\chiSq$ vs $\Pq$.  All of the quasars have $\Pq > 0.1$, where the quasars with the lowest probability happen to be the ones at the lowest redshift.  If $\Pq$ is close to $1$ but $\chiSq$ is large then this indicates that even though the database/calculated photometry agree very well with a quasar SED, the source either does not resemble a point-source or it has moved.  On the other hand, if $\Pq$ is close to $0$ and $\chiSq$ is low, then even though the source is a stationary point-source and follows quasar colours, it is much more likely that the source is a contaminant. 

We expect these thresholds will not change drastically between different data-sets and hence could, at least initially, be used `as is' in new surveys. That means, for application to future surveys we would not require a pre-labelled data-set.  Visual inspection would only be needed for testing the algorithm on a new data-set.  Thresholds can be set more conservatively to find the most likely quasar candidates even if this means that other quasars will be missed.

Of course better results will be achieved if the thresholds are optimized for each new survey. For instance, the better the survey PSF can fit a point-source, the lower/larger the chi-squared values will be for positive/negative candidates.  In addition, a more realistic PSF will be able to measure the photometry more accurately, resulting in a better choice of SED template.  Therefore, if other surveys use different PSFs to the ones described in Section~\ref{sec:manual_survey_data}, the thresholds should be adjusted accordingly. Similarly, the threshold for the average chi-squared value would need to be adjusted depending on the relative numbers of bands blueward/redward of the Ly$\alpha$ break.


\subsection{Classification results}
\label{sec:sim_results}

\begin{figure*}
\centering
\includegraphics[width=15cm]{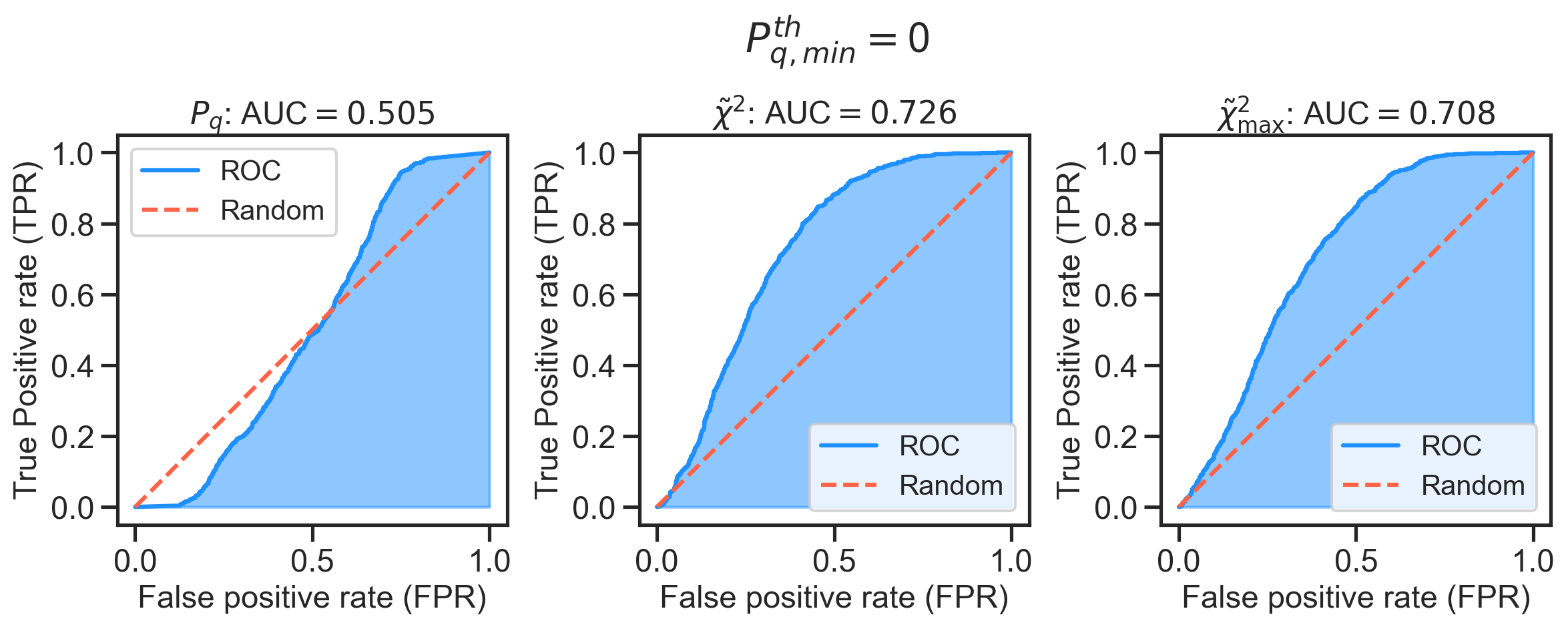}
\includegraphics[width=15cm]{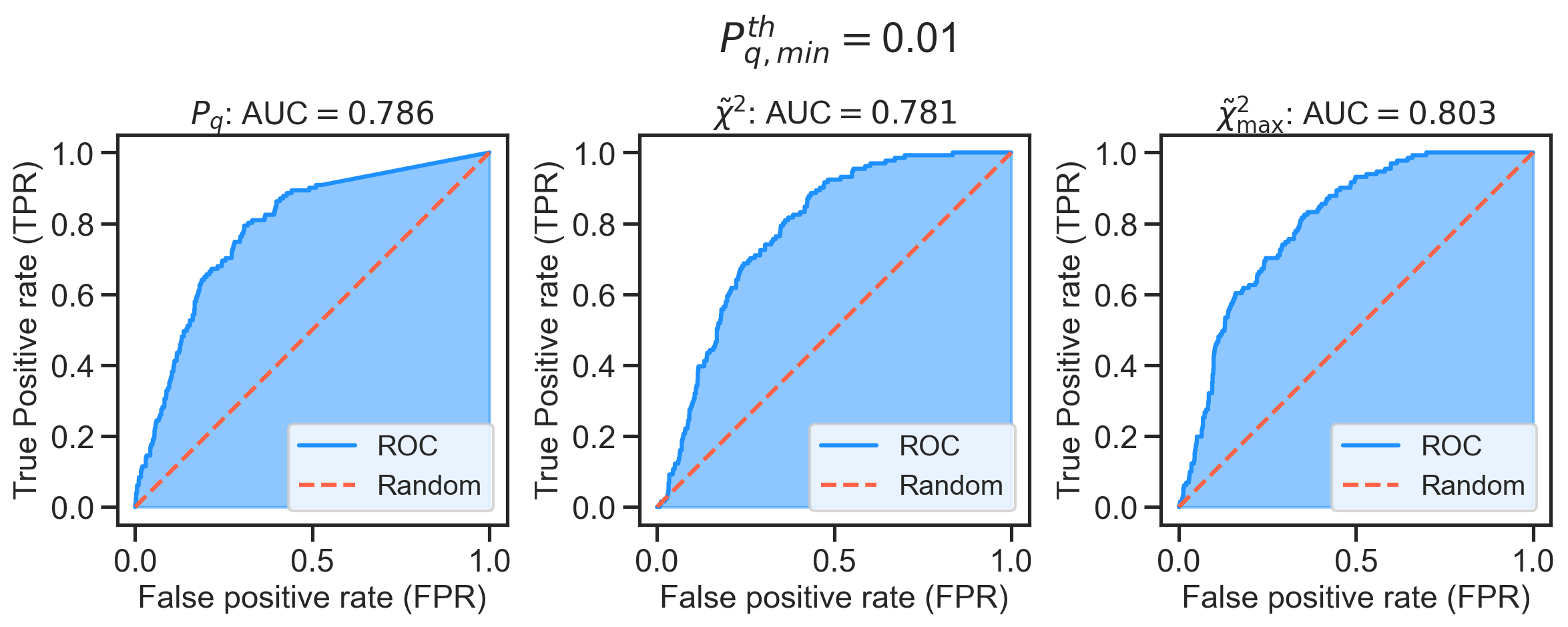}
\includegraphics[width=15cm]{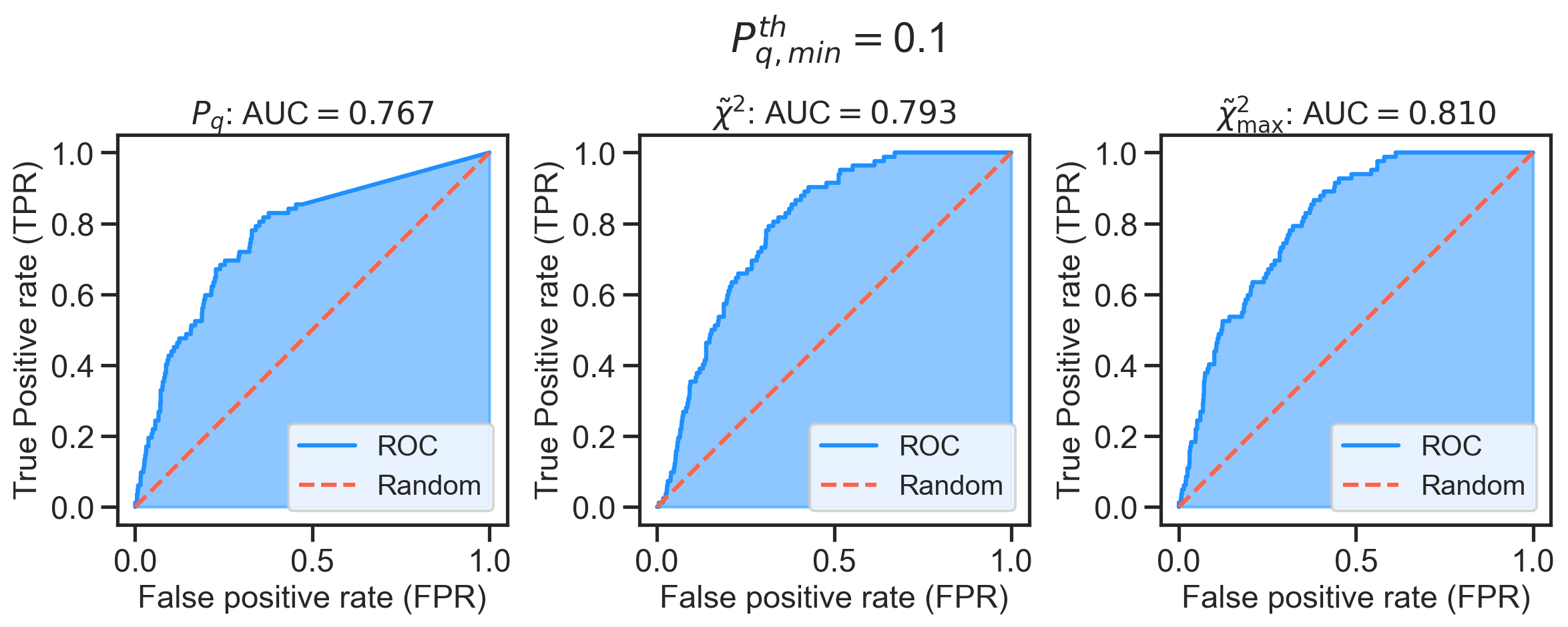}
\caption{ROC curves showing the FPR vs.\ TPR for individual features $\Pq$ (left column), $\chiSq$ (middle column) and $\chiSqMax$ (right column). This is done for three probability cuts: $\cutPq=0$ (top row); $\cutPq=0.01$ (middle row); and $\cutPq=0.1$ (bottom row).}
\label{fig:roc}
\end{figure*}

\begin{figure*}
\centering
    \includegraphics[scale=0.32]{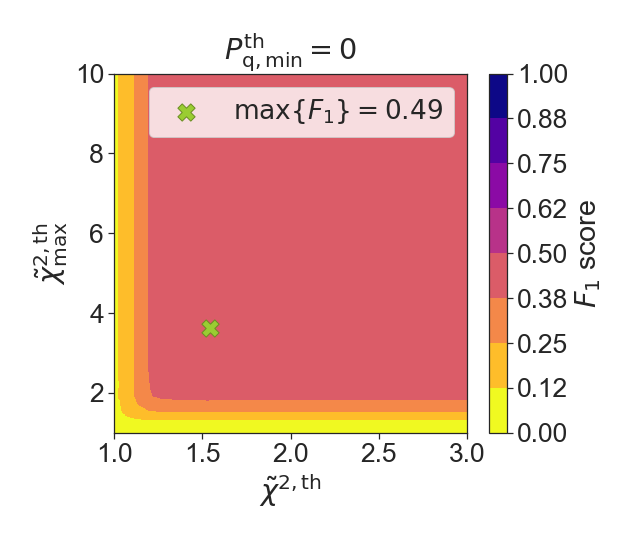}
    \includegraphics[scale=0.32]{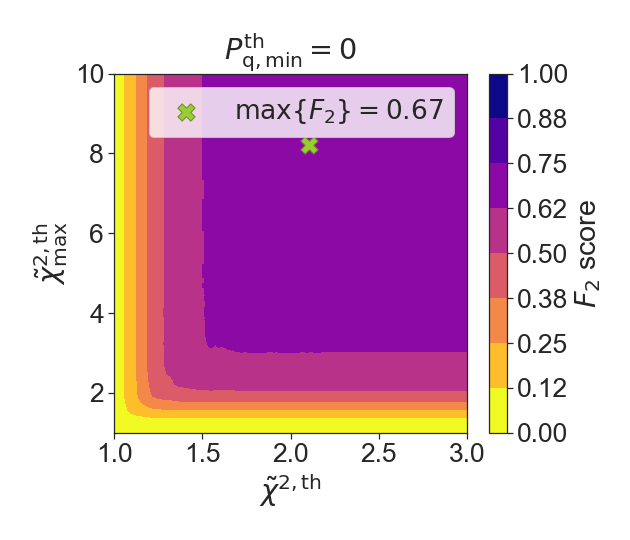}
    \includegraphics[scale=0.32]{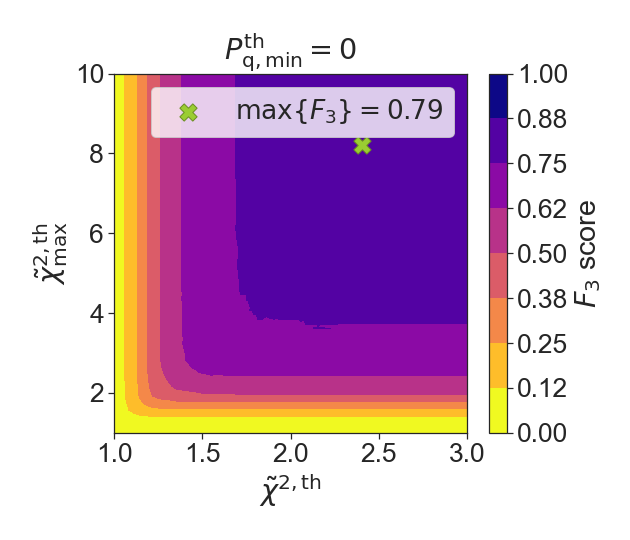}
    \includegraphics[scale=0.32]{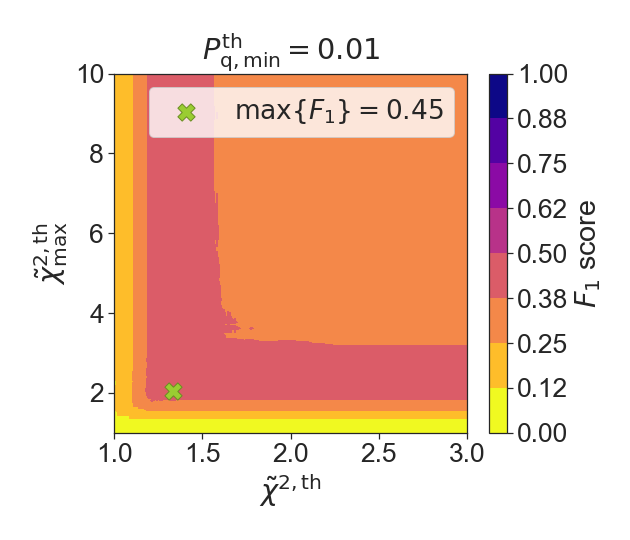}
    \includegraphics[scale=0.32]{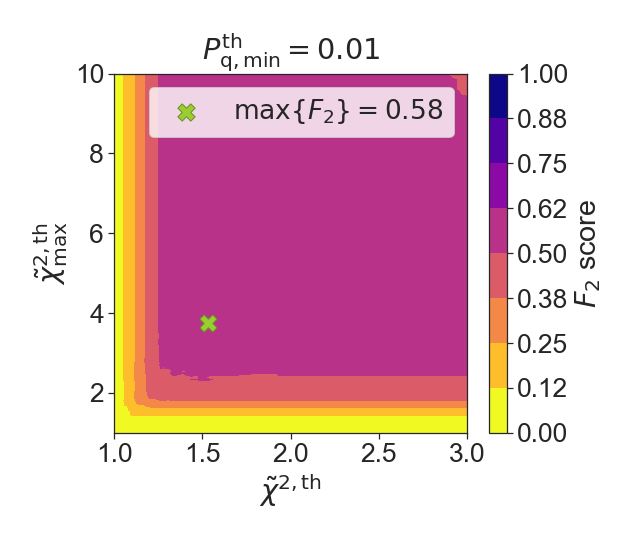}
    \includegraphics[scale=0.32]{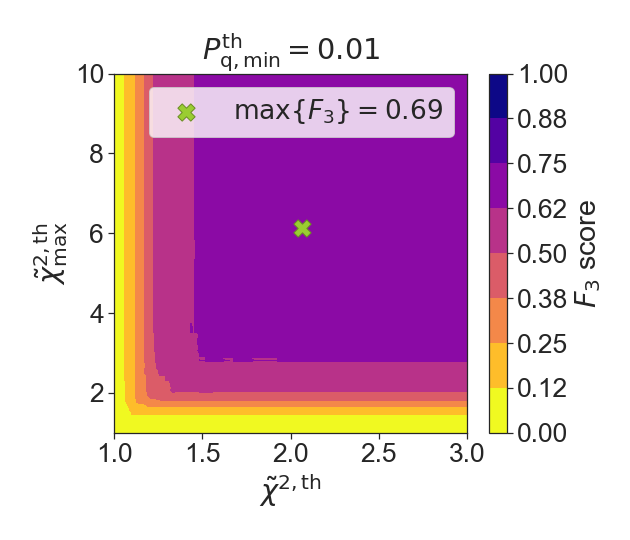}
    \includegraphics[scale=0.32]{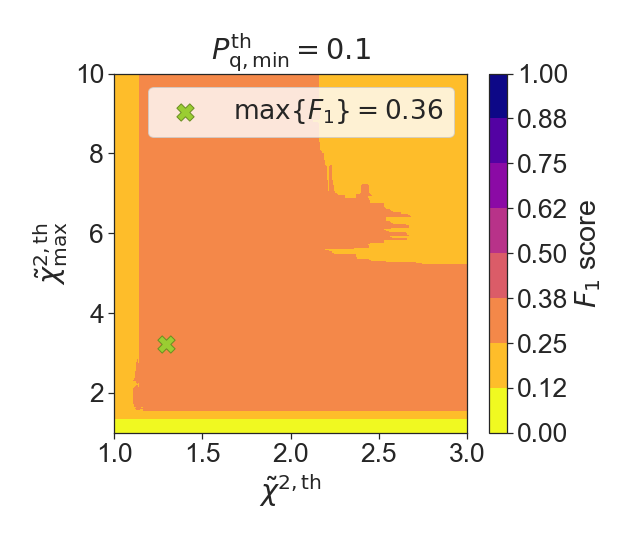}
    \includegraphics[scale=0.32]{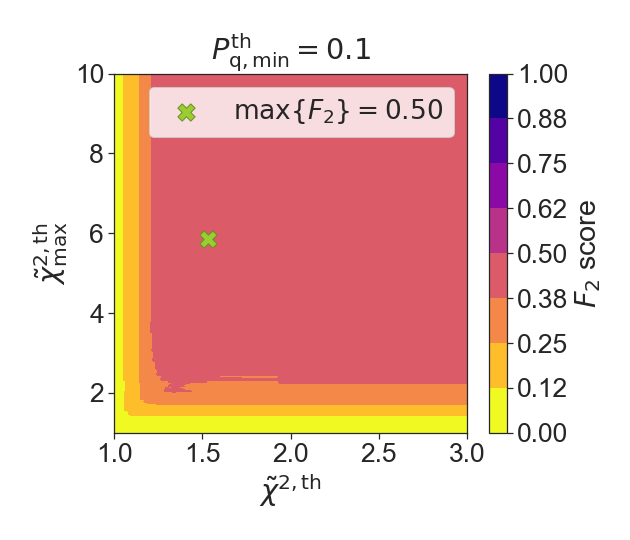}
    \includegraphics[scale=0.32]{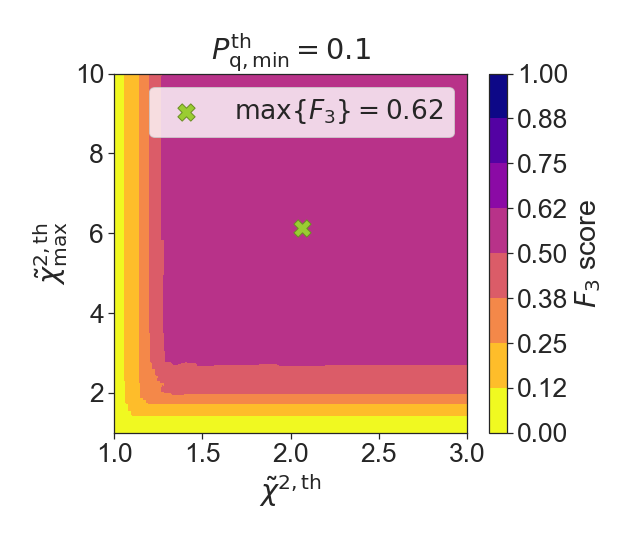}
\caption{The $F_{\beta}$ score of the manual threshold classifier as a function of the thresholds in the average and maximum residual chi-squared for $\beta=1$ (left column), $\beta=2$ (middle column) and $\beta=3$ (right column) as applied to the complete validation set \ie, $\Pq \geq 0$ (top row), the subset of the validation set where $\Pq \geq 0.1$ (middle row) and the subset of the validation set where $\Pq \geq 0.01$ (bottom row). The results are summarised in Table \ref{tab:fbeta_table}.}
\label{fig:manual_threshold}
\end{figure*}

\begin{figure*}[t]
\centering
    \includegraphics[scale=0.15]{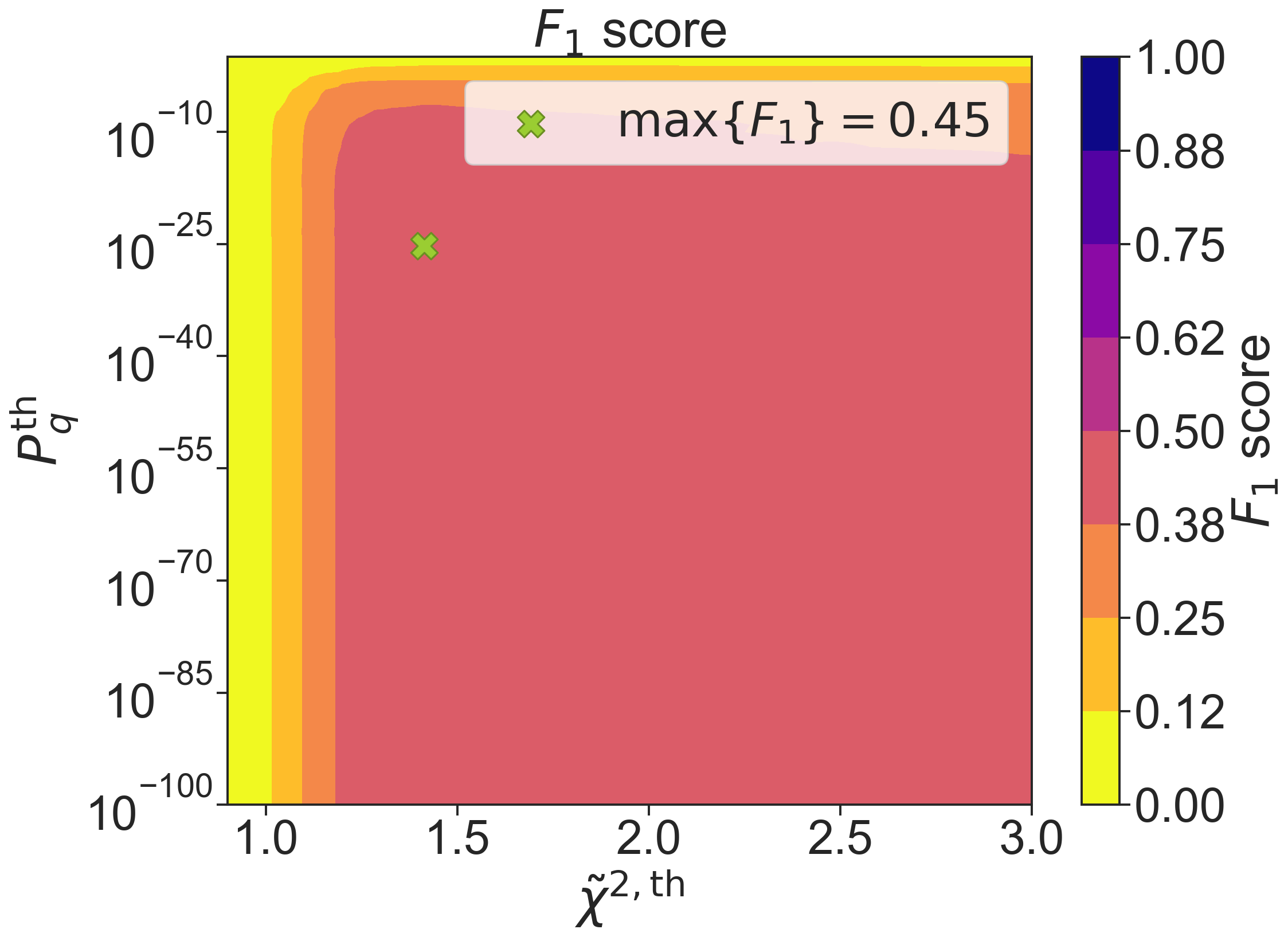}
    \includegraphics[scale=0.15]{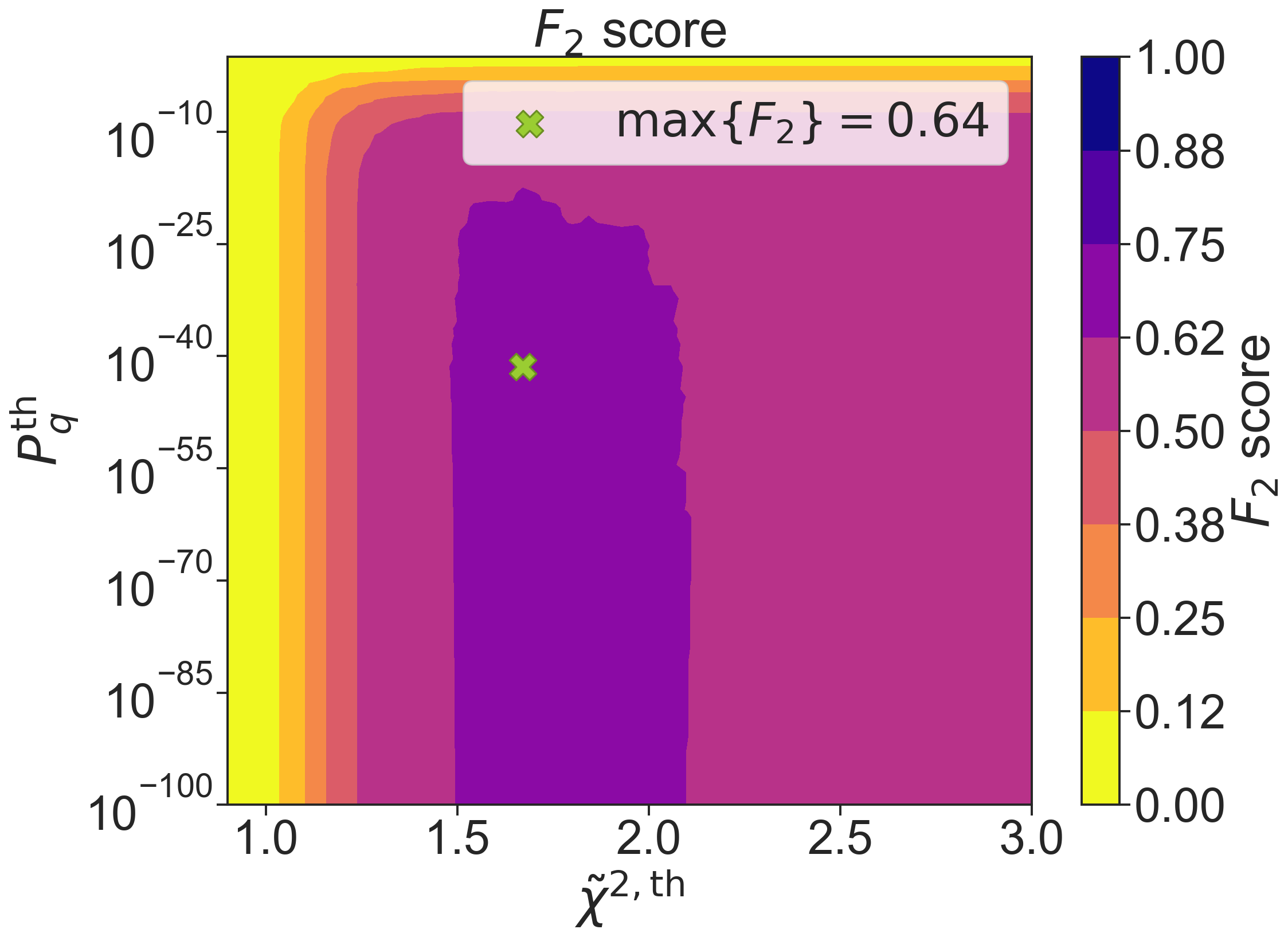}
    \includegraphics[scale=0.15]{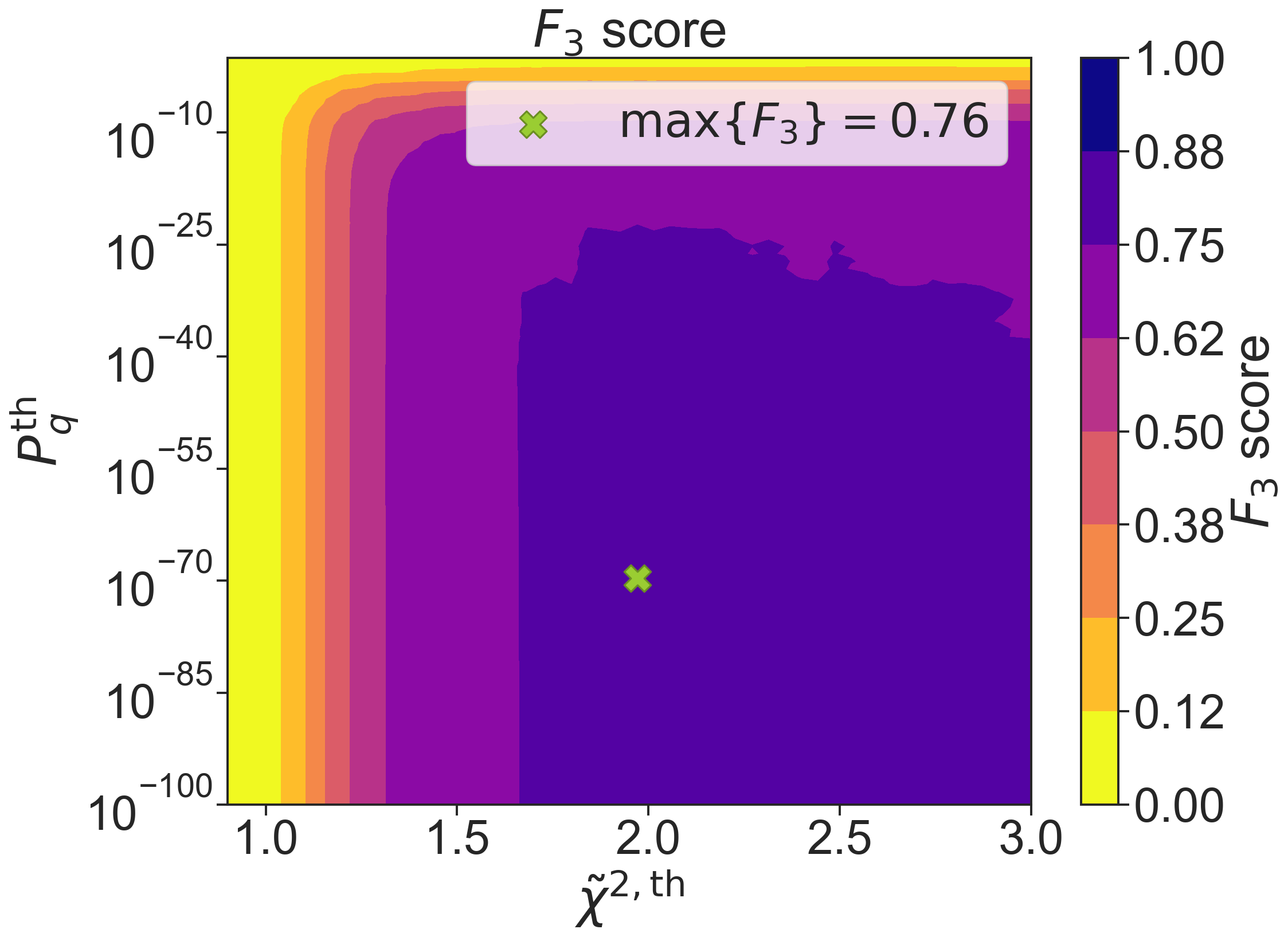}
\caption{The $F_{\beta}$ score of the manual threshold classifier as a function of the thresholds in the average residual chi-squared and the quasar probability for $\beta=1$ (left column), $\beta=2$ (middle column) and $\beta=3$ (right column) as applied to the complete validation set ($\Pq \geq 0$).}
\label{fig:fbeta_avgchi2_pq}
\end{figure*}

\begin{figure*}[t]
\centering
    \includegraphics[scale=0.15]{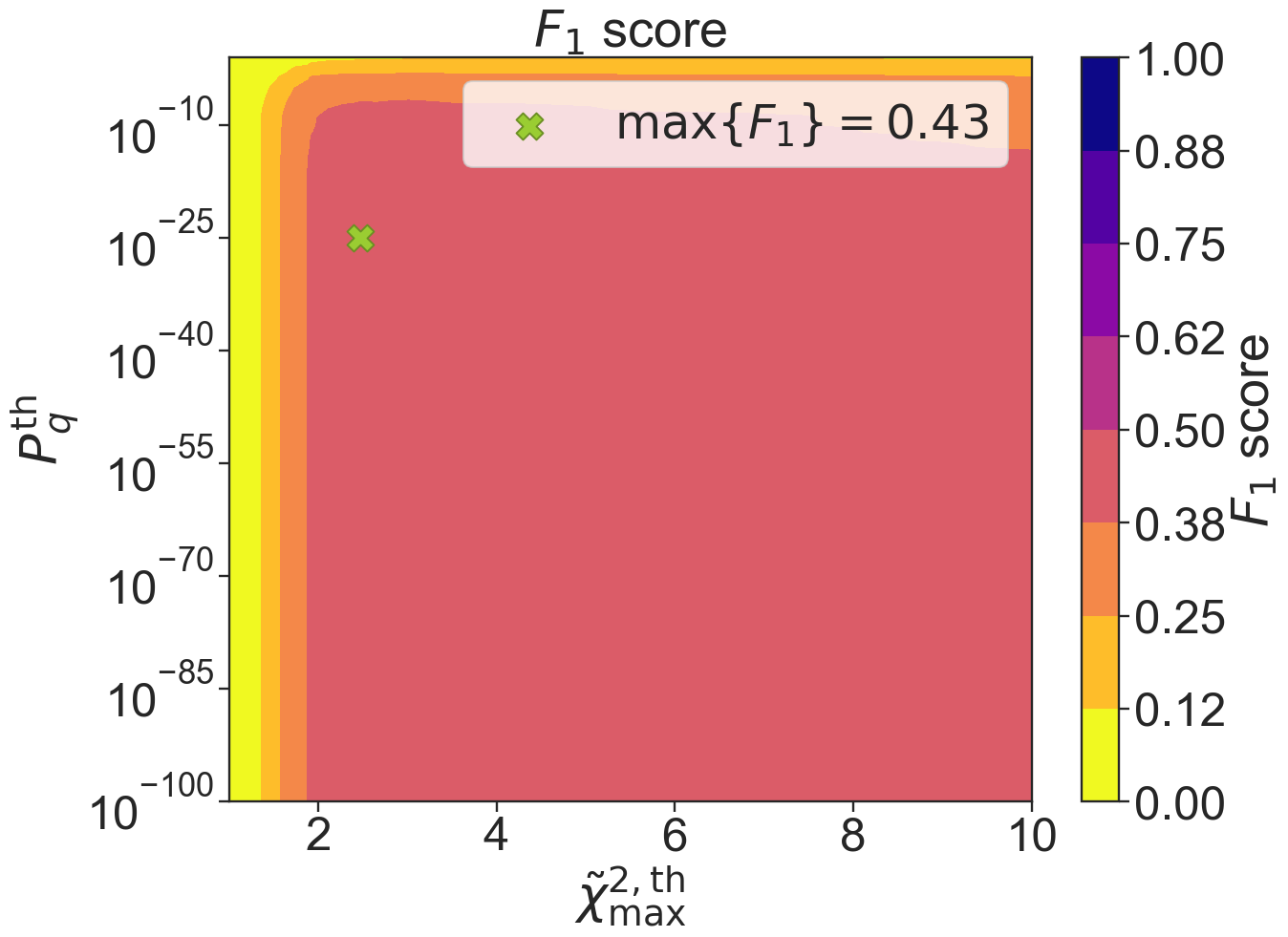}
    \includegraphics[scale=0.15]{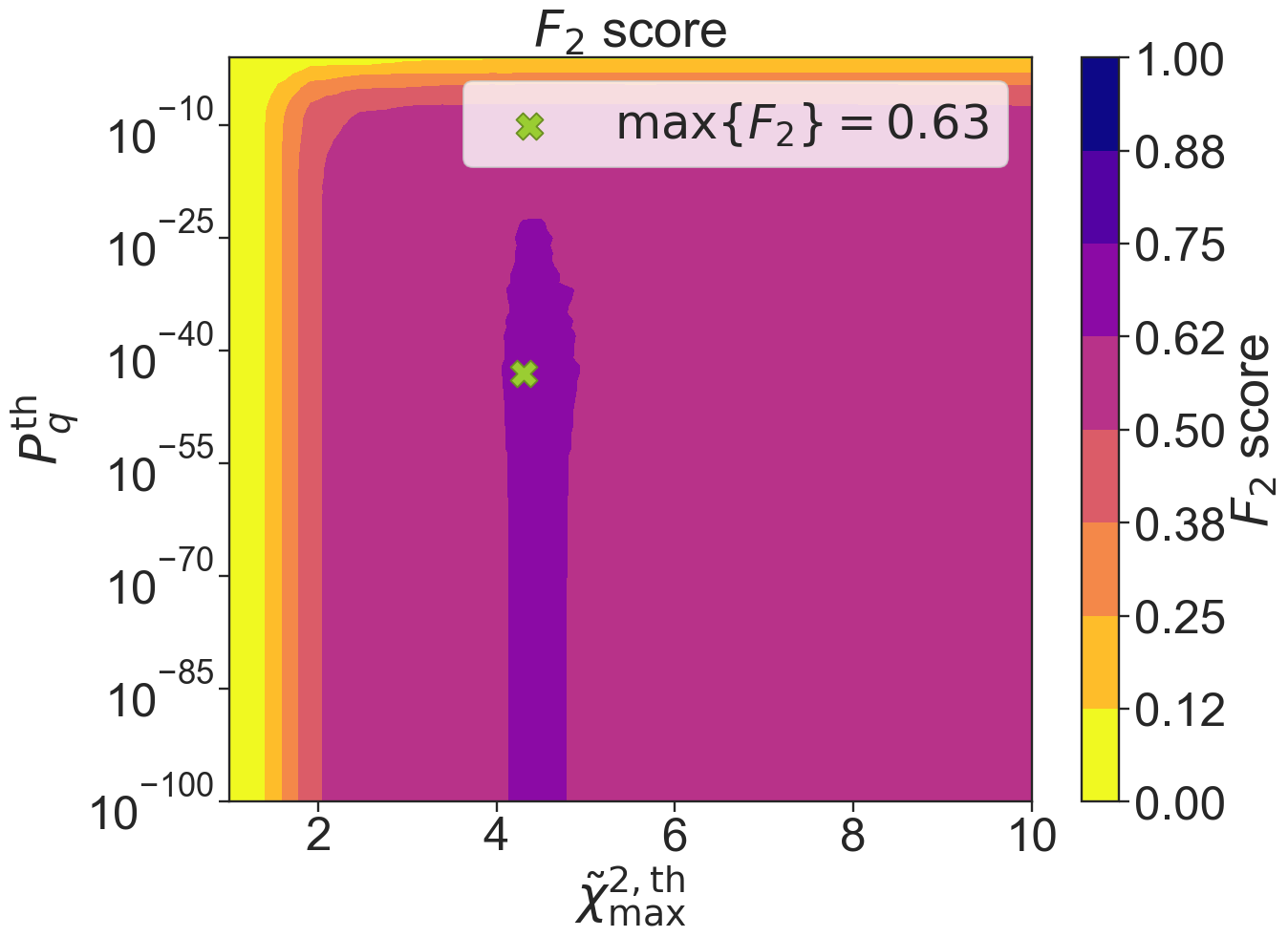}
    \includegraphics[scale=0.15]{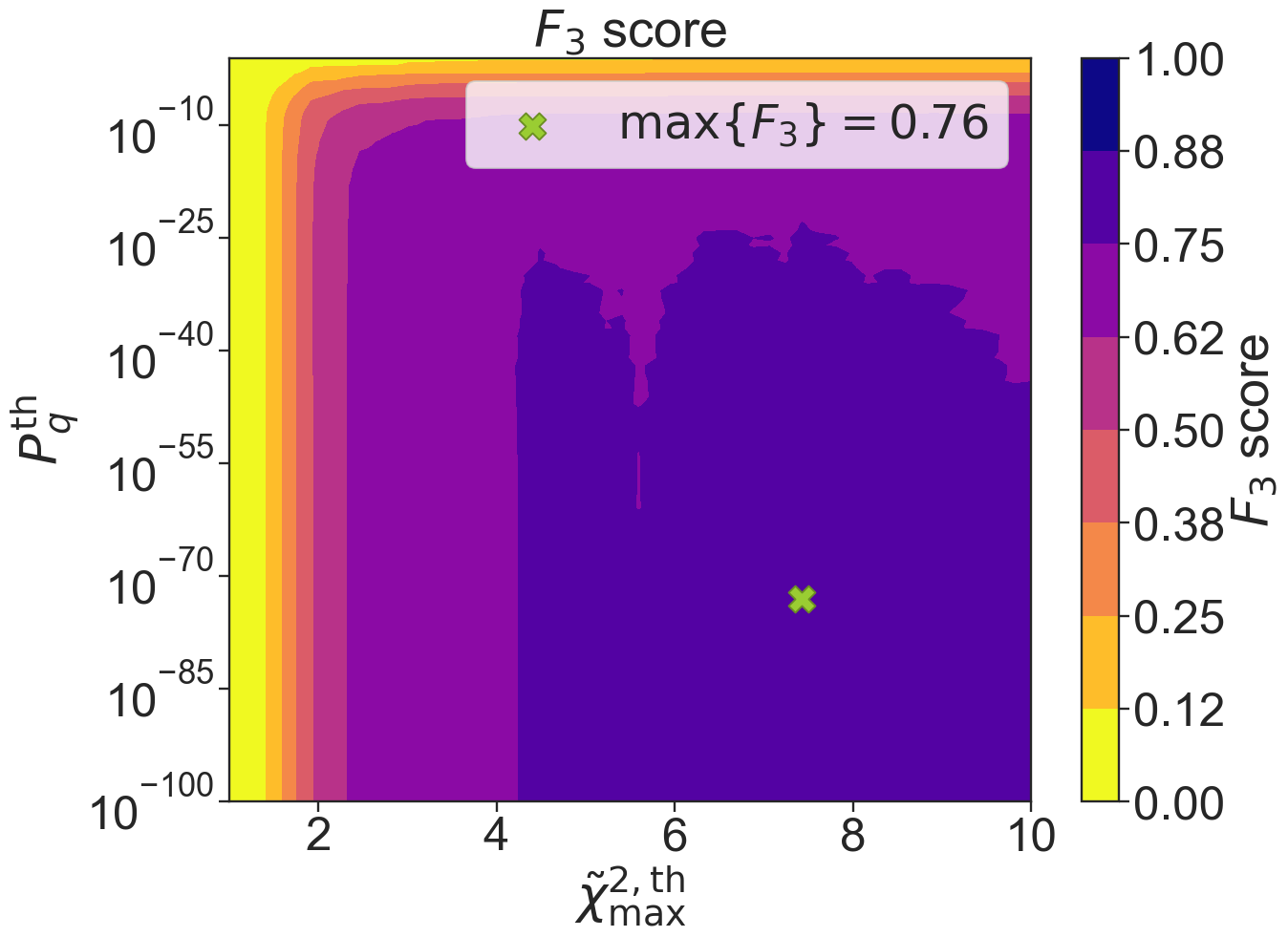}
\caption{The $F_{\beta}$ score of the manual threshold classifier as a function of the thresholds in the maximum residual chi-squared and the quasar probability for $\beta=1$ (left column), $\beta=2$ (middle column) and $\beta=3$ (right column) as applied to the complete validation set ($\Pq \geq 0$).}
\label{fig:fbeta_maxchi2_pq}
\end{figure*}

We present the classification results of our validation set using the various performance measures discussed in Section~\ref{sec:classification}.

We start by showing ROC curves for the three features $\Pq$, $\chiSq$ and $\chiSqMax$ individually in Figure \ref{fig:roc}.  These demonstrate reasonable performance, while also not being optimal as the joint discriminatory power of the features is not exploited.  We achieve the highest individual AUC score of $0.810$ when we use $\chiSqMax$ as our single metric together with $\cutPq=0.1$. Generally, similar AUC scores are achieved at probability thresholds $\cutPq=0.01$ and $\cutPq=0.1$ for all three metrics.  Only at $\cutPq=0$ does the metric $\Pq$ lead to a low AUC score of $0.505$, giving a large FPR for a TPR of $\geq 0.5$.

Better performance is achieved by combining these features, as shown in  \ref{fig:manual_threshold}, \ref{fig:fbeta_avgchi2_pq} and \ref{fig:fbeta_maxchi2_pq}, which show contour plots of the $F_{\beta}$ score for $\beta=1$, $\beta=2$ and $\beta=3$ as a function of threshold in two metrics each. We can see that the best $F_{\beta}$ scores are achieved when using the average and maximum residual chi-squared values $\cutChiSq, \cutChiSqMax$ for different $\Pq$ thresholds (Fig.~\ref{fig:manual_threshold}). There, the manual threshold classifier is applied on subsets of the validation set for quasar probability thresholds of $\cutPq=0$, $\cutPq=0.01$ and $\cutPq=0.1$ respectively. Notice that $\cutPq=0$ includes the full validation set, whereas $\cutPq=0.01$ and $\cutPq=0.1$ lead to subsets of the validation set.  These results are summarised in Table~\ref{tab:fbeta_table}.  The overall largest $F_{\beta}$ score is achieved by $\beta=3$ and $\cutPq=0$. It is given by $F_3=0.79$ and uses the thresholds $\cutChiSq=2.39$ and $\cutChiSqMax=6.83$. Comparing \ref{fig:fbeta_avgchi2_pq} and \ref{fig:fbeta_maxchi2_pq} to Fig.~\ref{fig:scatter_visual} suggests that a large number of our visually accepted candidates in the training set are more likely to be contaminants.

For all three probability cuts the maximum $F_{\beta}$ score increases with increasing $\beta$, for $\beta=1,2,3$, such that it is at its highest value at $\beta=3$.  Since a larger $\beta$ means that we increase the weighting on the recall, an increasing $F_{\beta}$ score with increasing $\beta$ means that the recall is larger than the precision. 

Further, the maximum scores of $F_1$, $F_2$ and $F_3$ decrease with increasing probability cut $\cutPq$.  $F_1$ has its highest value of $F_1=0.48$ at $\cutPq=0$. The value is low compared to the maximum $F_3$ score of $F_3=0.79$, also at $\cutPq=0$, suggesting that we have quite a lot of false positives in our sample set.  The high fraction of false positives is confirmed by very low precision values for all $F_{\beta}$ values.  The precision decreases with increasing $\cutPq$ for $F_2$ and $F_3$.  For $F_1$. there is a small increase in precision of $0.01$ between $\cutPq=0.01$ and $\cutPq=0.1$. 

In addition to the precision, the recall also decreases with increasing $\cutPq$.  A drop in recall with increasing $\cutPq$ means that the recall is larger for sources with $\Pq < \cutPq$ than it is for sources with $\Pq \geq \cutPq$.  Even though the $F_{\beta}$ scores are smaller for non-zero thresholds in the probability than they are for $\cutPq=0$, the visual classification labels will be more accurate for the set of $\cutPq > 0$.  If we assume that the majority of candidates with $\Pq < 0.01$ are contaminants, then any source with a positive label as per visual classification is actually much more likely to be a contaminant (see Section \ref{sec:validation_set}).

As the validation set gets smaller with increasing quasar probability threshold $\cutPq$, the regions of maximum $F_{\beta}$ score narrow down to smaller areas, and hence a smaller subset of possible thresholds in the residual average and maximum chi-squared $\cutChiSq, \cutChiSqMax$ remains.  Generally, as $\cutPq$ increases, the cuts in the average and maximum residual chi-squared are decreasing overall for each $\beta$ individually. 

At $\cutPq=0.1$, only $\sim\! 330$ sources are left in the validation set (see Section~\ref{sec:validation_set}). At such a low source count we can use $\beta=3$ which gives us a score of $F_3=0.72$ using the thresholds of $\cutChiSq=1.70$ and $\cutChiSqMax=5.96$. Even though we get a higher value of $F_3=0.79$ at $\cutChiSq=0$, the number of accepted candidates would include much more contaminants.

We also look at the case in which all three features are jointly optimised, which gives the results shown in Table~\ref{tab:fbeta_combined_table}.  The $F_\beta$ scores are slightly worse in this case due to the high number of visually accepted candidates with low $\Pq$ values, confirming that it is better to set this more interpretable threshold manually.

So far our focus has been on optimising the selection thresholds based on fairly general classification metrics (\ie, AUC or $F_\beta$), but a more pragmatic approach is to take an astronomical perspective by considering how one might operate a search for high-redshift quasars.  Given their rarity and scientific value (see Section~\ref{sec:intro}), it would be most natural to fix the recall at a high value and then determine how many follow-up observations would be required per discovery, which is $\sim\!1$/precision.  For the data-set analysed here, and with $\cutPq$ set to our standard value of 0.1 (\cf\ \citealt{Mortlock_etal:2012}), we find that fixing the recall to be 0.90 gives a maximum precision of 0.15 (obtained with $\chiSq = 1.75$ and $\chiSqMax = 4.41$), implying around six candidates would have to be followed up per discovery.  The search could be made more efficient (\eg, reducing the recall to 0.5 increases the precision to 0.27), but the loss of valuable scientific sources would be too great to make this a viable option.

\begin{table} 
\renewcommand{\arraystretch}{1.4}
    \centering
    \caption{Chi-squared thresholds which give the highest $F_\beta$ scores.}
    \label{tab:fbeta_table}
    \begin{tabular}{|c|c|c|c|c|c|c|}
    \hline
    $\beta$ & $\cutPq$ & $\cutChiSq$ & $\cutChiSqMax$ & precision & recall & $F_{\beta}$ \\
    \hline
    1 & 0.00 & 1.54 & 3.62 & 0.35 & 0.80 & 0.49 \\
    1 & 0.01 & 1.33 & 2.04 & 0.39 & 0.52 & 0.45 \\
    1 & 0.10 & 1.29 & 3.21 & 0.27 & 0.58 & 0.36 \\
    \hline 
    2 & 0.00 & 2.11 & 8.21 & 0.31 & 0.95 & 0.67 \\
    2 & 0.01 & 1.53 & 3.76 & 0.27 & 0.81 & 0.58 \\
    2 & 0.10 & 1.53 & 5.85 & 0.20 & 0.78 & 0.50 \\
    \hline 
    3 & 0.00 & 2.41 & 8.21 & 0.30 & 0.97 & 0.79 \\
    3 & 0.01 & 2.07 & 6.14 & 0.20 & 0.95 & 0.69 \\
    3 & 0.10 & 2.07 & 6.14 & 0.16 & 0.91 & 0.62 \\
    \hline 
    \end{tabular}
\end{table}

\begin{table} 
\renewcommand{\arraystretch}{1.4}
    \centering
    \caption{Combined thresholds which give the highest $F_\beta$ scores.}
    \label{tab:fbeta_combined_table}
    \begin{tabular}{|c|c|c|c|c|c|c|}
    \hline
    $\beta$ & $P_{\rm q}^{\rm th}$ & $\cutChiSq$ & $\cutChiSqMax$ & precision & recall & $F_{\beta}$ \\
    \hline
    1 & $10^{-23}$ & 1.42 & 3.06 & 0.32 & 0.77 & 0.46 \\
    \hline 
    2 & $10^{-43}$ & 1.64 & 4.58 & 0.29 & 0.91 & 0.64 \\
    \hline 
    3 & $10^{-92}$ & 2.96 & 7.46 & 0.25 & 0.98 & 0.76 \\
    \hline 
    \end{tabular}
\end{table}


\subsection{Individual objects}
\label{sec:discussion}

While we are interested primarily in the overall performance of our classifier (as described above), it is instructive to look at its performance on individual objects.


\begin{figure*}
    \centering
    \includegraphics[trim={2.75cm 1.1cm 2.2cm 1.5cm}, clip=True]{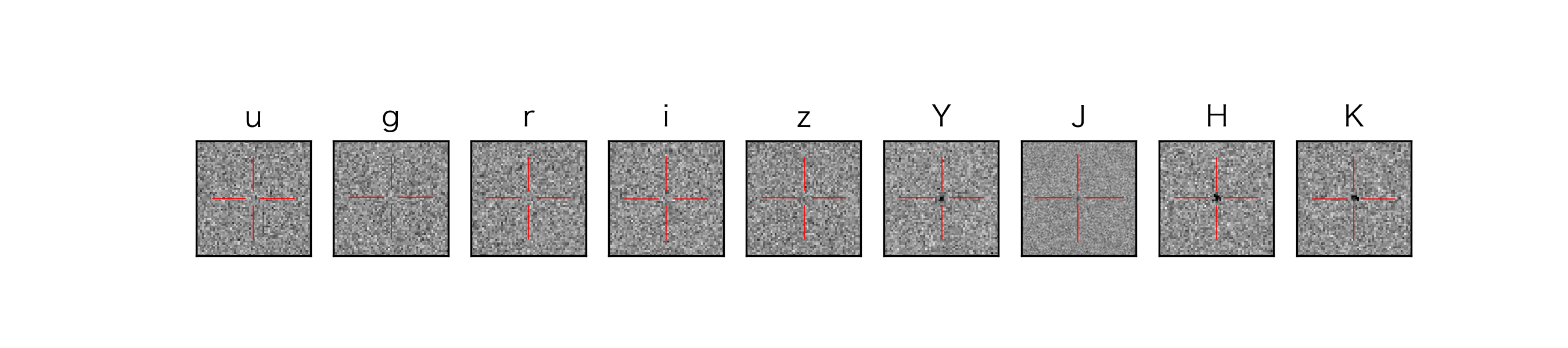}
    \includegraphics[trim={2.75cm 1.3cm 2.2cm 2.0cm}, clip=True]{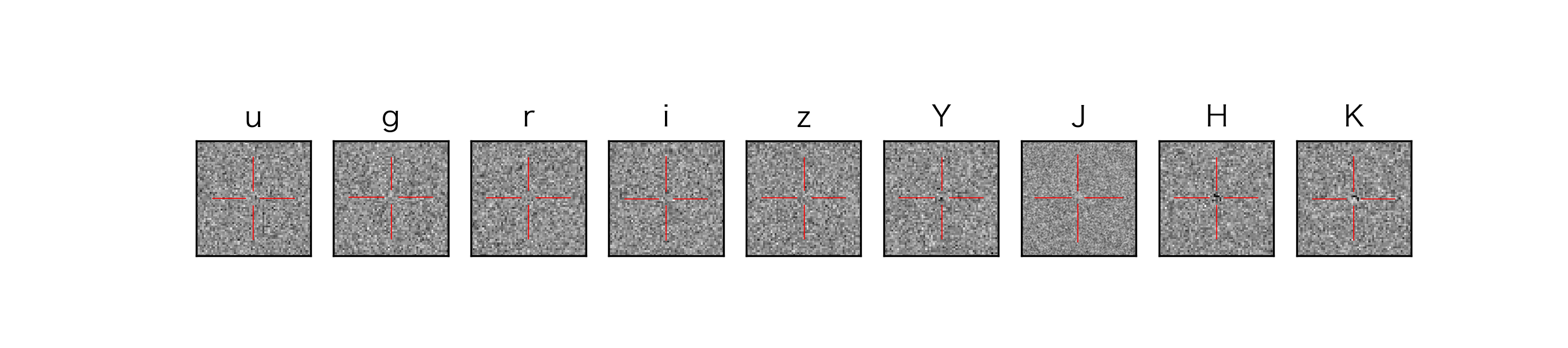}
    \caption{J000042.15$+$072229.9, a likely ETG, rejected as a candidate both due to both flux measurements and images \figcite}
    \label{fig:reconstruct_residual_etg}
\end{figure*}

{\bf J000042.15$+$072229.9:} shown in Fig.~\ref{fig:reconstruct_residual_etg}, has a high probability of being an ETG ($P_{\mathrm{g}}=1.00$).  The fact that this source is very unlikely to be a quasar is confirmed by a large maximum chi-squared value of $\chiSqMax=6.54$. The residuals show remaining flux, indicating that the best-fit quasar SED template does not reproduce the observed fluxes very well.


\begin{figure*}
\centering
    \includegraphics[trim={2.75cm 1.1cm 2.2cm 1.5cm}, clip=True]{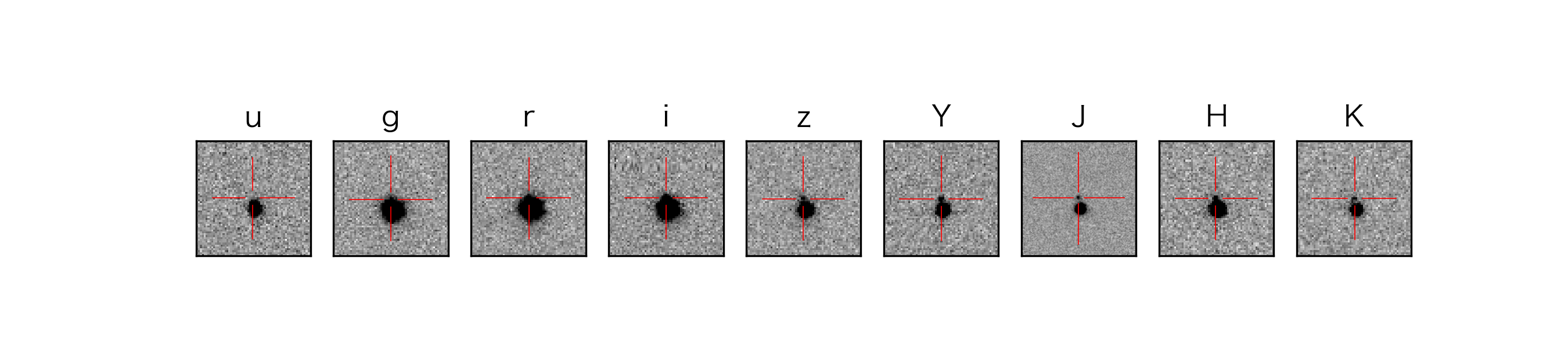}
    \includegraphics[trim={2.75cm 1.3cm 2.2cm 2.0cm}, clip=True]{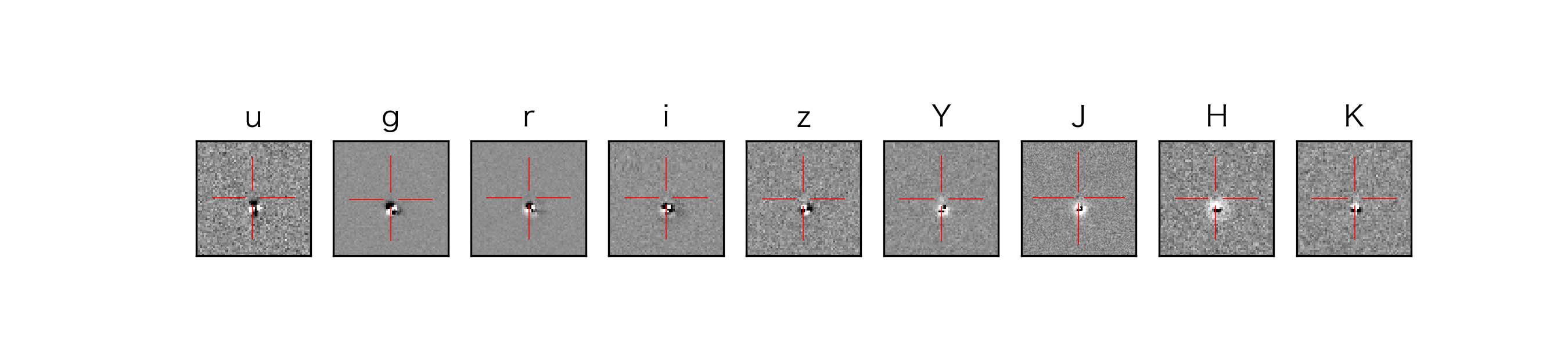}
\caption{J000226.75$-$005604.5, rejected due to the presence of a nearby bright star \figcite}
\label{fig:bright_star}
\end{figure*}

{\bf J000226.75$-$005604.5:} shown in Fig.~\ref{fig:bright_star} has a bright star in the vicinity of the source.  As the residual does not consist of noisy images only, this gives a low quasar probability $\Pq=0.00$ and a high residual chi-squared value $\chiSq =23.86$ as expected.


\begin{figure*}
    \centering
    \includegraphics[trim={2.75cm 1.1cm 2.2cm 1.5cm}, clip=True]{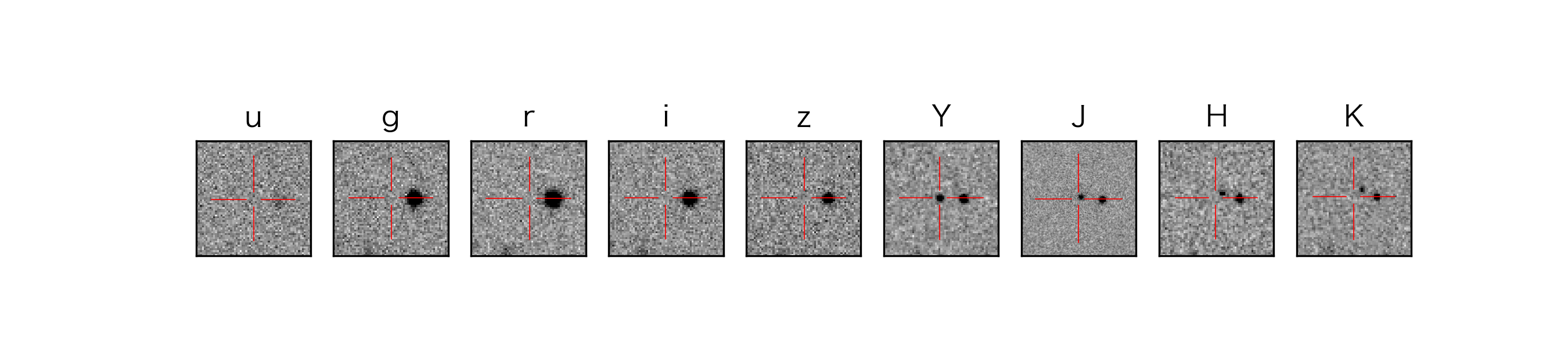}
    \includegraphics[trim={2.75cm 1.1cm 2.2cm 1.5cm}, clip=True]{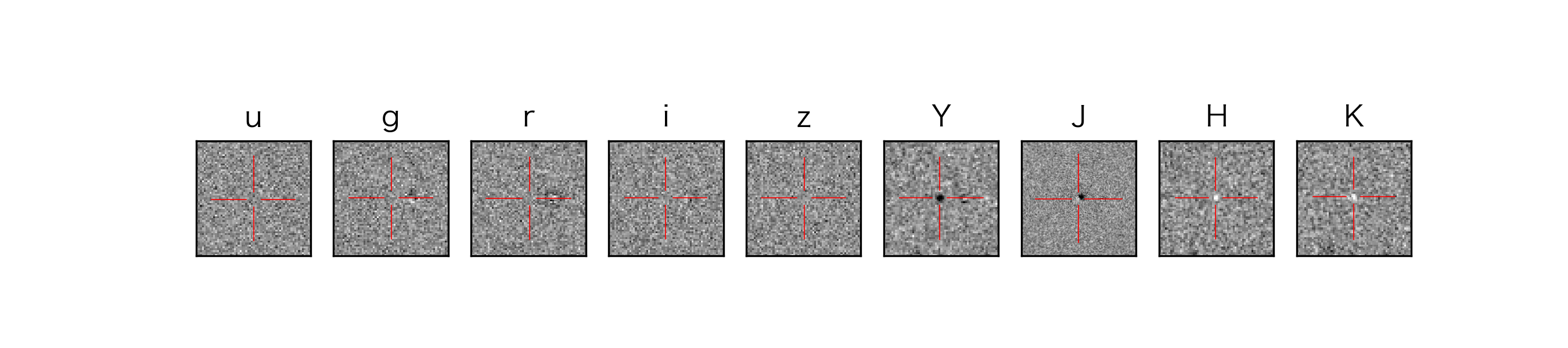}
    \caption{J002000.71--003412.8, a candidate which had incorrectly been selected following visual inspection but was (correctly) rejected by our method \figcite}
    \label{fig:moving_source}
\end{figure*}

{\bf J002000.71--003412.8:} shown in Fig.~\ref{fig:moving_source},  had been accepted based on visual inspection but should have been rejected as it clearly has an appreciable proper motion.  Our automated classifier correctly rejected this candidate due to large $\chi^2$ values in the $Y$, $H$ and $K$ bands.


\begin{figure*}
\centering
    \includegraphics[trim={2.75cm 1.1cm 2.2cm 1.5cm}, clip=True]{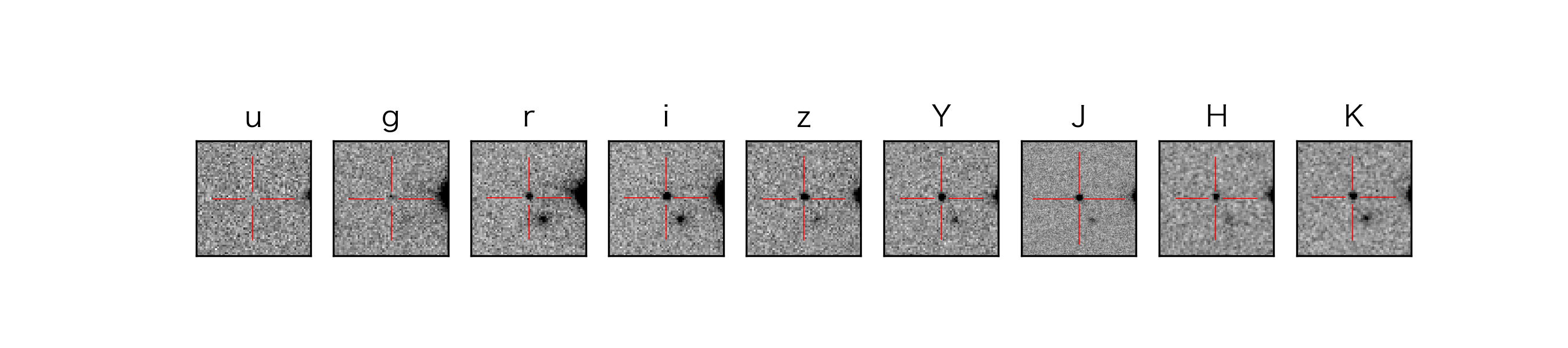}
    \includegraphics[trim={2.75cm 1.3cm 2.2cm 2.0cm}, clip=True]{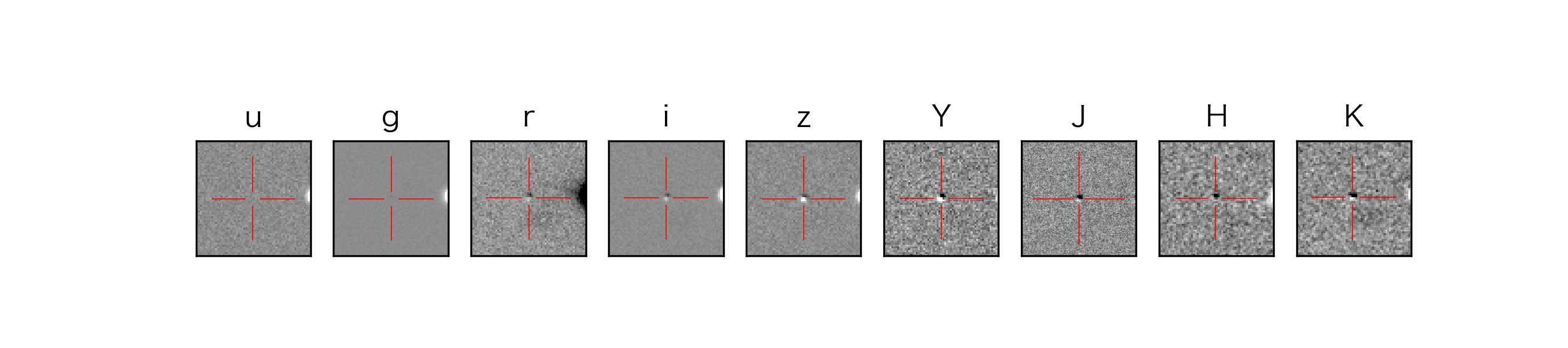}
\caption{J002101.25+004352.4, rejected due to the presence of a close companion \figcite}
\label{fig:pair}
\end{figure*}

{\bf J002101.25+004352.4:} shown in Fig.~\ref{fig:pair} was rejected due to the presence of a close companion. In addition, it has a bright source nearby. The residual, with its characteristic `dipolar' form, indicates that either the source is not exactly at the centre or the source is not point-like or rotationally-symmetric.


\begin{figure*}
\centering
    \includegraphics[trim={2.75cm 1.1cm 2.2cm 1.5cm}, clip=True]{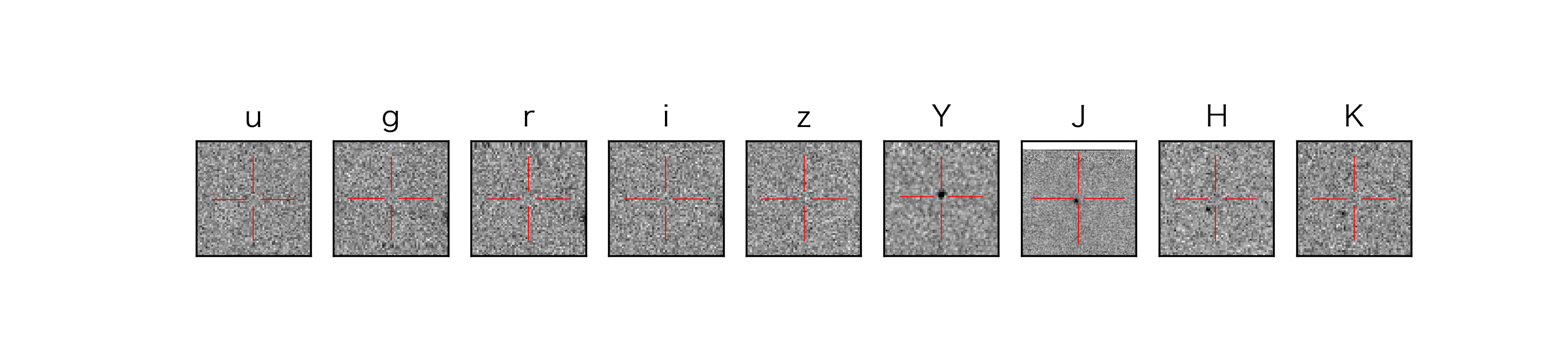}
    \includegraphics[trim={2.75cm 1.3cm 2.2cm 2.0cm}, clip=True]{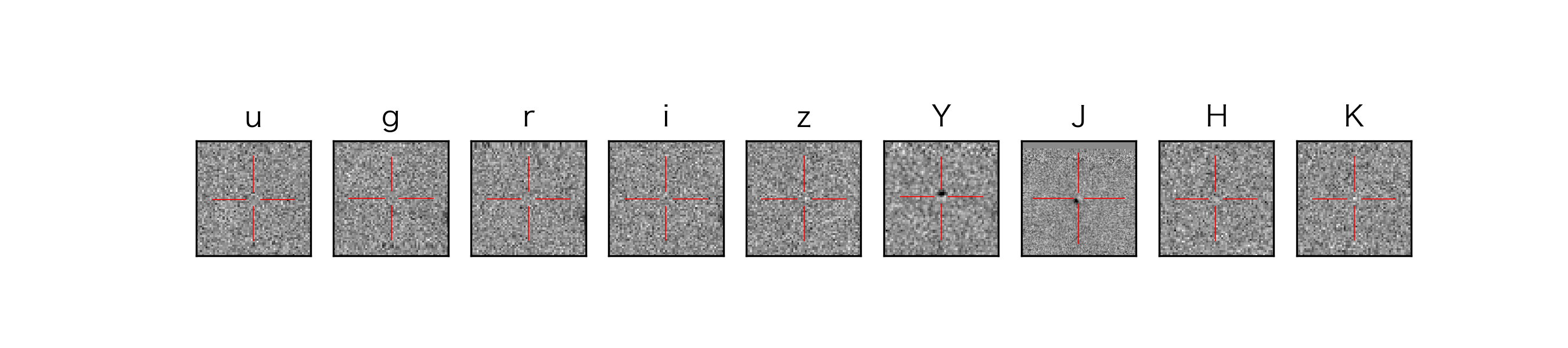}
\caption{J002318.32+000435.8, rejected as its position changed between the SDSS and UKIDSS observations \figcite}
\label{fig:moving}
\end{figure*}

{\bf J002318.32+000435.8:} shown in Fig.~\ref{fig:moving}, was rejected as a moving object because the with $\chiSq=2.72$.


\begin{figure*}[ht]
    \centering
    \includegraphics[trim={2.75cm 1.1cm 2.2cm 1.5cm}, clip=True]{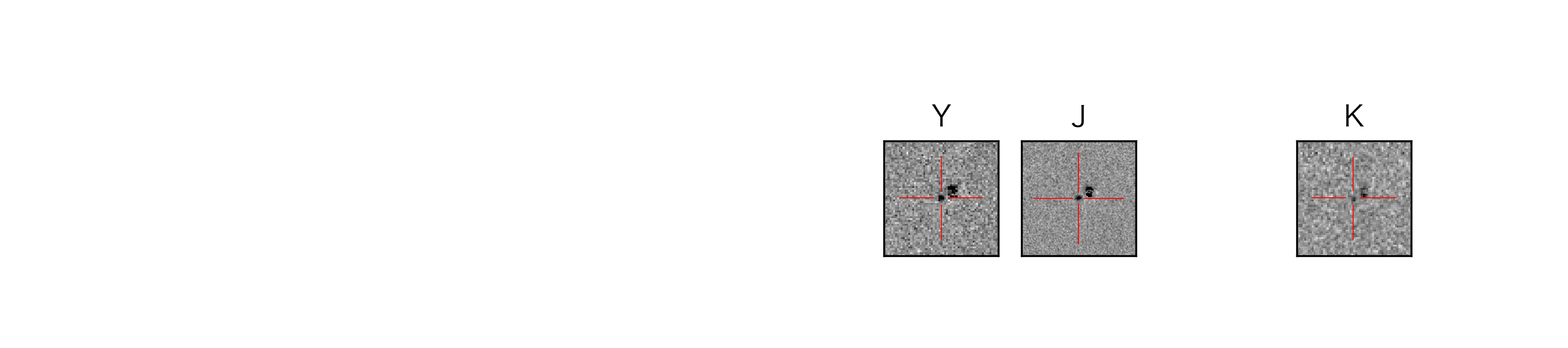}
    \includegraphics[trim={2.75cm 1.1cm 2.2cm 1.5cm}, clip=True]{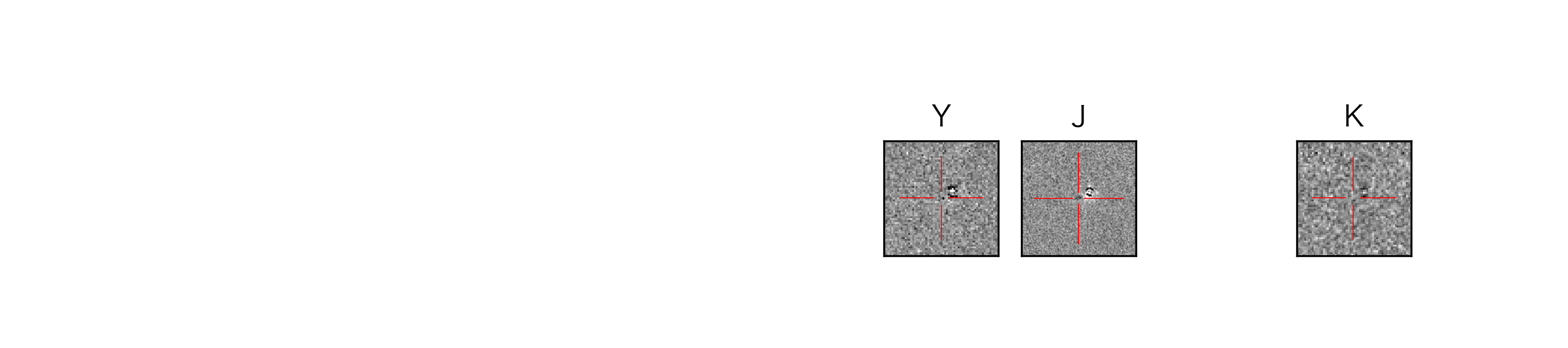}
    \caption{J033404.44$+$001003.4, which was rejected as UKIDSS cross-talk but not (always) rejected by the classifier \figcite}
    \label{fig:crosstalk}
\end{figure*}

{\bf J033404.44$+$001003.4:} shown in Fig.~\ref{fig:crosstalk}, was initially and correctly rejected as being an example of UKIDSS cross-talk \citep{Dye_etal:2006}, but is sometimes -- depending on thresholds -- selected by the automated classifier, as the deviation from the expected PSF of an astronomical point-source is small.  The candidate highlights the importance of using external information, beyond the pixel values, when available.


\begin{figure*}[ht]
\centering
    \includegraphics[trim={2.75cm 1.1cm 2.2cm 1.5cm}, clip=True]{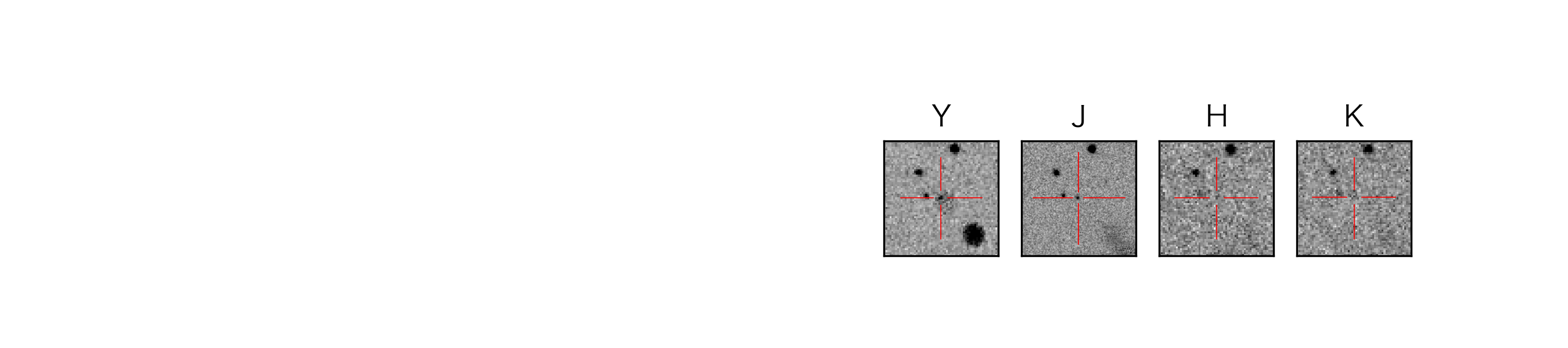}
    \includegraphics[trim={2.75cm 1.3cm 2.2cm 2.0cm}, clip=True]{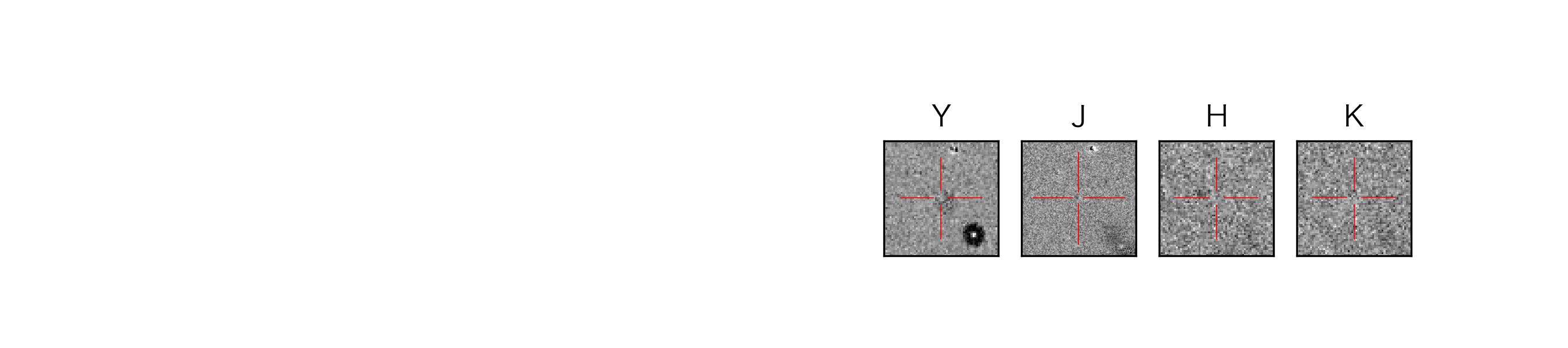}
\caption{J073749.93+261844.9, rejected due to persistence \figcite}
\label{fig:persistence}
\end{figure*}

{\bf J073749.93+261844.9:} shown in Fig.~\ref{fig:persistence}, is a persistence image. The SDSS images have been neglected as they have data processing issues flagged in the database (\verb|PSF_FLUX_INTERP| and in some \verb|COSMIC_RAY|). The pipeline fails to identify this source as it does conform to our model of a stationary point-source that follows the SED of one of our quasar templates. Thus, it gives a residual chi-squared of $\chiSq=1.00$ and a quasar probability of $\Pq=0.06$.


\begin{figure*}
\centering
    \includegraphics[trim={2.75cm 1.1cm 2.2cm 1.5cm}, clip=True]{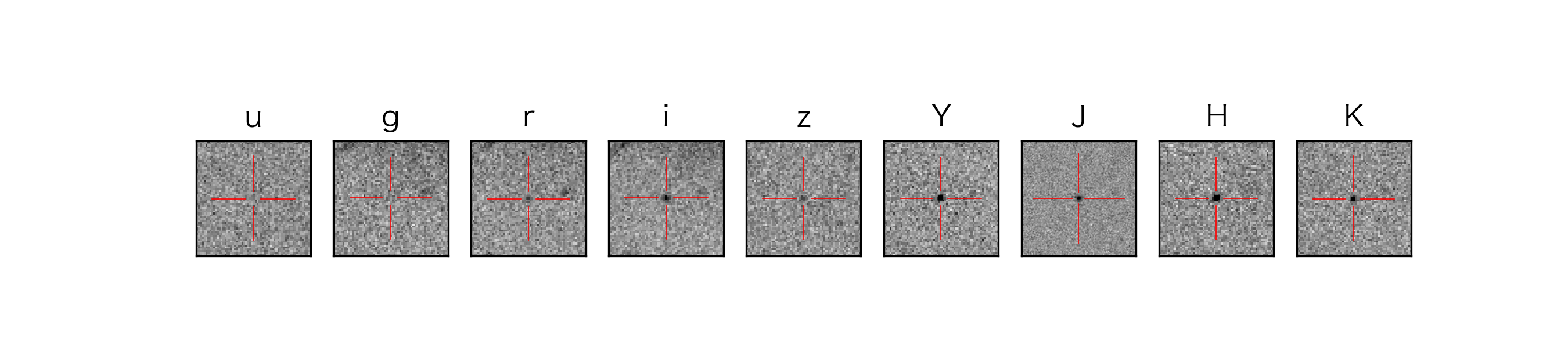}
    \includegraphics[trim={2.75cm 1.3cm 2.2cm 2.0cm}, clip=True]{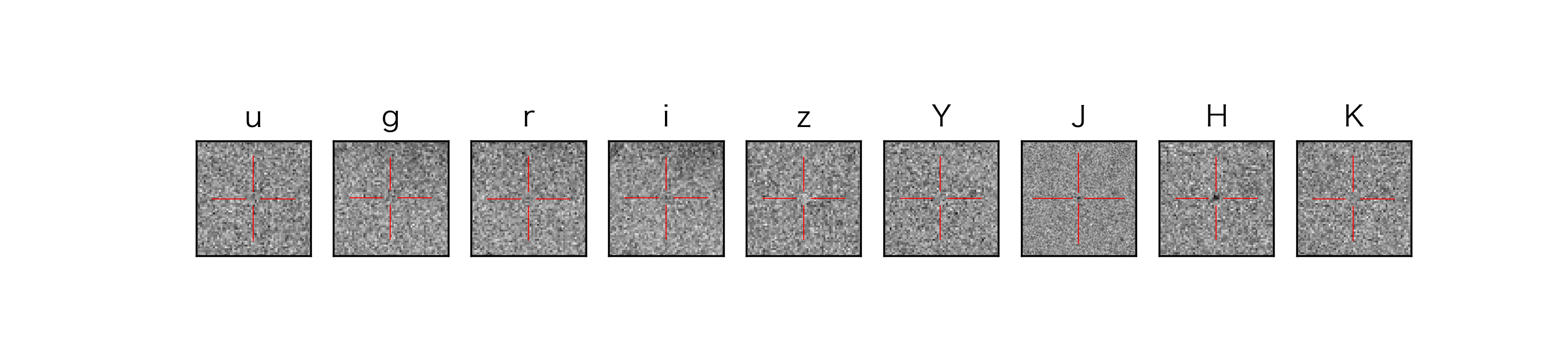}
\caption{J091108.35+092732.0, rejected due to its proximity to a diffraction spike \figcite}
\label{fig:diffraction}
\end{figure*}

{\bf J091108.35+092732.0:} shown in Fig.~\ref{fig:diffraction}, is close to a diffraction spike. In this case the model is a good representation again since the diffraction spikes do not occur within the cross-hair (where the calculation is made). The largest chi-squared value occurs in the $H$-band with $\chi^2_H=2.86$. However, the overall reduced chi-squared is low at $\chiSq=1.18$. Nevertheless, this source has a zero quasar probability $\Pq=0.00$ and hence would be rejected with a cut in the quasar probability.


\begin{figure*}
    \centering
    \includegraphics[trim={2.75cm 1.1cm 2.2cm 1.5cm}, clip=True]{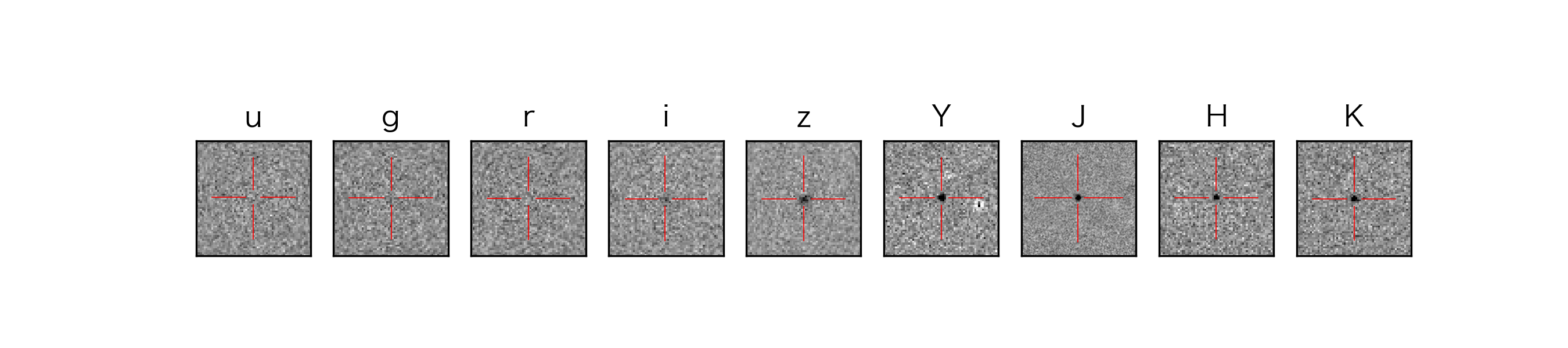}
    \includegraphics[trim={2.75cm 1.1cm 2.2cm 1.5cm}, clip=True]{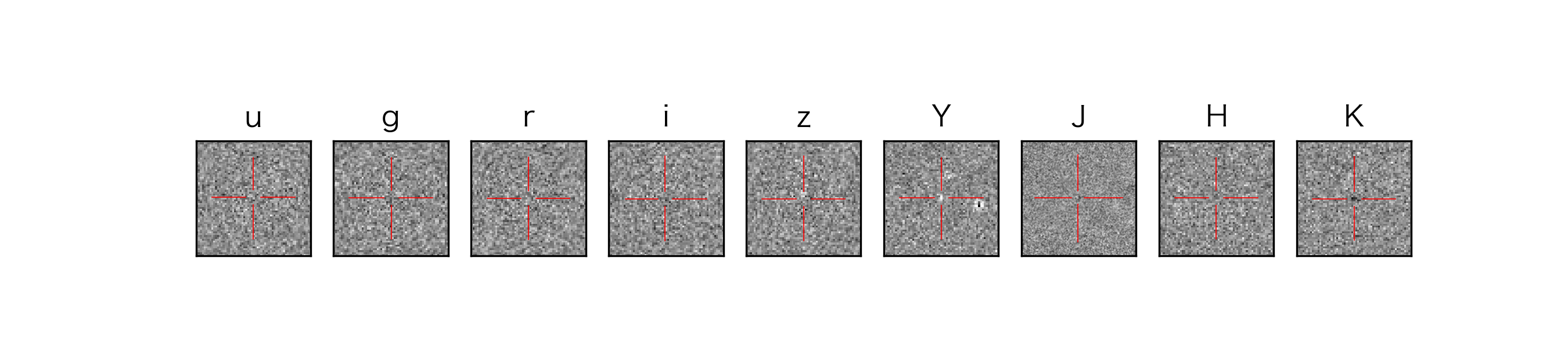}
    \protect\caption{J104433.04-012501.9, a redshift $z = 5.74$ quasar \citep{Fan_etal:2000} that was incorrectly rejected due to poor quasar SED fits \figcite}
    \label{fig:rejected_quasar}
\end{figure*}

{\bf J104433.04-012501.9:} shown in Fig.~\ref{fig:rejected_quasar}, is a redshift $z = 5.74$ quasar \citep{Fan_etal:2000,Goodrich_etal:2001} which was incorrectly rejected by our classifier if a cut of $\cutPq=0.01$ is adopted.  It has a reasonable best-fit redshift of $\hat{z}=6.00$ but, due to its due its $Y$-band flux it has a minimum chi-squared value of $\chi^2_Y=\chiSqMax=4.56$.  None of our nine quasar SED templates are a good match for this quasar, indicating that a possible improvement would be to add more quasar SED templates. 


{\bf J114803.28+070208.3:} is a redshift $z = 6.34$ quasar \citep{Mortlock_etal:2012,Jiang_etal:2016} which was rejected by our classifier as it had $\Pq = 1.22 \times 10^{-6}$.  This is because this source had an unwanted flag ({\tt COSMIC\_RAY}) in the SDSS database and so our pipeline did not use the SDSS photometry. Upon including the SDSS images as well, the probability increased $\Pq=1.00$. 


\begin{figure*}
    \centering
    \includegraphics[trim={2.75cm 1.1cm 2.2cm 1.5cm}, clip=True]{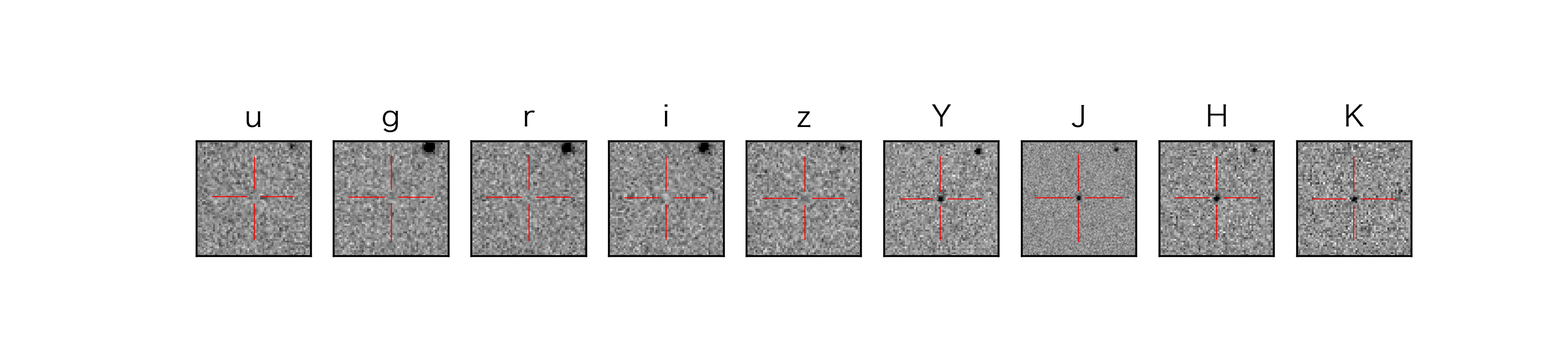}
    \includegraphics[trim={2.75cm 1.1cm 2.2cm 1.5cm}, clip=True]{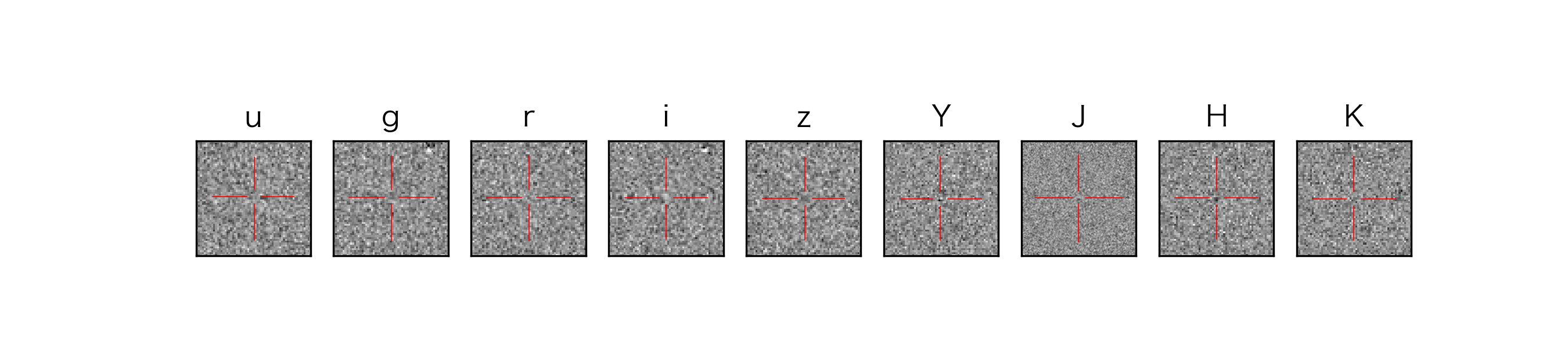}
    \caption{J115132.21$+$07518.8, a candidate which had incorrectly been rejected following visual inspection but was (correctly) selected by our method \figcite}
    \label{fig:good_source_rejected_visually}
\end{figure*}

{\bf J115132.21+07518.8:} shown in Fig.~\ref{fig:good_source_rejected_visually}, had originally been rejected based on visual inspection, in part because the $H$ and $K$ band data were not available at that time (2011).  Our pipeline, which automatically downloaded the new $H$ and $K$ images, gave this candidate a quasar probability of $\Pq=0.56$ and $\chiSq=1.24$ and so it was selected.  Fortunately, a definitive classification was possible without the need for follow-up observations as it is has $i=23.00 \pm 0.19$ and $z=21.33\pm 0.08$ in Pan-STARRS; including these data reduced its quasar probability to $\Pq=0.01$.


\section{Conclusions}
\label{sec:conclusion}

We have developed an automated method to identify candidate high-redshift ($z \gtrsim 7$) quasars in broad-band optical-NIR surveys.  This combines Bayesian model comparison of data from photometric catalogues, building on the methods presented previously by \citet{Mortlock_etal:2012}, \citet{Barnett_etal:2019} and \citet{Barnett_etal:2021}, with goodness-of-fit tests on image-level data. The former exploits the prior knowledge of the target and contaminating astrophysical populations; the latter exploits the fact that it is reasonable to reject any source which is not stationary and point-like without the need for explicit modelling of the huge number of possible contaminants (including non-astronomical artefacts).

We first tested our methods on simulations, using the $F_\beta$ statistic to assess performance so that the imbalanced training data and preference for completeness over efficiency could be properly incorporated.  We then embedded the classification software in an automated pipeline to test it on on real image data from cross-matched SDSS and UKIDSS catalogues. Our method improved on visual inspection both in terms of overall performance and in identifying several sources which were previously mis-classified, presumably due to data-entry errors.  Of course, any automated classifier is limited by the models it uses (for a traditional statistical method) or the training data (for ML methods), and there are several unusual objects which the method incorrectly rejects, so a potential improvement would be to adopt a more diverse set of quasar SED templates.

Looking ahead, the next step is to deploy these methods on data from surveys like {\em Euclid} and LSST.  Aside from the obvious need for new pipelines for these very different databases, it will also be necessary to to tune the selection thresholds as appropriate for the depth, wavelength coverage and noise properties of these data-sets.  Even though these catalogues will be too large for visual inspection of all candidates, it should still be plausible to inspect a representative sample of sources for this purpose.  One option, not explored here, would then be to use an automated ML classifier to set selection boundaries, which would then yield a fully automated, and hence repeatable and objective, method for finding the most distant quasars.


\section*{Acknowledgments}

We thank Steve Warren for providing numerous insights throughout this project, in particular regarding the tests on the SDSS-UKIDSS data.  We thank the anonymous referee for their careful reading of the paper and constructive suggestions.


\section*{Data Availability}

No new data was generated 


\bibliographystyle{mnras}
\bibliography{references}

\begin{thebibliography}{}
\makeatletter
\relax
\def\mn@urlcharsother{\let\do\@makeother \do\$\do\&\do\#\do\^\do\_\do\%\do\~}
\def\mn@doi{\begingroup\mn@urlcharsother \@ifnextchar [ {\mn@doi@}
  {\mn@doi@[]}}
\def\mn@doi@[#1]#2{\def\@tempa{#1}\ifx\@tempa\@empty \href
  {http://dx.doi.org/#2} {doi:#2}\else \href {http://dx.doi.org/#2} {#1}\fi
  \endgroup}
\def\mn@eprint#1#2{\mn@eprint@#1:#2::\@nil}
\def\mn@eprint@arXiv#1{\href {http://arxiv.org/abs/#1} {{\tt arXiv:#1}}}
\def\mn@eprint@dblp#1{\href {http://dblp.uni-trier.de/rec/bibtex/#1.xml}
  {dblp:#1}}
\def\mn@eprint@#1:#2:#3:#4\@nil{\def\@tempa {#1}\def\@tempb {#2}\def\@tempc
  {#3}\ifx \@tempc \@empty \let \@tempc \@tempb \let \@tempb \@tempa \fi \ifx
  \@tempb \@empty \def\@tempb {arXiv}\fi \@ifundefined
  {mn@eprint@\@tempb}{\@tempb:\@tempc}{\expandafter \expandafter \csname
  mn@eprint@\@tempb\endcsname \expandafter{\@tempc}}}

\bibitem[\protect\citeauthoryear{{Akeson} et~al.,}{{Akeson}
  et~al.}{2019}]{Akeson_etal:2019}
{Akeson} R.,  et~al., 2019, \mn@doi [arXiv e-prints]
  {10.48550/arXiv.1902.05569}, \href
  {https://ui.adsabs.harvard.edu/abs/2019arXiv190205569A} {p. arXiv:1902.05569}

\bibitem[\protect\citeauthoryear{{Almeida}, {Baugh}  \& {Lacey}}{{Almeida}
  et~al.}{2007}]{Almeida_etal:2007}
{Almeida} C.,  {Baugh} C.~M.,   {Lacey} C.~G.,  2007, \mn@doi [\mnras]
  {10.1111/j.1365-2966.2007.11530.x}, \href
  {https://ui.adsabs.harvard.edu/abs/2007MNRAS.376.1711A} {376, 1711}

\bibitem[\protect\citeauthoryear{{Anders}, {Gieles}  \& {de Grijs}}{{Anders}
  et~al.}{2006}]{Anders_etal:2006}
{Anders} P.,  {Gieles} M.,   {de Grijs} R.,  2006, \mn@doi [\aap]
  {10.1051/0004-6361:20054175}, \href
  {https://ui.adsabs.harvard.edu/abs/2006A&A...451..375A} {451, 375}

\bibitem[\protect\citeauthoryear{{Artigau}, {Bouchard}, {Doyon}  \&
  {Lafreni{\`e}re}}{{Artigau} et~al.}{2009}]{Artigau_etal:2009}
{Artigau} {\'E}.,  {Bouchard} S.,  {Doyon} R.,   {Lafreni{\`e}re} D.,  2009,
  \mn@doi [\apj] {10.1088/0004-637X/701/2/1534}, \href
  {https://ui.adsabs.harvard.edu/abs/2009ApJ...701.1534A} {701, 1534}

\bibitem[\protect\citeauthoryear{{Ba{\~n}ados} et~al.,}{{Ba{\~n}ados}
  et~al.}{2016}]{Banados_etal:2016}
{Ba{\~n}ados} E.,  et~al., 2016, \mn@doi [\apjs] {10.3847/0067-0049/227/1/11},
  \href {https://ui.adsabs.harvard.edu/abs/2016ApJS..227...11B} {227, 11}

\bibitem[\protect\citeauthoryear{{Ba{\~n}ados} et~al.,}{{Ba{\~n}ados}
  et~al.}{2018}]{Banados_etal:2018}
{Ba{\~n}ados} E.,  et~al., 2018, \mn@doi [\nat] {10.1038/nature25180}, \href
  {https://ui.adsabs.harvard.edu/abs/2018Natur.553..473B} {553, 473}

\bibitem[\protect\citeauthoryear{{Ba{\~n}ados} et~al.,}{{Ba{\~n}ados}
  et~al.}{2021}]{Banados_etal:2021}
{Ba{\~n}ados} E.,  et~al., 2021, \mn@doi [\apj] {10.3847/1538-4357/abe239},
  \href {https://ui.adsabs.harvard.edu/abs/2021ApJ...909...80B} {909, 80}

\bibitem[\protect\citeauthoryear{{Baeza-Yates} \& {Ribeiro-Neto}}{{Baeza-Yates}
  \& {Ribeiro-Neto}}{2011}]{Baeza-Yates_Ribeiro-Neto:2011}
{Baeza-Yates} Y.,  {Ribeiro-Neto} B.,  2011, Modern Information Retrieval: The
  Concepts and Technology Behind Search.
Harlow: Addison-Wesley

\bibitem[\protect\citeauthoryear{{Bailer-Jones}, {Smith}, {Tiede}, {Sordo}  \&
  {Vallenari}}{{Bailer-Jones} et~al.}{2008}]{Bailer-Jones_etal:2008}
{Bailer-Jones} C.~A.~L.,  {Smith} K.~W.,  {Tiede} C.,  {Sordo} R.,
  {Vallenari} A.,  2008, \mn@doi [\mnras] {10.1111/j.1365-2966.2008.13983.x},
  \href {https://ui.adsabs.harvard.edu/abs/2008MNRAS.391.1838B} {391, 1838}

\bibitem[\protect\citeauthoryear{{Bailer-Jones}, {Fouesneau}  \&
  {Andrae}}{{Bailer-Jones} et~al.}{2019}]{Bailer-Jones_etal:2019}
{Bailer-Jones} C. A.~L.,  {Fouesneau} M.,   {Andrae} R.,  2019, \mn@doi
  [\mnras] {10.1093/mnras/stz2947}, \href
  {https://ui.adsabs.harvard.edu/abs/2019MNRAS.490.5615B} {490, 5615}

\bibitem[\protect\citeauthoryear{{Barnett}, {Warren}, {Cross}, {Mortlock},
  {Fan}, {Wang}  \& {Hewett}}{{Barnett} et~al.}{2021}]{Barnett_etal:2021}
{Barnett} R.,  {Warren} S.~J.,  {Cross} N.~J.~G.,  {Mortlock} D.~J.,  {Fan} X.,
   {Wang} F.,   {Hewett} P.~C.,  2021, \mn@doi [\mnras]
  {10.1093/mnras/staa3808}, \href
  {https://ui.adsabs.harvard.edu/abs/2021MNRAS.501.1663B} {501, 1663}

\bibitem[\protect\citeauthoryear{{Bernstein}}{{Bernstein}}{2002}]{Bernstein:2002}
{Bernstein} G.,  2002, \mn@doi [\pasp] {10.1086/337997}, \href
  {https://ui.adsabs.harvard.edu/abs/2002PASP..114...98B} {114, 98}

\bibitem[\protect\citeauthoryear{{Bertin} \& {Arnouts}}{{Bertin} \&
  {Arnouts}}{1996}]{Bertin_Arnouts:1996}
{Bertin} E.,  {Arnouts} S.,  1996, \mn@doi [\aaps] {10.1051/aas:1996164}, \href
  {https://ui.adsabs.harvard.edu/abs/1996A&AS..117..393B} {117, 393}

\bibitem[\protect\citeauthoryear{{Bogd{\'a}n} et~al.,}{{Bogd{\'a}n}
  et~al.}{2024}]{Bogdan_etal:2024}
{Bogd{\'a}n} {\'A}.,  et~al., 2024, \mn@doi [Nature Astronomy]
  {10.1038/s41550-023-02111-9}, \href
  {https://ui.adsabs.harvard.edu/abs/2024NatAs...8..126B} {8, 126}

\bibitem[\protect\citeauthoryear{{Bovy}, {Hogg}  \& {Roweis}}{{Bovy}
  et~al.}{2011a}]{Bovy_etal:2011}
{Bovy} J.,  {Hogg} D.~W.,   {Roweis} S.~T.,  2011a, \mn@doi [Annals of Applied
  Statistics] {10.1214/10-AOAS439}, 5, 1657

\bibitem[\protect\citeauthoryear{{Bovy} et~al.,}{{Bovy}
  et~al.}{2011b}]{Bovy_etal:2011b}
{Bovy} J.,  et~al., 2011b, \mn@doi [\apj] {10.1088/0004-637X/729/2/141}, \href
  {https://ui.adsabs.harvard.edu/abs/2011ApJ...729..141B} {729, 141}

\bibitem[\protect\citeauthoryear{{Bovy} et~al.,}{{Bovy}
  et~al.}{2012}]{Bovy_etal:2012}
{Bovy} J.,  et~al., 2012, \mn@doi [\apj] {10.1088/0004-637X/749/1/41}, \href
  {https://ui.adsabs.harvard.edu/abs/2012ApJ...749...41B} {749, 41}

\bibitem[\protect\citeauthoryear{{Bradley} et~al.,}{{Bradley}
  et~al.}{2020}]{Bradley_etal:2020}
{Bradley} L.,  et~al., 2020, {astropy/photutils: 1.0.0},
  \mn@doi{10.5281/zenodo.4044744}

\bibitem[\protect\citeauthoryear{{Bruzual} \& {Charlot}}{{Bruzual} \&
  {Charlot}}{2003}]{Bruzual_Charlot:2003}
{Bruzual} G.,  {Charlot} S.,  2003, \mn@doi [\mnras]
  {10.1046/j.1365-8711.2003.06897.x}, \href
  {https://ui.adsabs.harvard.edu/abs/2003MNRAS.344.1000B} {344, 1000}

\bibitem[\protect\citeauthoryear{Burnham}{Burnham}{2002}]{Burnham:2002}
Burnham K.~P.,  2002, Model Selection and Multimodel Inference: A Practical
  Information-Theoretic Approach, 2nd ed. edn.
Springer, New York

\bibitem[\protect\citeauthoryear{{Byrne}, {Meyer}, {Farina}, {Ba{\~n}ados},
  {Walter}, {Decarli}, {Belladitta}  \& {Loiacono}}{{Byrne}
  et~al.}{2024}]{Byrne_etal:2024}
{Byrne} X.,  {Meyer} R.~A.,  {Farina} E.~P.,  {Ba{\~n}ados} E.,  {Walter} F.,
  {Decarli} R.,  {Belladitta} S.,   {Loiacono} F.,  2024, \mn@doi [\mnras]
  {10.1093/mnras/stae902}, \href
  {https://ui.adsabs.harvard.edu/abs/2024MNRAS.tmp..910B} {p. in press}

\bibitem[\protect\citeauthoryear{Canny}{Canny}{1986}]{Canny:1968}
Canny J.,  1986, \mn@doi [Pattern Analysis and Machine Intelligence, IEEE
  Transactions on] {10.1109/TPAMI.1986.4767851}, PAMI-8, 679

\bibitem[\protect\citeauthoryear{Carballo, González-Serrano, Benn  \&
  Jiménez-Luján}{Carballo et~al.}{2008}]{Carballo_etal:2008}
Carballo R.,  González-Serrano J.~I.,  Benn C.~R.,   Jiménez-Luján F.,
  2008, \mn@doi [Monthly Notices of the Royal Astronomical Society]
  {10.1111/j.1365-2966.2008.13896.x}, 391, 369–382

\bibitem[\protect\citeauthoryear{{Carrasco} et~al.,}{{Carrasco}
  et~al.}{2015}]{Carrasco_etal:2015}
{Carrasco} D.,  et~al., 2015, \mn@doi [\aap] {10.1051/0004-6361/201525752},
  \href {https://ui.adsabs.harvard.edu/abs/2015A&A...584A..44C} {584, A44}

\bibitem[\protect\citeauthoryear{{Clarke}, {Scaife}, {Greenhalgh}  \&
  {Griguta}}{{Clarke} et~al.}{2020}]{Clarke_etal:2020}
{Clarke} A.~O.,  {Scaife} A.~M.~M.,  {Greenhalgh} R.,   {Griguta} V.,  2020,
  \mn@doi [\aap] {10.1051/0004-6361/201936770}, \href
  {https://ui.adsabs.harvard.edu/abs/2020A&A...639A..84C} {639, A84}

\bibitem[\protect\citeauthoryear{{Doert} \& {Errando}}{{Doert} \&
  {Errando}}{2014}]{Doert_Errando:2014}
{Doert} M.,  {Errando} M.,  2014, \mn@doi [\apj] {10.1088/0004-637X/782/1/41},
  \href {https://ui.adsabs.harvard.edu/abs/2014ApJ...782...41D} {782, 41}

\bibitem[\protect\citeauthoryear{{Dye} et~al.,}{{Dye}
  et~al.}{2006}]{Dye_etal:2006}
{Dye} S.,  et~al., 2006, \mn@doi [\mnras] {10.1111/j.1365-2966.2006.10928.x},
  \href {https://ui.adsabs.harvard.edu/abs/2006MNRAS.372.1227D} {372, 1227}

\bibitem[\protect\citeauthoryear{{Euclid Collaboration: Barnett}
  et~al.,}{{Euclid Collaboration: Barnett} et~al.}{2019}]{Barnett_etal:2019}
{Euclid Collaboration: Barnett} R.,  et~al., 2019, \mn@doi [\aap]
  {10.1051/0004-6361/201936427}, \href
  {https://ui.adsabs.harvard.edu/abs/2019A&A...631A..85E} {631, A85}

\bibitem[\protect\citeauthoryear{{Euclid Collaboration: Mellier}
  et~al.}{{Euclid Collaboration: Mellier} et~al.}{2024}]{Mellier_etal:2024}
{Euclid Collaboration: Mellier} Y.,  et~al., 2024, \mn@doi [arXiv e-prints]
  {10.48550/arXiv.2405.13491}, \href
  {https://ui.adsabs.harvard.edu/abs/2024arXiv240513491E} {p. arXiv:2405.13491}

\bibitem[\protect\citeauthoryear{{Euclid Collaboration: Scaramella}
  et~al.}{{Euclid Collaboration: Scaramella}
  et~al.}{2022}]{Scaramella_etal:2022}
{Euclid Collaboration: Scaramella} R.,  et~al., 2022, \mn@doi [\aap]
  {10.1051/0004-6361/202141938}, \href
  {https://ui.adsabs.harvard.edu/abs/2022A&A...662A.112E} {662, A112}

\bibitem[\protect\citeauthoryear{{Fan} et~al.,}{{Fan}
  et~al.}{2000}]{Fan_etal:2000}
{Fan} X.,  et~al., 2000, \mn@doi [\aj] {10.1086/301534}, 120, 1167

\bibitem[\protect\citeauthoryear{{Fan} et~al.,}{{Fan}
  et~al.}{2001}]{Fan_etal:2001}
{Fan} X.,  et~al., 2001, \mn@doi [\aj] {10.1086/324111}, \href
  {https://ui.adsabs.harvard.edu/abs/2001AJ....122.2833F} {122, 2833}

\bibitem[\protect\citeauthoryear{{Fan} et~al.,}{{Fan}
  et~al.}{2004}]{Fan_etal:2004}
{Fan} X.,  et~al., 2004, \mn@doi [\aj] {10.1086/422434}, \href
  {https://ui.adsabs.harvard.edu/abs/2004AJ....128..515F} {128, 515}

\bibitem[\protect\citeauthoryear{{Fan} et~al.,}{{Fan}
  et~al.}{2019a}]{Fan_etal:2019b}
{Fan} X.,  et~al., 2019a, \baas, \href
  {https://ui.adsabs.harvard.edu/abs/2019BAAS...51c.121F} {51, 121}

\bibitem[\protect\citeauthoryear{{Fan} et~al.,}{{Fan}
  et~al.}{2019b}]{Fan_etal:2019a}
{Fan} X.,  et~al., 2019b, \mn@doi [\apjl] {10.3847/2041-8213/aaeffe}, \href
  {https://ui.adsabs.harvard.edu/abs/2019ApJ...870L..11F} {870, L11}

\bibitem[\protect\citeauthoryear{{Fan}, {Ba{\~n}ados}  \& {Simcoe}}{{Fan}
  et~al.}{2023}]{Fan_etal:2023}
{Fan} X.,  {Ba{\~n}ados} E.,   {Simcoe} R.~A.,  2023, \mn@doi [\araa]
  {10.1146/annurev-astro-052920-102455}, \href
  {https://ui.adsabs.harvard.edu/abs/2023ARA&A..61..373F} {61, 373}

\bibitem[\protect\citeauthoryear{{Glikman}, {Eigenbrod}, {Djorgovski},
  {Meylan}, {Thompson}, {Mahabal}  \& {Courbin}}{{Glikman}
  et~al.}{2008}]{Glikman_etal:2008}
{Glikman} E.,  {Eigenbrod} A.,  {Djorgovski} S.~G.,  {Meylan} G.,  {Thompson}
  D.,  {Mahabal} A.,   {Courbin} F.,  2008, \mn@doi [\aj]
  {10.1088/0004-6256/136/3/954}, \href
  {https://ui.adsabs.harvard.edu/abs/2008AJ....136..954G} {136, 954}

\bibitem[\protect\citeauthoryear{{Gloudemans} et~al.,}{{Gloudemans}
  et~al.}{2022}]{Gloudemans_etal:2022}
{Gloudemans} A.~J.,  et~al., 2022, \mn@doi [\aap]
  {10.1051/0004-6361/202244763}, 668, A27

\bibitem[\protect\citeauthoryear{{Goodrich} et~al.,}{{Goodrich}
  et~al.}{2001}]{Goodrich_etal:2001}
{Goodrich} R.~W.,  et~al., 2001, \mn@doi [\apjl] {10.1086/324466}, \href
  {https://ui.adsabs.harvard.edu/abs/2001ApJ...561L..23G} {561, L23}

\bibitem[\protect\citeauthoryear{{Goto}}{{Goto}}{2006}]{Goto:2006}
{Goto} T.,  2006, \mn@doi [\mnras] {10.1111/j.1365-2966.2006.10702.x}, \href
  {https://ui.adsabs.harvard.edu/abs/2006MNRAS.371..769G} {371, 769}

\bibitem[\protect\citeauthoryear{{Goulding} et~al.,}{{Goulding}
  et~al.}{2012}]{Goulding_etal:2012}
{Goulding} N.~T.,  et~al., 2012, \mn@doi [\mnras]
  {10.1111/j.1365-2966.2012.21932.x}, \href
  {https://ui.adsabs.harvard.edu/abs/2012MNRAS.427.3358G} {427, 3358}

\bibitem[\protect\citeauthoryear{{Graham}, {Djorgovski}, {Drake}, {Mahabal},
  {Chang}, {Stern}, {Donalek}  \& {Glikman}}{{Graham}
  et~al.}{2014}]{Graham_etal:2014}
{Graham} M.~J.,  {Djorgovski} S.~G.,  {Drake} A.~J.,  {Mahabal} A.~A.,  {Chang}
  M.,  {Stern} D.,  {Donalek} C.,   {Glikman} E.,  2014, \mn@doi [\mnras]
  {10.1093/mnras/stt2499}, \href
  {https://ui.adsabs.harvard.edu/abs/2014MNRAS.439..703G} {439, 703}

\bibitem[\protect\citeauthoryear{{Gunn} \& {Peterson}}{{Gunn} \&
  {Peterson}}{1965}]{Gunn_Peterson:1965}
{Gunn} J.~E.,  {Peterson} B.~A.,  1965, \mn@doi [\apj] {10.1086/148444}, \href
  {https://ui.adsabs.harvard.edu/abs/1965ApJ...142.1633G} {142, 1633}

\bibitem[\protect\citeauthoryear{{Heintz}, {Fynbo}, {H{\o}g}, {M{\o}ller},
  {Krogager}, {Geier}, {Jakobsson}  \& {Christensen}}{{Heintz}
  et~al.}{2018}]{Heintz_etal:2018}
{Heintz} K.~E.,  {Fynbo} J.~P.~U.,  {H{\o}g} E.,  {M{\o}ller} P.,  {Krogager}
  J.~K.,  {Geier} S.,  {Jakobsson} P.,   {Christensen} L.,  2018, \mn@doi
  [\aap] {10.1051/0004-6361/201833396}, \href
  {https://ui.adsabs.harvard.edu/abs/2018A&A...615L...8H} {615, L8}

\bibitem[\protect\citeauthoryear{{Hernitschek} et~al.,}{{Hernitschek}
  et~al.}{2016}]{Hernitschek_etal:2016}
{Hernitschek} N.,  et~al., 2016, \mn@doi [\apj] {10.3847/0004-637X/817/1/73},
  \href {https://ui.adsabs.harvard.edu/abs/2016ApJ...817...73H} {817, 73}

\bibitem[\protect\citeauthoryear{{Hewett}, {Warren}, {Leggett}  \&
  {Hodgkin}}{{Hewett} et~al.}{2006}]{Hewett_etal:2006}
{Hewett} P.,  {Warren} S.~J.,  {Leggett} S.~K.,   {Hodgkin} S.,  2006, \mn@doi
  [\mnras] {10.1111/j.1365-2966.2005.09969.x}, \href
  {https://ui.adsabs.harvard.edu/abs/2006MNRAS.367..454H} {367, 454}

\bibitem[\protect\citeauthoryear{{Hopkins}, {Hernquist}, {Cox}, {Di Matteo},
  {Robertson}  \& {Springel}}{{Hopkins} et~al.}{2006}]{Hopkins_etal:2006}
{Hopkins} P.~F.,  {Hernquist} L.,  {Cox} T.~J.,  {Di Matteo} T.,  {Robertson}
  B.,   {Springel} V.,  2006, \mn@doi [\apjs] {10.1086/499298}, \href
  {https://ui.adsabs.harvard.edu/abs/2006ApJS..163....1H} {163, 1}

\bibitem[\protect\citeauthoryear{Hough}{Hough}{1959}]{Hough:1959}
Hough P. V.~C.,  1959, Conf. Proc. C, 590914, 554

\bibitem[\protect\citeauthoryear{{Ivezi{\'c}} et~al.,}{{Ivezi{\'c}}
  et~al.}{2008}]{Ivezic_etal:2008}
{Ivezi{\'c}} {\v Z}.,  et~al., 2008, \mn@doi [Serbian Astronomical Journal]
  {10.2298/SAJ0876001I}, 176, 1

\bibitem[\protect\citeauthoryear{{Jiang} et~al.,}{{Jiang}
  et~al.}{2016}]{Jiang_etal:2016}
{Jiang} L.,  et~al., 2016, \mn@doi [\apj] {10.3847/1538-4357/833/2/222}, \href
  {https://ui.adsabs.harvard.edu/abs/2016ApJ...833..222J} {833, 222}

\bibitem[\protect\citeauthoryear{{Jin}, {Zhang}, {Zhang}, {Zhao}, {Wu}  \&
  {Fan}}{{Jin} et~al.}{2019}]{Xin_etal:2019}
{Jin} X.,  {Zhang} Y.,  {Zhang} J.,  {Zhao} Y.,  {Wu} X.-b.,   {Fan} D.,  2019,
  \mn@doi [\mnras] {10.1093/mnras/stz680}, \href
  {https://ui.adsabs.harvard.edu/abs/2019MNRAS.485.4539J} {485, 4539}

\bibitem[\protect\citeauthoryear{{Kang}, {Fan}, {Mao}, {Wu}, {Feng}  \&
  {Yin}}{{Kang} et~al.}{2019}]{Kang_etal:2019}
{Kang} S.-J.,  {Fan} J.-H.,  {Mao} W.,  {Wu} Q.,  {Feng} J.,   {Yin} Y.,  2019,
  \mn@doi [\apj] {10.3847/1538-4357/ab0383}, \href
  {https://ui.adsabs.harvard.edu/abs/2019ApJ...872..189K} {872, 189}

\bibitem[\protect\citeauthoryear{{Kellermann}, {Sramek}, {Schmidt}, {Shaffer}
  \& {Green}}{{Kellermann} et~al.}{1989}]{Kellerman_etal:1989}
{Kellermann} K.~I.,  {Sramek} R.,  {Schmidt} M.,  {Shaffer} D.~B.,   {Green}
  R.,  1989, \mn@doi [\aj] {10.1086/115207}, \href
  {https://ui.adsabs.harvard.edu/abs/1989AJ.....98.1195K} {98, 1195}

\bibitem[\protect\citeauthoryear{{Kirkpatrick}, {Schlegel}, {Ross}, {Myers},
  {Hennawi}, {Sheldon}, {Schneider}  \& {Weaver}}{{Kirkpatrick}
  et~al.}{2011}]{Kirkpatrick_etal:2011}
{Kirkpatrick} J.~A.,  {Schlegel} D.~J.,  {Ross} N.~P.,  {Myers} A.~D.,
  {Hennawi} J.~F.,  {Sheldon} E.~S.,  {Schneider} D.~P.,   {Weaver} B.~A.,
  2011, \mn@doi [\apj] {10.1088/0004-637X/743/2/125}, \href
  {https://ui.adsabs.harvard.edu/abs/2011ApJ...743..125K} {743, 125}

\bibitem[\protect\citeauthoryear{{Koo} \& {Kron}}{{Koo} \&
  {Kron}}{1982}]{Koo_Kron:1982}
{Koo} D.~C.,  {Kron} R.~G.,  1982, \aap, \href
  {https://ui.adsabs.harvard.edu/abs/1982A&A...105..107K} {105, 107}

\bibitem[\protect\citeauthoryear{Krolik}{Krolik}{1999}]{Krolik:1999}
Krolik J.~H.,  1999, Active Galactic Nuclei: From the Central Black Holes to
  the Galactic Environment.
Princton University Press, Princeton

\bibitem[\protect\citeauthoryear{{Lang}, {Hogg}, {Jester}  \& {Rix}}{{Lang}
  et~al.}{2009}]{Lang_etal:2009}
{Lang} D.,  {Hogg} D.~W.,  {Jester} S.,   {Rix} H.-W.,  2009, \mn@doi [\aj]
  {10.1088/0004-6256/137/5/4400}, \href
  {https://ui.adsabs.harvard.edu/abs/2009AJ....137.4400L} {137, 4400}

\bibitem[\protect\citeauthoryear{{Laureijs} et~al.,}{{Laureijs}
  et~al.}{2011}]{Laureijs_etal:2011}
{Laureijs} R.,  et~al., 2011, arXiv e-prints, \href
  {https://ui.adsabs.harvard.edu/abs/2011arXiv1110.3193L} {p. arXiv:1110.3193}

\bibitem[\protect\citeauthoryear{{Lawrence} et~al.}{{Lawrence}
  et~al.}{2007}]{Lawrence_etal:2007}
{Lawrence} A.,  et~al., 2007, \mn@doi [\mnras]
  {10.1111/j.1365-2966.2007.12040.x}, 379, 1599

\bibitem[\protect\citeauthoryear{{Le}, {Pak}, {Jaffe}, {Kaplan}, {Lee}, {Im}
  \& {Seifahrt}}{{Le} et~al.}{2015}]{Le_etal:2015}
{Le} H. A.~N.,  {Pak} S.,  {Jaffe} D.~T.,  {Kaplan} K.,  {Lee} J.-J.,  {Im} M.,
    {Seifahrt} A.,  2015, \mn@doi [Advances in Space Research]
  {10.1016/j.asr.2015.03.007}, \href
  {https://ui.adsabs.harvard.edu/abs/2015AdSpR..55.2509L} {55, 2509}

\bibitem[\protect\citeauthoryear{{Lynden-Bell}}{{Lynden-Bell}}{1969}]{Lynden_Bell_1969}
{Lynden-Bell} D.,  1969, \mn@doi [\nat] {10.1038/223690a0}, \href
  {https://ui.adsabs.harvard.edu/abs/1969Natur.223..690L} {223, 690}

\bibitem[\protect\citeauthoryear{{Ma} et~al.,}{{Ma}
  et~al.}{2019}]{Ma_etal:2019}
{Ma} Z.,  et~al., 2019, \mn@doi [\apjs] {10.3847/1538-4365/aaf9a2}, \href
  {https://ui.adsabs.harvard.edu/abs/2019ApJS..240...34M} {240, 34}

\bibitem[\protect\citeauthoryear{{MacLeod} et~al.,}{{MacLeod}
  et~al.}{2012}]{MacLeod_etal:2012}
{MacLeod} C.~L.,  et~al., 2012, \mn@doi [\apj] {10.1088/0004-637X/753/2/106},
  \href {https://ui.adsabs.harvard.edu/abs/2012ApJ...753..106M} {753, 106}

\bibitem[\protect\citeauthoryear{{Maiolino} et~al.,}{{Maiolino}
  et~al.}{2024}]{Maiolino_etal:2024}
{Maiolino} R.,  et~al., 2024, \mn@doi [\nat] {10.1038/s41586-024-07052-5},
  \href {https://ui.adsabs.harvard.edu/abs/2024Natur.627...59M} {627, 59}

\bibitem[\protect\citeauthoryear{{Matsuoka} et~al.,}{{Matsuoka}
  et~al.}{2016}]{Matsuoka_etal:2016}
{Matsuoka} Y.,  et~al., 2016, \mn@doi [\apj] {10.3847/0004-637X/828/1/26},
  \href {https://ui.adsabs.harvard.edu/abs/2016ApJ...828...26M} {828, 26}

\bibitem[\protect\citeauthoryear{{Matsuoka} et~al.,}{{Matsuoka}
  et~al.}{2018}]{Matsuoka_etal:2018}
{Matsuoka} Y.,  et~al., 2018, \mn@doi [\apj] {10.3847/1538-4357/aaee7a}, \href
  {https://ui.adsabs.harvard.edu/abs/2018ApJ...869..150M} {869, 150}

\bibitem[\protect\citeauthoryear{{Matthews} \& {Sandage}}{{Matthews} \&
  {Sandage}}{1963}]{Matthews_etal:1963}
{Matthews} T.~A.,  {Sandage} A.~R.,  1963, \mn@doi [\apj] {10.1086/147615},
  \href {https://ui.adsabs.harvard.edu/abs/1963ApJ...138...30M} {138, 30}

\bibitem[\protect\citeauthoryear{{McGreer}, {Becker}, {Helfand}  \&
  {White}}{{McGreer} et~al.}{2006}]{McGreer_etal:2006}
{McGreer} I.~D.,  {Becker} R.~H.,  {Helfand} D.~J.,   {White} R.~L.,  2006,
  \mn@doi [\apj] {10.1086/507767}, \href
  {https://ui.adsabs.harvard.edu/abs/2006ApJ...652..157M} {652, 157}

\bibitem[\protect\citeauthoryear{{McLure} \& {Dunlop}}{{McLure} \&
  {Dunlop}}{2004}]{Mclure_Dunlop:2004}
{McLure} R.~J.,  {Dunlop} J.~S.,  2004, \mn@doi [\mnras]
  {10.1111/j.1365-2966.2004.08034.x}, \href
  {https://ui.adsabs.harvard.edu/abs/2004MNRAS.352.1390M} {352, 1390}

\bibitem[\protect\citeauthoryear{{Meusinger}, {Hinze}  \& {de
  Hoon}}{{Meusinger} et~al.}{2011}]{Meusinger_etal:2011}
{Meusinger} H.,  {Hinze} A.,   {de Hoon} A.,  2011, \mn@doi [\aap]
  {10.1051/0004-6361/201015520}, \href
  {https://ui.adsabs.harvard.edu/abs/2011A&A...525A..37M} {525, A37}

\bibitem[\protect\citeauthoryear{{Moffat}}{{Moffat}}{1969}]{Moffat:1969}
{Moffat} A.~F.~J.,  1969, \aap, \href
  {https://ui.adsabs.harvard.edu/abs/1969A&A.....3..455M} {3, 455}

\bibitem[\protect\citeauthoryear{{Mortlock}}{{Mortlock}}{2016}]{Mortlock:2016}
{Mortlock} D.,  2016, in {Mesinger} A.,  ed.,  Astrophysics and Space Science
  Library Vol. 423, Understanding the Epoch of Cosmic Reionization: Challenges
  and Progress. p.~187

\bibitem[\protect\citeauthoryear{{Mortlock} et~al.,}{{Mortlock}
  et~al.}{2009}]{Mortlock_etal:2009}
{Mortlock} D.~J.,  et~al., 2009, \mn@doi [\aap] {10.1051/0004-6361/200811161},
  \href {https://ui.adsabs.harvard.edu/abs/2009A&A...505...97M} {505, 97}

\bibitem[\protect\citeauthoryear{{Mortlock} et~al.,}{{Mortlock}
  et~al.}{2011}]{Mortlock_etal:2011}
{Mortlock} D.~J.,  et~al., 2011, \mn@doi [\nat] {10.1038/nature10159}, \href
  {https://ui.adsabs.harvard.edu/abs/2011Natur.474..616M} {474, 616}

\bibitem[\protect\citeauthoryear{{Mortlock}, {Patel}, {Warren}, {Hewett},
  {Venemans}, {McMahon}  \& {Simpson}}{{Mortlock}
  et~al.}{2012}]{Mortlock_etal:2012}
{Mortlock} D.~J.,  {Patel} M.,  {Warren} S.~J.,  {Hewett} P.~C.,  {Venemans}
  B.~P.,  {McMahon} R.~G.,   {Simpson} C.,  2012, \mn@doi [\mnras]
  {10.1111/j.1365-2966.2011.19710.x}, \href
  {https://ui.adsabs.harvard.edu/abs/2012MNRAS.419..390M} {419, 390}

\bibitem[\protect\citeauthoryear{{Nanni}, {Hennawi}, {Wang}, {Yang},
  {Schindler}  \& {Fan}}{{Nanni} et~al.}{2022}]{Nanni_etal:2022}
{Nanni} R.,  {Hennawi} J.~F.,  {Wang} F.,  {Yang} J.,  {Schindler} J.-T.,
  {Fan} X.,  2022, \mn@doi [\mnras] {10.1093/mnras/stac1944}, \href
  {https://ui.adsabs.harvard.edu/abs/2022MNRAS.515.3224N} {515, 3224}

\bibitem[\protect\citeauthoryear{{Neath}, {Flores}  \& {Cavanaugh}}{{Neath}
  et~al.}{2017}]{Neath_etal:2017}
{Neath} A.,  {Flores} J.,   {Cavanaugh} J.,  2017, \mn@doi [Wiley
  Interdisciplinary Reviews: Computational Statistics] {10.1002/wics.1420}, 10,
  e1420

\bibitem[\protect\citeauthoryear{{Oke} \& {Gunn}}{{Oke} \&
  {Gunn}}{1983}]{Oke_Gunn:1983}
{Oke} J.~B.,  {Gunn} J.~E.,  1983, \mn@doi [\apj] {10.1086/160817}, \href
  {https://ui.adsabs.harvard.edu/abs/1983ApJ...266..713O} {266, 713}

\bibitem[\protect\citeauthoryear{{Onken}, {Bian}, {Fan}, {Wang}, {Wolf}  \&
  {Yang}}{{Onken} et~al.}{2020}]{Onken_etal:2020}
{Onken} C.~A.,  {Bian} F.,  {Fan} X.,  {Wang} F.,  {Wolf} C.,   {Yang} J.,
  2020, \mn@doi [\mnras] {10.1093/mnras/staa1635}, \href
  {https://ui.adsabs.harvard.edu/abs/2020MNRAS.496.2309O} {496, 2309}

\bibitem[\protect\citeauthoryear{{Padovani}}{{Padovani}}{2011}]{Padovani:2011}
{Padovani} P.,  2011, \mn@doi [\mnras] {10.1111/j.1365-2966.2010.17789.x},
  \href {https://ui.adsabs.harvard.edu/abs/2011MNRAS.411.1547P} {411, 1547}

\bibitem[\protect\citeauthoryear{Padovani et~al.,}{Padovani
  et~al.}{2017}]{Padovani:2017}
Padovani P.,  et~al., 2017, \mn@doi [The Astronomy and Astrophysics Review]
  {10.1007/s00159-017-0102-9}, 25

\bibitem[\protect\citeauthoryear{{Peters} \& {Richards}}{{Peters} \&
  {Richards}}{2015}]{Peters_Richards:2015}
{Peters} C.~M.,  {Richards} G.,  2015, in IAU General Assembly. p. 2257231

\bibitem[\protect\citeauthoryear{{Peterson}}{{Peterson}}{1997}]{Peterson:1997}
{Peterson} B.~M.,  1997, An Introduction to Active Galactic Nuclei.
Cambridge University Press, \mn@doi{10.1017/CBO9781139170901}

\bibitem[\protect\citeauthoryear{Pipien, Cuby, Basa, Willott, Cuillandre,
  Arnouts  \& Hudelot}{Pipien et~al.}{2018}]{Pipien_etal:2018}
Pipien S.,  Cuby J.-G.,  Basa S.,  Willott C.~J.,  Cuillandre J.-C.,  Arnouts
  S.,   Hudelot P.,  2018, \mn@doi [Astronomy \& Astrophysics]
  {10.1051/0004-6361/201833488}, 617, A127

\bibitem[\protect\citeauthoryear{{Planck Collaboration}}{{Planck
  Collaboration}}{2020}]{Planck:2020}
{Planck Collaboration} 2020, \mn@doi [\aap] {10.1051/0004-6361/201833910},
  \href {https://ui.adsabs.harvard.edu/abs/2020A&A...641A...6P} {641, A6}

\bibitem[\protect\citeauthoryear{{Portillo}, {Speagle}  \&
  {Finkbeiner}}{{Portillo} et~al.}{2020}]{Portillo_etal:2020}
{Portillo} S. K.~N.,  {Speagle} J.~S.,   {Finkbeiner} D.~P.,  2020, \mn@doi
  [\aj] {10.3847/1538-3881/ab76ba}, \href
  {https://ui.adsabs.harvard.edu/abs/2020AJ....159..165P} {159, 165}

\bibitem[\protect\citeauthoryear{{Potts} \& {Villforth}}{{Potts} \&
  {Villforth}}{2021}]{Potts_etal:2021}
{Potts} B.,  {Villforth} C.,  2021, \mn@doi [\aap]
  {10.1051/0004-6361/202140597}, \href
  {https://ui.adsabs.harvard.edu/abs/2021A&A...650A..33P} {650, A33}

\bibitem[\protect\citeauthoryear{{Rankine}, {Matthews}, {Hewett}, {Banerji},
  {Morabito}  \& {Richards}}{{Rankine} et~al.}{2021}]{Rankine_etal:2021}
{Rankine} A.~L.,  {Matthews} J.~H.,  {Hewett} P.~C.,  {Banerji} M.,  {Morabito}
  L.~K.,   {Richards} G.~T.,  2021, \mn@doi [\mnras] {10.1093/mnras/stab302},
  \href {https://ui.adsabs.harvard.edu/abs/2021MNRAS.502.4154R} {502, 4154}

\bibitem[\protect\citeauthoryear{{Reed} et~al.,}{{Reed}
  et~al.}{2015}]{Reed_etal:2015}
{Reed} S.~L.,  et~al., 2015, \mn@doi [\mnras] {10.1093/mnras/stv2031}, \href
  {https://ui.adsabs.harvard.edu/abs/2015MNRAS.454.3952R} {454, 3952}

\bibitem[\protect\citeauthoryear{{Reed} et~al.,}{{Reed}
  et~al.}{2017}]{Reed_etal:2017}
{Reed} S.~L.,  et~al., 2017, \mn@doi [\mnras] {10.1093/mnras/stx728}, \href
  {https://ui.adsabs.harvard.edu/abs/2017MNRAS.468.4702R} {468, 4702}

\bibitem[\protect\citeauthoryear{{Reed} et~al.,}{{Reed}
  et~al.}{2019}]{Reed_etal:2019}
{Reed} S.~L.,  et~al., 2019, \mn@doi [\mnras] {10.1093/mnras/stz1341}, \href
  {https://ui.adsabs.harvard.edu/abs/2019MNRAS.487.1874R} {487, 1874}

\bibitem[\protect\citeauthoryear{{Richards} et~al.,}{{Richards}
  et~al.}{2002}]{Richards_etal:2002}
{Richards} G.~T.,  et~al., 2002, \mn@doi [\aj] {10.1086/340187}, \href
  {https://ui.adsabs.harvard.edu/abs/2002AJ....123.2945R} {123, 2945}

\bibitem[\protect\citeauthoryear{{Richards} et~al.,}{{Richards}
  et~al.}{2004}]{Richards_etal:2004}
{Richards} G.~T.,  et~al., 2004, \mn@doi [\apjs] {10.1086/425356}, \href
  {https://ui.adsabs.harvard.edu/abs/2004ApJS..155..257R} {155, 257}

\bibitem[\protect\citeauthoryear{{Richards} et~al.,}{{Richards}
  et~al.}{2009}]{Richards_etal:2009}
{Richards} G.~T.,  et~al., 2009, \mn@doi [\aj] {10.1088/0004-6256/137/4/3884},
  \href {https://ui.adsabs.harvard.edu/abs/2009AJ....137.3884R} {137, 3884}

\bibitem[\protect\citeauthoryear{{Richards} et~al.,}{{Richards}
  et~al.}{2015}]{Richards_etal:2015}
{Richards} G.~T.,  et~al., 2015, \mn@doi [\apjs] {10.1088/0067-0049/219/2/39},
  \href {https://ui.adsabs.harvard.edu/abs/2015ApJS..219...39R} {219, 39}

\bibitem[\protect\citeauthoryear{{Ruan} et~al.,}{{Ruan}
  et~al.}{2016}]{Ruan_etal:2016}
{Ruan} J.~J.,  et~al., 2016, \mn@doi [\apj] {10.3847/0004-637X/826/2/188},
  \href {https://ui.adsabs.harvard.edu/abs/2016ApJ...826..188R} {826, 188}

\bibitem[\protect\citeauthoryear{{Sandage} \& {Wyndham}}{{Sandage} \&
  {Wyndham}}{1965}]{Sandage_Wyndham:1965}
{Sandage} A.,  {Wyndham} J.~D.,  1965, \mn@doi [\apj] {10.1086/148125}, \href
  {https://ui.adsabs.harvard.edu/abs/1965ApJ...141..328S} {141, 328}

\bibitem[\protect\citeauthoryear{{Schindler} et~al.,}{{Schindler}
  et~al.}{2019}]{Schindler_etal:2019}
{Schindler} J.-T.,  et~al., 2019, \mn@doi [\apjs] {10.3847/1538-4365/ab20d0},
  \href {https://ui.adsabs.harvard.edu/abs/2019ApJS..243....5S} {243, 5}

\bibitem[\protect\citeauthoryear{{Schmidt}}{{Schmidt}}{1969}]{Schmidt:1969}
{Schmidt} M.,  1969, \mn@doi [\araa] {10.1146/annurev.aa.07.090169.002523},
  \href {https://ui.adsabs.harvard.edu/abs/1969ARA&A...7..527S} {7, 527}

\bibitem[\protect\citeauthoryear{{Shu}, {Koposov}, {Evans}, {Belokurov},
  {McMahon}, {Auger}  \& {Lemon}}{{Shu} et~al.}{2019}]{Shu_etal:2019}
{Shu} Y.,  {Koposov} S.~E.,  {Evans} N.~W.,  {Belokurov} V.,  {McMahon} R.~G.,
  {Auger} M.~W.,   {Lemon} C.~A.,  2019, \mn@doi [\mnras]
  {10.1093/mnras/stz2487}, \href
  {https://ui.adsabs.harvard.edu/abs/2019MNRAS.489.4741S} {489, 4741}

\bibitem[\protect\citeauthoryear{{Sivia} \& {Skilling}}{{Sivia} \&
  {Skilling}}{2006}]{Sivia_Skilling:2006}
{Sivia} D.~S.,  {Skilling} J.,  2006, Data Analysis: A Bayesian tutorial, 2nd
  edn.
Oxford University Press, Oxford

\bibitem[\protect\citeauthoryear{{Smith} \& {Hoffleit}}{{Smith} \&
  {Hoffleit}}{1963}]{Smith_Hoffleit:1963}
{Smith} H.~J.,  {Hoffleit} D.,  1963, \mn@doi [\aj] {10.1086/109152}, \href
  {https://ui.adsabs.harvard.edu/abs/1963AJ.....68S.292S} {68, 292}

\bibitem[\protect\citeauthoryear{{Stern} et~al.,}{{Stern}
  et~al.}{2007}]{Stern_etal:2007}
{Stern} D.,  et~al., 2007, \mn@doi [\apj] {10.1086/516833}, \href
  {https://ui.adsabs.harvard.edu/abs/2007ApJ...663..677S} {663, 677}

\bibitem[\protect\citeauthoryear{{Tantalo}, {Chinellato}, {Merlin}, {Piovan}
  \& {Chiosi}}{{Tantalo} et~al.}{2010}]{Tantalo_etal:2010}
{Tantalo} R.,  {Chinellato} S.,  {Merlin} E.,  {Piovan} L.,   {Chiosi} C.,
  2010, \mn@doi [\aap] {10.1051/0004-6361/200912709}, \href
  {https://ui.adsabs.harvard.edu/abs/2010A&A...518A..43T} {518, A43}

\bibitem[\protect\citeauthoryear{{Tee}, {Fan}, {Wang}, {Yang}, {Malhotra}  \&
  {Rhoads}}{{Tee} et~al.}{2023}]{Tee_etal:2023}
{Tee} W.~L.,  {Fan} X.,  {Wang} F.,  {Yang} J.,  {Malhotra} S.,   {Rhoads}
  J.~E.,  2023, \mn@doi [\apj] {10.3847/1538-4357/acf12d}, \href
  {https://ui.adsabs.harvard.edu/abs/2023ApJ...956...52T} {956, 52}

\bibitem[\protect\citeauthoryear{{Temple}, {Hewett}  \& {Banerji}}{{Temple}
  et~al.}{2021}]{Temple_etal:2021}
{Temple} M.~J.,  {Hewett} P.~C.,   {Banerji} M.,  2021, \mn@doi [\mnras]
  {10.1093/mnras/stab2586}, \href
  {https://ui.adsabs.harvard.edu/abs/2021MNRAS.508..737T} {508, 737}

\bibitem[\protect\citeauthoryear{{Vanden Berk} et~al.,}{{Vanden Berk}
  et~al.}{2004}]{Vanden_Berk_etal:2004}
{Vanden Berk} D.~E.,  et~al., 2004, \mn@doi [\apj] {10.1086/380563}, \href
  {https://ui.adsabs.harvard.edu/abs/2004ApJ...601..692V} {601, 692}

\bibitem[\protect\citeauthoryear{{Venemans}, {McMahon}, {Warren},
  {Gonzalez-Solares}, {Hewett}, {Mortlock}, {Dye}  \& {Sharp}}{{Venemans}
  et~al.}{2007}]{Venemans_etal:2007}
{Venemans} B.~P.,  {McMahon} R.~G.,  {Warren} S.~J.,  {Gonzalez-Solares} E.~A.,
   {Hewett} P.~C.,  {Mortlock} D.~J.,  {Dye} S.,   {Sharp} R.~G.,  2007,
  \mn@doi [\mnras] {10.1111/j.1745-3933.2007.00290.x}, \href
  {https://ui.adsabs.harvard.edu/abs/2007MNRAS.376L..76V} {376, L76}

\bibitem[\protect\citeauthoryear{{Venemans} et~al.,}{{Venemans}
  et~al.}{2013}]{Venemans_etal:2013}
{Venemans} B.~P.,  et~al., 2013, \mn@doi [\apj] {10.1088/0004-637X/779/1/24},
  \href {https://ui.adsabs.harvard.edu/abs/2013ApJ...779...24V} {779, 24}

\bibitem[\protect\citeauthoryear{{Wagenveld}, {Saxena}, {Duncan},
  {R{\"o}ttgering}  \& {Zhang}}{{Wagenveld} et~al.}{2022}]{Wagenveld_etal:2022}
{Wagenveld} J.~D.,  {Saxena} A.,  {Duncan} K.~J.,  {R{\"o}ttgering} H.~J.~A.,
  {Zhang} M.,  2022, \mn@doi [\aap] {10.1051/0004-6361/202142445}, \href
  {https://ui.adsabs.harvard.edu/abs/2022A&A...660A..22W} {660, A22}

\bibitem[\protect\citeauthoryear{Wang et~al.,}{Wang
  et~al.}{2016}]{Wang_etal:2016}
Wang F.,  et~al., 2016, \mn@doi [The Astrophysical Journal]
  {10.3847/0004-637x/819/1/24}, 819, 24

\bibitem[\protect\citeauthoryear{{Wang} et~al.,}{{Wang}
  et~al.}{2021}]{Wang_etal:2021}
{Wang} F.,  et~al., 2021, \mn@doi [\apjl] {10.3847/2041-8213/abd8c6}, \href
  {https://ui.adsabs.harvard.edu/abs/2021ApJ...907L...1W} {907, L1}

\bibitem[\protect\citeauthoryear{{Warren}, {Hewett}, {Irwin}, {McMahon}  \&
  {Bridgeland}}{{Warren} et~al.}{1987}]{Warren_etal:1987}
{Warren} S.~J.,  {Hewett} P.~C.,  {Irwin} M.~J.,  {McMahon} R.~G.,
  {Bridgeland} M.~T.,  1987, \mn@doi [\nat] {10.1038/325131a0}, \href
  {https://ui.adsabs.harvard.edu/abs/1987Natur.325..131W} {325, 131}

\bibitem[\protect\citeauthoryear{{Warren}, {Hewett}  \& {Osmer}}{{Warren}
  et~al.}{1994}]{Warren_etal:1994}
{Warren} S.~J.,  {Hewett} P.~C.,   {Osmer} P.~S.,  1994, \mn@doi [\apj]
  {10.1086/173660}, \href
  {https://ui.adsabs.harvard.edu/abs/1994ApJ...421..412W} {421, 412}

\bibitem[\protect\citeauthoryear{Weedman}{Weedman}{1986}]{Weedman:1986}
Weedman D.~W.,  1986, Quasar Astronomy.
Cambridge University Press, \mn@doi{10.1017/CBO9780511600173}

\bibitem[\protect\citeauthoryear{{Wenzl} et~al.,}{{Wenzl}
  et~al.}{2021}]{Wenzl_etal:2021}
{Wenzl} L.,  et~al., 2021, \mn@doi [\aj] {10.3847/1538-3881/ac0254}, \href
  {https://ui.adsabs.harvard.edu/abs/2021AJ....162...72W} {162, 72}

\bibitem[\protect\citeauthoryear{{Williams}}{{Williams}}{2018}]{Williams:2018}
{Williams} P. K.~G.,  2018, in {Deeg} H.~J.,  {Belmonte} J.~A.,  eds, ,
  Handbook of Exoplanets.
Springer, p.~171

\bibitem[\protect\citeauthoryear{{Willott} et~al.,}{{Willott}
  et~al.}{2007}]{Willott_etal:2007}
{Willott} C.~J.,  et~al., 2007, \mn@doi [\aj] {10.1086/522962}, \href
  {https://ui.adsabs.harvard.edu/abs/2007AJ....134.2435W} {134, 2435}

\bibitem[\protect\citeauthoryear{{Wilson} \& {Colbert}}{{Wilson} \&
  {Colbert}}{1995}]{Wilson_Colbert:1995}
{Wilson} A.~S.,  {Colbert} E.~J.~M.,  1995, \mn@doi [\apj] {10.1086/175054},
  \href {https://ui.adsabs.harvard.edu/abs/1995ApJ...438...62W} {438, 62}

\bibitem[\protect\citeauthoryear{{Wolf}, {Lai}, {Onken}, {Amrutha}, {Bian},
  {Hon}, {Tisserand}  \& {Webster}}{{Wolf} et~al.}{2024}]{Wolf_etal:2024}
{Wolf} C.,  {Lai} S.,  {Onken} C.~A.,  {Amrutha} N.,  {Bian} F.,  {Hon} W.~J.,
  {Tisserand} P.,   {Webster} R.~L.,  2024, \mn@doi [Nature Astronomy]
  {10.1038/s41550-024-02195-x}, \href
  {https://ui.adsabs.harvard.edu/abs/2024NatAs...8..520W} {8, 520}

\bibitem[\protect\citeauthoryear{{Wyithe} \& {Loeb}}{{Wyithe} \&
  {Loeb}}{2002}]{Wyithe_Loeb:2002}
{Wyithe} J. S.~B.,  {Loeb} A.,  2002, \mn@doi [\apj] {10.1086/342181}, \href
  {https://ui.adsabs.harvard.edu/abs/2002ApJ...577...57W} {577, 57}

\bibitem[\protect\citeauthoryear{{Yang} et~al.,}{{Yang}
  et~al.}{2020}]{Yang_etal:2020}
{Yang} J.,  et~al., 2020, \mn@doi [\apjl] {10.3847/2041-8213/ab9c26}, \href
  {https://ui.adsabs.harvard.edu/abs/2020ApJ...897L..14Y} {897, L14}

\bibitem[\protect\citeauthoryear{{Y{\`e}che} et~al.,}{{Y{\`e}che}
  et~al.}{2010}]{Yeche_etal:2010}
{Y{\`e}che} C.,  et~al., 2010, \mn@doi [\aap] {10.1051/0004-6361/200913508},
  \href {https://ui.adsabs.harvard.edu/abs/2010A&A...523A..14Y} {523, A14}

\bibitem[\protect\citeauthoryear{{York} et~al.}{{York}
  et~al.}{2000}]{York_etal:2000}
{York} D.~G.,  et~al., 2000, \mn@doi [\aj] {10.1086/301513}, 120, 1579

\makeatother
\end{thebibliography}


\appendix


\section{Astronomical population models}
\label{sec:models}

Both the Bayesian model comparison calculation in Section~\ref{sec:bayesian_model_comparison} and the goodness-of-fit test in Section~\ref{sec:pixel} use models of the quasar, MLTY dwarf and ETG populations.  These are based on the models fully described in \cite{Mortlock_etal:2012}, \cite{Barnett_etal:2019} and \cite{Barnett_etal:2021}, but with some refinements as will be required for {\em Euclid} and LSST.  The models are also augmented to include morphology and proper motion as required for the full image-level data model described in Section~\ref{sec:data_model}.


\subsection{High-redshift quasars}
\label{section:model_q}

High-redshift quasars have largely self-similar spectra, so can be parametrised primarily by redshift, $z$, and absolute magnitude, $M_{1450}$; but we do consider a range of nine SED templates with different continuum slopes and emission line strengths taken from \citet{Hewett_etal:2006} and updated by \cite{Temple_etal:2021}. Our SED models are consistent with those employed by \cite{Temple_etal:2021}, which reproduce the optical through NIR photometric properties of the $\sim$750,000 quasars in the SDSS DR16 quasar catalogue to within 0.1~mag for redshifts $0 < z < 5$.
This leads to the parametrisation $\params_{\textrm{q}} = (z, M_{1450}, T)$, where $T$ indexes the discrete SEDs and is averaged over when evaluating Eq.~\ref{eq:likelihood_fluxes}.  The population model used combines the best-fit luminosity function from \cite{Matsuoka_etal:2018} with the redshift evolution measured by \cite{Jiang_etal:2016}.

All high-redshift quasars are assumed to be point-sources. This point-source assumption could reasonably be relaxed to search for gravitationally-lensed sources \citep{Fan_etal:2019a, Byrne_etal:2024}.

All high-redshift quasars are assumed to be stationary sources.


\subsection{MLTY dwarfs}
\label{section:model_s}

MLTY dwarfs have long been recognised as the main contaminant for high-redshift quasar searches \citep{Fan_etal:2001} and there are hence well-developed population models for their numbers and colours at optical and NIR wavelengths.  Here we adopt the population model described by \cite{Barnett_etal:2019}, with MLTY dwarfs parameterized by the $J$-band apparent magnitude and spectral type, $T$, so that $\params_{\textrm{s}} = (J, T)$.

All MLTY dwarfs are assumed to be point-sources.  

No physical model is adopted for the proper motion of MLTY dwarfs as this would hugely complicate the analysis, coupling survey cadences with a Galactic model.  Rather, the simulations described in Section~\ref{sec:simulations} are emperical, with a range of different angular offsets chosen directly.


\subsection{Early-type galaxies}
\label{section:model_g}

It is only ETGs in the redshift range $1 \lesssim z \lesssim 2$ that have red optical-NIR colours similar to high-redshift quasars, so we require a model only of this restricted part of the galaxy population.  The population model, given in Eq.~B1 of \cite{Barnett_etal:2021}, is empirical and parameterized by apparent $J$-band magnitude, $J$, and redshift, $z$, so $\params_{\textrm{g}} = (J, z)$.  The colours of the ETGs are obtained by adopting the evolutionary models of \cite{Bruzual_Charlot:2003} with formation redshifts of $z_{\textrm{f}} = 2$ and $z_{\textrm{f}} = 10$, as detailed further in \cite{Barnett_etal:2021}.

The ETGs with similar colours to $z \gtrsim 7$ quasars will be partly resolved in the {\em Euclid} data \citep{Barnett_etal:2019}, so we simulate extended contaminants in Section~\ref{sec:simulations}.  Given the moderate level to which they are expected to be resolved there is no need to convolve realistic intrinsic light profiles with the PSF; we instead model their PSF-convolved profiles as a mixture of two Gaussian functions according to
\begin{equation}
\psf(r; F_1, \sigma_1, \sigma_2)  
  = \frac{F_1}{\pi \, \sigma_1^2} 
  \, e^{- (r / \sigma_1)^2}
  + 
  \frac{1 - F_1}{\pi \, \sigma_2^2}
  \, e^{-(r / \sigma_2)^2}
  .
\end{equation}
For {\em Euclid}, for which the PSF in the NIR bands has a characteristic scale of $\sim\!0\farcs3$ \citep{Mellier_etal:2024}, we adopt default parameters $\sigma_1=0\farcs4$, $\sigma_2=0\farcs6$ and $F_1=0.5$; these would need to be adjusted for different surveys. 

All ETGs are assumed to be stationary sources.  


\end{document}